\documentclass[twocolumn]{aastex62}

\usepackage{capt-of}
\usepackage{tabularx,colortbl}
\usepackage{enumitem}

\begin{document}

\title{THE NANOGRAV 11-YEAR DATA SET: PULSE PROFILE VARIABILITY}

\correspondingauthor{Paul Brook}
\email{paul.brook@nanograv.org}

\author{P.~R.~Brook}
\affiliation{Department of Physics and Astronomy, West Virginia University, Morgantown, WV 26506, USA}
\affiliation{Center for Gravitational Waves and Cosmology, West Virginia University, Chestnut Ridge Research Building, Morgantown, WV 26505}
\author{A.~Karastergiou}
\affiliation{Astrophysics, University of Oxford, Denys Wilkinson Building, Keble Road, Oxford OX1 3RH, UK}
\affiliation{Physics Department, University of the Western Cape, Cape Town 7535, South Africa}
\affiliation{Department of Physics and Electronics, Rhodes University, PO Box 94, Grahamstown 6140, South Africa}
\author{M.~A.~McLaughlin}
\affiliation{Department of Physics and Astronomy, West Virginia University, Morgantown, WV 26506, USA}
\affiliation{Center for Gravitational Waves and Cosmology, West Virginia University, Chestnut Ridge Research Building, Morgantown, WV 26505}
\author{M.~T.~Lam}
\altaffiliation{NANOGrav Physics Frontiers Center Postdoctoral Fellow}
\affiliation{Department of Physics and Astronomy, West Virginia University, Morgantown, WV 26506, USA}
\affiliation{Center for Gravitational Waves and Cosmology, West Virginia University, Chestnut Ridge Research Building, Morgantown, WV 26505}
\author{Z.~Arzoumanian}
\affiliation{X-Ray Astrophysics Laboratory, NASA Goddard Space Flight Center, Code 662, Greenbelt, MD 20771, USA}
\author{S.~Chatterjee}
\affiliation{Department of Astronomy and Cornell Center for Astrophysics and Planetary Science, Cornell University, Ithaca, NY 14853, USA}
\author{J.~M.~Cordes}
\affiliation{Department of Astronomy and Cornell Center for Astrophysics and Planetary Science, Cornell University, Ithaca, NY 14853, USA}
\author{K.~Crowter}
\affiliation{Department of Physics and Astronomy, University of British Columbia, 6224 Agricultural Road, Vancouver, BC V6T 1Z1, Canada}
\author{M.~DeCesar}
\affiliation{Department of Physics, Lafayette College, Easton, PA 18042, USA}
\author{P.~B.~Demorest}
\affiliation{National Radio Astronomy Observatory, 1003 Lopezville Road, Socorro, NM, 87801, USA}
\author{T.~Dolch}
\affiliation{Department of Physics, Hillsdale College, 33 E. College Street, Hillsdale, MI 49242, USA}
\author{J.~A.~Ellis}
\affiliation{Department of Physics and Astronomy, West Virginia University, Morgantown, WV 26506, USA}
\affiliation{Center for Gravitational Waves and Cosmology, West Virginia University, Chestnut Ridge Research Building, Morgantown, WV 26505}
\author{R.~D.~Ferdman}
\affiliation{Department of Physics, University of East Anglia, Norwich, NR4 7TJ, UK}
\author{E.~Ferrara}
\affiliation{NASA Goddard Space Flight Center, Greenbelt, MD 20771, USA}
\author{E.~Fonseca}
\affiliation{Department of Physics, McGill University, 3600 rue Universite, Montreal, QC H3A 2T8, Canada}
\author{P.~A.~Gentile}
\affiliation{Department of Physics and Astronomy, West Virginia University, Morgantown, WV 26506, USA}
\affiliation{Center for Gravitational Waves and Cosmology, West Virginia University, Chestnut Ridge Research Building, Morgantown, WV 26505}
\author{G.~Jones}
\affiliation{Center for Gravitation, Cosmology and Astrophysics, Department of Physics, University of Wisconsin-Milwaukee, P.O. Box 413, Milwaukee, WI 53201, USA}
\author{M.~L.~Jones}
\affiliation{Department of Physics and Astronomy, West Virginia University, Morgantown, WV 26506, USA}
\affiliation{Center for Gravitational Waves and Cosmology, West Virginia University, Chestnut Ridge Research Building, Morgantown, WV 26505}
\author{T.~J.~W.~Lazio}
\affiliation{Jet Propulsion Laboratory, California Institute of Technology, 4800 Oak Grove Dr. Pasadena CA, 91109, USA}
\affiliation{Theoretical AstroPhysics Including Relativity (TAPIR), MC 350-17, California Institute of Technology, Pasadena, California 91125, USA}
\author{L.~Levin}
\affiliation{Jodrell Bank Centre for Astrophysics, Alan Turing Building, School of Physics and Astronomy, The University of Manchester, Oxford Road, Manchester, M13 9PL, UK}
\author{D.~R.~Lorimer}
\affiliation{Department of Physics and Astronomy, West Virginia University, Morgantown, WV 26506, USA}
\affiliation{Center for Gravitational Waves and Cosmology, West Virginia University, Chestnut Ridge Research Building, Morgantown, WV 26505}
\author{R.~S.~Lynch}
\affiliation{Green Bank Observatory, P.O. Box 2, Green Bank, WV 24944, USA}
\author{C.~Ng}
\affiliation{Department of Physics and Astronomy, University of British Columbia, 6224 Agricultural Road, Vancouver, BC V6T 1Z1, Canada}
\affiliation{Dunlap Institute for Astronomy \& Astrophysics, University of Toronto, 50 St. George Street, Toronto, ON M5S 3H4, Canada}
\author{D.~J.~Nice}
\affiliation{Department of Physics, Lafayette College, Easton, PA 18042, USA}
\author{T.~T.~Pennucci}
\affiliation{E\"otv\"os Lor\'and University, Budapest, Hungary}
\affiliation{Hungarian Academy of Sciences MTA-ELTE Extragalactic Astrophysics Research Group, 1117 Budapest, Hungary}
\author{S.~M.~Ransom}
\affiliation{National Radio Astronomy Observatory, 520 Edgemont Road, Charlottesville, VA 22903, USA}
\author{P. S. Ray}
\affiliation{Naval Research Laboratory, Washington DC 20375, USA}
\author{R. Spiewak}
\affiliation{Center for Gravitation, Cosmology and Astrophysics, Department of Physics, University of Wisconsin-Milwaukee, P.O. Box 413, Milwaukee, WI 53201, USA}
\affiliation{Centre for Astrophysics and Supercomputing, Swinburne University of Technology, PO Box 218, Hawthorn VIC 3122, Australia}
\author{I.~H.~Stairs}
\affiliation{Department of Physics and Astronomy, University of British Columbia, 6224 Agricultural Road, Vancouver, BC V6T 1Z1, Canada}
\author{D.~R.~Stinebring}
\affiliation{Department of Physics \& Astronomy, Oberlin College, Oberlin, OH 44074, USA}
\author{K.~Stovall}
\affiliation{National Radio Astronomy Observatory, 1003 Lopezville Road, Socorro, NM, 87801, USA}
\affiliation{Department of Physics and Astronomy, University of New Mexico, Albuquerque, NM, 87131, USA}
\author{J.~K.~Swiggum}
\altaffiliation{NANOGrav Physics Frontiers Center Postdoctoral Fellow}
\affiliation{Center for Gravitation, Cosmology and Astrophysics, Department of Physics, University of Wisconsin-Milwaukee, P.O. Box 413, Milwaukee, WI 53201, USA}
\author{W.~W.~Zhu}
\affiliation{National Astronomical Observatories, Chinese Academy of Science, 20A Datun Road, Chaoyang District, Beijing 100012, China}
\affiliation{Max-Planck-Institut f\"{u}r Radioastronomie, Auf dem H\"{u}gel 69, D-53121, Bonn, Germany}  

\begin{abstract}
  Access to 50 years of data has led to the discovery of pulsar
  emission and rotation variability on timescales of months and
  years. Most of this long-term variability has been seen in
  long-period pulsars, with relatively little focus on recycled
  millisecond pulsars. We have analyzed a 38-pulsar subset of the
  45 millisecond pulsars in the NANOGrav 11-year data set, in order to
  review their pulse profile stability. The most variability, on any
  timescale, is seen in PSRs~J1713$+$0747, B1937$+$21 and
  J2145$-$0750. The strongest evidence for long-timescale pulse
  profile changes is seen in PSRs~B1937$+$21 and J1643$-$1224. We have
  focused our analyses on these four pulsars in an attempt to
  elucidate the causes of their profile variability. Effects of
  scintillation seem to be responsible for the profile modifications
  of PSR~J2145$-$0750. We see evidence that imperfect polarization
  calibration contributes to the profile variability of
  PSRs~J1713$+$0747 and B1937$+$21, along with radio frequency
  interference around 2~GHz, but find that propagation effects also
  have an influence. The changes seen in PSR~J1643$-$1224 have been
  reported previously, yet elude explanation beyond their
  astrophysical nature. Regardless of cause, unmodeled pulse profile
  changes are detrimental to the accuracy of pulsar timing and must be
  incorporated into the timing models where possible.
\end{abstract}

\keywords{ISM: general - pulsars: general - pulsars: individual (J1643$-$1224, J1713$+$0747, B1937$+$21, J2145$-$0750) - stars: neutron}

\section{Introduction}
\label{intro}
The radio emission from a pulsar can vary over a wide range of
timescales. In practically all pulsars, the rotational phase, shape,
and amplitude of individual radio pulses are known to vary
considerably from one to the next with each rotation
\citep[e.g.][]{1971MNRAS.153..337L,1975ApJ...195..513T}. The average
shape of a few thousand pulses, however, is typically very stable and
known as the \emph{pulse profile} \citep[e.g.][]{1975ApJ...198..661H,1995ApJ...452..814R}. Soon after pulsars were discovered,
however, changes in some pulse profiles were seen on short timescales
in the form of \emph{mode-changing} and \emph{nulling} \citep{1970Natur.228...42B,1970Natur.228.1297B}. Mode-changing
is a phenomenon in which pulsars switch between two or more
quasi-stable emission states on timescales ranging from a few pulse
periods to hours and days. In a nulling pulsar, one of these states
shows little or no emission.

Pulsar data have now been collected for over 50 years. This allows
us to also identify longer-term pulse profile variability. In 2006, the first known \emph{intermittent pulsar} was
identified by \citet{2006Sci...312..549K}. Intermittent pulsars go through
a quasi-periodic cycle between phases in which radio emission is, and
is not, detected. The timescale of this behavior ranges from months to
years \citep{2012ApJ...746...63C,
  2012ApJ...758..141L,2017ApJ...834...72L}. In intermittent pulsars,
each state is associated with a different rate of
rotational velocity loss, known as the \emph{spindown rate}
$\dot{\nu}$. Kramer et al.\ attribute $\dot{\nu}$ variations to global
changes in magnetospheric particle currents; changing numbers of
charged particles at the polar cap would simultaneously affect the
pulsar's radio emission. More links between pulse profile and rotation
were provided in \citet{2010Sci...329..408L}, an analysis that showed
six pulsars for which $\dot{\nu}$ is correlated with changes in pulse shape over months and years. Further notable
examples of long-term variability, including pulse profile and
spindown correlation, continue to be found
\citep[e.g.][]{2014ApJ...780L..31B, 2016MNRAS.456.1374B}. All of the
examples of long-term pulse profile variability given above are found
in long-period pulsars (typically defined as those with spin periods
above around 30~ms and those that have not been spun up or \emph{recycled} through the
accretion of matter from a companion star); relatively little work has been done
regarding the long-term pulse profile variability of millisecond
pulsars (MSPs). The issue of stability for MSPs is particularly
important, however, as they are employed as high-precision timing
tools that can facilitate fundamental studies of physics. For example,
MSPs are used in \emph{pulsar timing arrays} in an attempt to detect
gravitational waves at nanohertz frequencies
\citep{2013CQGra..30v4007H, 2013CQGra..30v4009K,
  2013CQGra..30v4008M}. MSPs are suitable for this role as they are
known to be more rotationally stable than long-period pulsars
due to their high angular momentum.
Additionally, the time of arrival (TOA)
of a pulse from an MSP can be measured with more precision than that
of a long-period pulsar, as the uncertainty is proportional to
the temporal width of a pulsar's pulse profile.

A pulse TOA is measured by a process of \emph{template matching}
\citep{Taylor117, 2006ApJ...642.1004V}. The stability of a pulse profile
at a given frequency permits the cross-correlation of an observed
profile with a high signal-to-noise ratio (S/N) template, to provide a
TOA of the former. The template is either the average of many previous
observations, or a noise-free model of this average. Therefore, any
unmodeled pulse profile changes will result in inaccurate pulse TOAs,
which are detrimental to an MSP's utility as a timing tool.

Pulse profile variability can be caused by any of the following:
intrinsic changes in the pulsar and/or its magnetosphere, geodetic
precession \citep{1998ApJ...509..856K,2005ApJ...624..906H},
torque-free precession \citep{2000Natur.406..484S},
propagation through the ionized interstellar medium (IISM), instrumental effects,
and radio frequency interference (RFI). As well as the potential
benefits for pulsar timing, understanding the
causes of pulse profile variability and the sometimes correlated
changes in rotational behavior may elucidate physical processes
intrinsic to pulsars and their magnetosphere and also constrain the
effects of pulse propagation.

As the long-term pulse profile variability of MSPs has not been well studied,
it has only previously been reported in the
MSP J1643$-$1224; \citet{2016ApJ...828L...1S} describe a sudden and
permanent broadband pulse profile modification, accompanied by changes
in timing.

In \citet{2016MNRAS.456.1374B}, new techniques were used to identify
pulse profile variability in long-period pulsar data collected by the
Parkes Telescope. In this work we apply similar techniques to a large
sample of MSPs using data recorded by the NANOGrav collaboration, with
the aim of uncovering and quantifying MSP pulse profile variability.
The NANOGrav collaboration produces TOAs by template matching the pulse
profile in each frequency channel (typically between 5 and 64 over the
observing band; \citealp{2015ApJ...813...65T}), thereby producing multiple TOAs for each
observation. The analysis done here, however, looks for changes in
pulse profiles that have been frequency-integrated over the observing
band. This is done to maximize the S/N to facilitate the principal aim
of characterizing the long-term profile behavior in the
pulsar. However, when integrating a pulsar signal over a wide
observing band, pulse profiles are more susceptible to variations
induced by propagation effects \citep[e.g.][]{2014ApJ...790...93P}.

In Section 2 we describe the NANOGrav data used for the variability
analysis outlined in Section 3. The results of the analysis are
presented in Section 4, followed by a discussion in Section
5. Conclusions are drawn in Section 6.
\section{Data}
\label{data}
The data analyzed in this paper are a subset of the NANOGrav 11-year
data set \citep{2018ApJS..235...37A}, collected by the Green Bank Telescope (GBT) and Arecibo
Observatory (AO). Since 2010, data collected by the GBT have been
recorded by the Green Bank Ultimate Pulsar Processing Instrument
\citep[GUPPI;][]{2008SPIE.7019E..1DD,2010SPIE.7740E..0AF}. The
observations are carried out at center frequencies around 820 and
1500~MHz. Since 2012, data collected at AO have been recorded by
the Puerto Rican Ultimate Pulsar Processing Instrument (PUPPI). The
observations are carried out at center frequencies around 327~MHz
(PSR~J2317$+$1439 only), 430~MHz, 1400~MHz and 2030~MHz. This
GUPPI/PUPPI subset was used, as the instruments process a bandwidth of
up to 800 MHz (divided into 1.5625~MHz frequency channels) depending
on the mode of operation. Details of frequency coverage are given in
Table 1 of \citet{2015ApJ...813...65T}. Earlier narrow-bandwidth data
in the NANOGrav data set were excluded from this analysis due to relatively
low S/N.

GUPPI and PUPPI performed coherent dedispersion and folding in
real-time. The  data were folded at the  dynamically calculated pulsar
period using a pre-computed ephemeris to produce the pulse profile,
consisting of 2048 phase bins. The pulsar signals were flux and
polarization calibrated, and narrow-band RFI was removed in the manner
of \citet{2018ApJS..235...37A}.

The polarization calibration was done via an injected calibration
signal that is generated by a local noise diode at 25~Hz. Preceding
each pulsar observation, the noise diode signal is split, coupled
into the two polarization paths and measured with the pulsar backends.
This permits calibration of the differential gain and phase between the two hands of
polarization. For a complete description of the instrumental response to
a polarized signal, one must compute the \emph{Mueller matrix}: a
frequency-dependent linear transformation from the intrinsic to
observed Stokes parameters
\citep{2001PASP..113.1274H,2004ApJS..152..129V}.
The Mueller matrix is determined by tracking a polarized source
over a wide range of parallactic angles and fitting the resulting
variation of the observed Stokes parameters as the feed rotates with
respect to the sky. This allows the determination of effects such as
the magnitude and phase of the cross coupling of the receiver arms.

While all the data sets have undergone noise diode calibration as
described, full Mueller matrix calibration has also been performed
on the 1500~MHz GUPPI data only. As this method provides more
accurate pulse profile information, the GUPPI 1500~MHz profiles
analyzed in this work have had full Mueller matrix calibration
applied, unless stated otherwise. Mueller matrix calibration has also
recently been applied to PUPPI data by \citet{2018ApJ...862...47G}, but
their results have not been included in this analysis.
The pulsed noise signals themselves were calibrated in on- and
off-source observations  of  unpolarized  continuum  radio sources on
a monthly basis.

Each final pulse profile analyzed here is the integration of
typically 20 to 30 minutes of observation across the entire frequency
band. Pulsars at declinations between 0 and +39 degrees were observed
at AO while all others were observed with the
GBT. PSRs~J1713$+$0747 and B1937$+$21 were observed with both
telescopes.

The dispersion measure (DM) is fit to the data at almost every observing
epoch and applies for a window of up to 14 days, though typically much
shorter.
In NANOGrav timing analysis, an additional timing delay $\Delta t_{\rm {FD}}$ is added to all timing models to compensate for
TOA perturbations induced by the frequency-dependence of pulse
profile shapes. DM and $\Delta t_{\rm {FD}}$ are covariant when
finding the best-fit timing model parameters for a pulsar, and so the
best-fit DM value is highly dependent on $\Delta t_{\rm {FD}}$.
For the purposes of creating the frequency-integrated pulse profiles
employed in this variability analysis, we have calculated the best-fit
DM parameters without the inclusion of $\Delta t_{\rm {FD}}$.
This is discussed further in Section~\ref{inaccurate_dm}.

Further details of the observations, data reduction and timing models
can be found in \citet{2018ApJS..235...37A} and references therein. 
\section{Analysis}
\label{analysis}
The most effective metric for quantifying pulse profile variability is
dependent on the timescales involved. Pulsar observations can often be
widely and irregularly spaced; smooth trends that occur on timescales
much longer than the time between these observations may not be
obvious when analyzing individual pulse profiles. Such trends can
instead be revealed when the variability is modeled and interpolated
across many epochs of observation. If the pulse profile variations
take place on timescales comparable to, or shorter than the span
between observations, then any variability may appear stochastic, and a
smooth trend (if one is present) may not be easily detected. The
analysis techniques used to uncover and quantify both of these
systematic and noisy types of pulse profile variability are described
in the following.

To quantify the amount of pulse profile variability in an individual
observation, we calculate the differences between the observed pulse
profile and a constant model; these differences are termed the
\emph{profile residuals} \citep{2016MNRAS.456.1374B}.
The model for a particular pulsar and observing frequency is a median
profile; a median value is calculated in each individual phase bin
using all observations in the pulsar data set. The median
was used so that the model would be minimally affected by any outlying
pulse profile shapes. The technique used to align the profiles before
constructing the median model is simple cross-correlation. We note that the shape of the model is not crucial, as we are interested in how the
observations change with time. The model merely defines the zero-point for the profile residuals.

Before the profile residuals can be calculated,
the observations are processed to ensure that the off-pulse baseline
is centered on zero. Any individual observations with highly irregular
pulse profiles are treated as the result of RFI or instrumental issues and removed from
further analysis. Additionally, the noisiest observations in a data
set are considered unreliable and also excluded; an observation is
removed if the standard deviation of the off-pulse region is more than
a factor of two larger than the median value taken from the off-pulse
regions across all epochs.

Pulse profile changes can manifest as a modulation of shape or as a
change in flux density across the profile as a whole. Large flux
density variations are observed in most of the pulsar data analyzed in
this work, and are thought to be attributable to refractive and
diffractive interstellar scintillation (RISS and DISS respectively;
\citealp[e.g.][]{1990ARA&A..28..561R}). To disentangle
the less common pulse profile shape changes, we must normalize the
flux density of all observations. Alignment of the profiles with the
constant model is also essential for the analysis that follows, as the
timeseries in each pulse phase bin are modeled independently. This
alignment is non-trivial; when profile deviations occur (either
intrinsically or due to effects of propagation, instrumentation or RFI), it
is possible that the alignment may be slightly biased in that
direction when simple cross-correlation is employed. In this
analysis, the flux density normalization and the phase alignment are
carried out simultaneously in the following way.
\subsection{Flux Density Normalization and Phase Alignment}
\label{align_scale}
\begin{figure*}[ht]
  \centering
  \begin{tabular}{@{}cc@{}}
    \includegraphics[width=165mm]{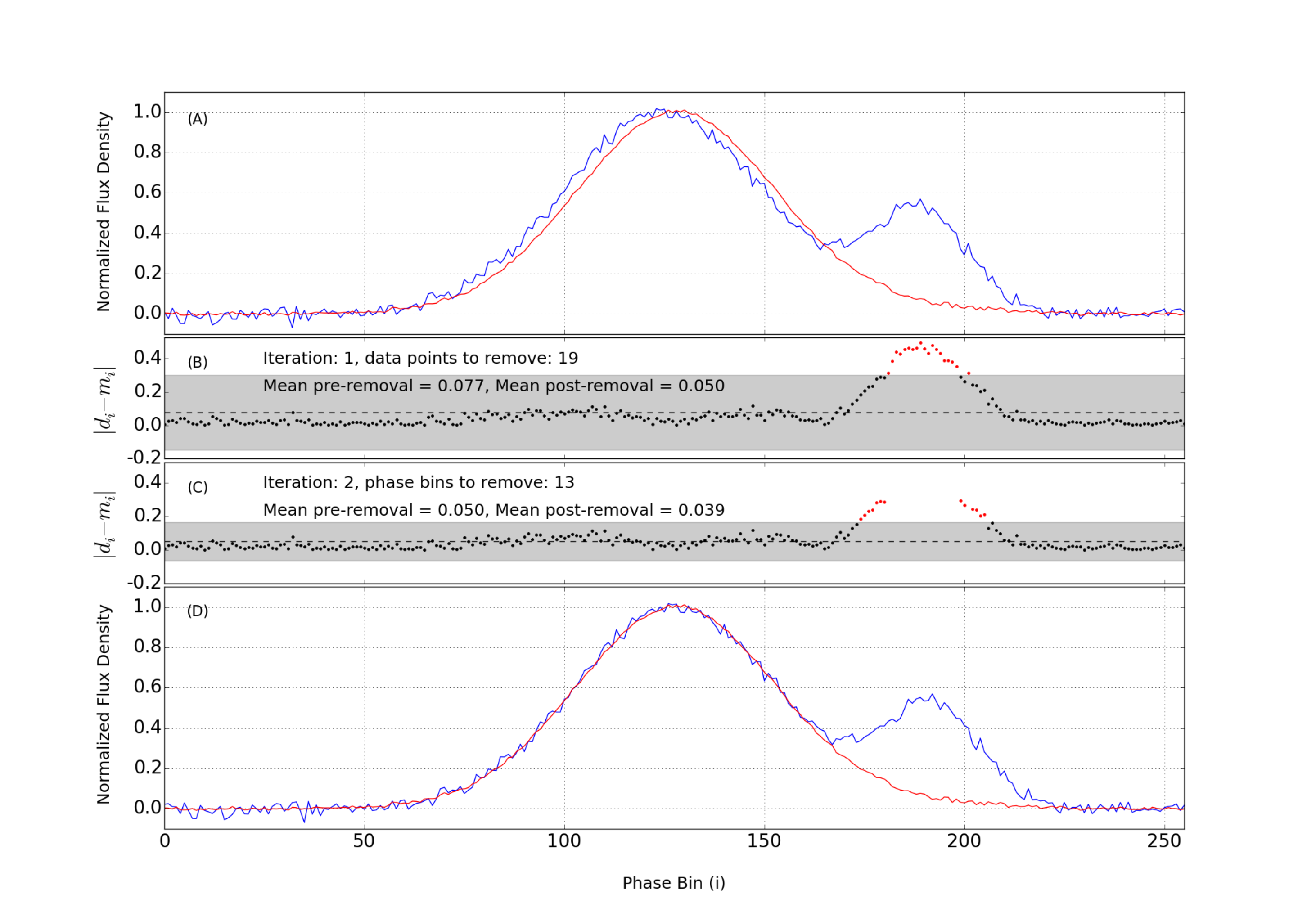}
  \end{tabular}
  \captionof{figure}{The method used to normalize and align pulse profiles. Panel
    A shows a simulated comparison of a model profile (red) and an observed
    profile (blue) with a trailing edge deviation. The alignment shown
    as an example, is the result of simple cross-correlation. For
    every alignment, a series of iterations, such as those illustrated in
    Panels B and C, are executed until a value for $\delta/n_{\rm f}$ is
    eventually determined (see text for details). In Panel B, the
    black dots show the absolute difference between the observed and model
    profiles in each phase bin ($|d_{i}- m_{i}|$). The dashed line indicates the mean and the gray band
    shows two standard deviations of the data; any points outside this
    band are removed before the standard deviation is recalculated
    and the process is repeated, as seen in Panel C. For each
    iteration, the mean of the data is calculated before and after the
    data removal. A change by less than 0.1\% signifies the end of the
    process (14 iterations were required for this outlier removal in these simulated data shown in Panel A). The last data to be removed are restored and the final
    mean $\delta$ of $|d_{i}- m_{i}|$ is calculated. This value, divided by the
    number of contributing phase bins $\delta/n_{\rm f}$, is minimized whilst adjusting
    the phase and normalized flux density of the observation (relative
    to the model). Panel D is the relative alignment with the lowest
    value of $\delta/n_{\rm f}$.}
  \label{align_exp}
\end{figure*}

In order to compare pulse profile shapes, we need to align them in
phase and normalize them in flux density as effectively as possible.
In many cases, the TOAs deviate enough from the pulsar timing model to
disqualify their use in the alignment of the observations.
Traditional profile alignment and normalization algorithms use $\chi^{2}$
minimization techniques and operate on all profile bins.  These
algorithms are susceptible to biases in cases when the two profiles
differ in shape over some range of pulse phase.  For this reason, we
employed the following robust fitting algorithm, which is less
susceptible to such biases. We characterize two pulse profiles as being correctly
normalized and aligned by maximizing the number of phase bins that are in
agreement; this is defined more formally below. Each observation, in
turn, is normalized and aligned relative to the constant model. The
observed profile is shifted in phase over the model.
For each of the 2048 phase
bin alignments, the scaling factor of the observation is varied over a
range defined such that the observation's profile peak is within 10\% of the
peak of the constant model. This range is sampled uniformly in 100 steps. The 10\% restriction will reduce
computation time while safely accommodating all realistic scaling
trials. For each combination of phase shift and scaling factor, the
absolute difference between model and observation is calculated in
each phase bin along with the mean,
\begin{equation}
  \delta = \frac{1}{n}\sum_{i}^{n}{|d_{i}- m_{i}|},
\end{equation}
where $d_{i}$ and $m_{i}$ are the values of the observational data and
the model (respectively) in phase bin $i$, and $n$ is the number of
phase bins in the calculation. For identical profile shapes, for example,
$\delta$ will be zero as the two profiles overlay exactly. We next
exclude any phase bins in which $|d_{i} - m_{i}|$ is more than
two standard deviations (2$\sigma$) away from $\delta$. After these outliers are removed,
$\delta$ is then recalculated. This step is repeated until the
recalculated mean $\delta$ changes by less than 0.1\% of its previous value, at
which stage the phase bin exclusion process is considered complete.
The final number of phase bins that have not been excluded is $n_{\rm f}$.
These steps are illustrated in Figure~\ref{align_exp}. All remaining
bins now have relatively comparable values of $|d_{i} - m_{i}|$. This
process is  performed so that only the stable parts of the profile
are used to align and scale, i.e. localized profile deviations that
appear in observations are not required to match the constant
model. To align and scale the profiles, we want to minimize the
differences between the non-excluded phase bins, but also want to
penalize fits in which only a small number of phase bins remain
after the exclusion process. In order to find the optimal fit, we minimize
$\delta$/$n_{\rm f}$.

In this analysis we calculate the variability of both
normalized and non-normalized pulse profiles. The latter are also aligned using
the technique above, but their flux density levels are restored at
the end of the process.

Precision timing of pulsars demands that observations are aligned to
fractions of a phase bin (under the assumption that the pulse profile
is unchanging). However, aligning in single bin increments (with 2048
bin resolution) is simple, and sufficient in this profile profile variability
analysis; any profile residuals produced by fractional phase bin
misalignment would typically be insignificant when compared to the amount of
noise in individual bins. If required, higher precision alignment could be implemented. 

Once the pulse profiles are correctly aligned and
normalized, we can proceed to calculate and analyze the profile
residuals.
\subsection{Visualizing Variability}
\begin{figure*}[ht]
  \begin{tabular}{@{}cc@{}}
    \includegraphics[width=\textwidth]{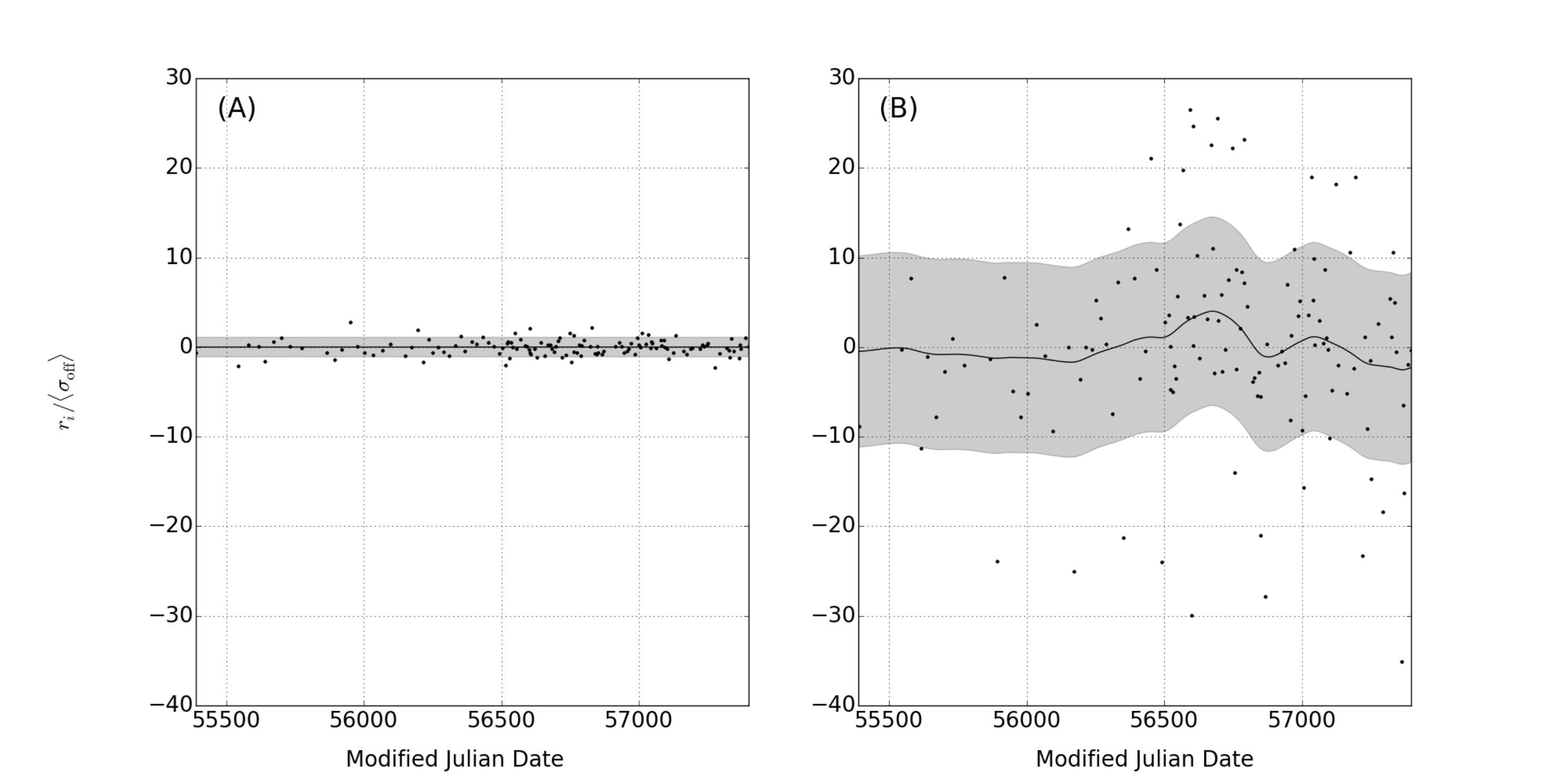} \\
    \includegraphics[width=\textwidth]{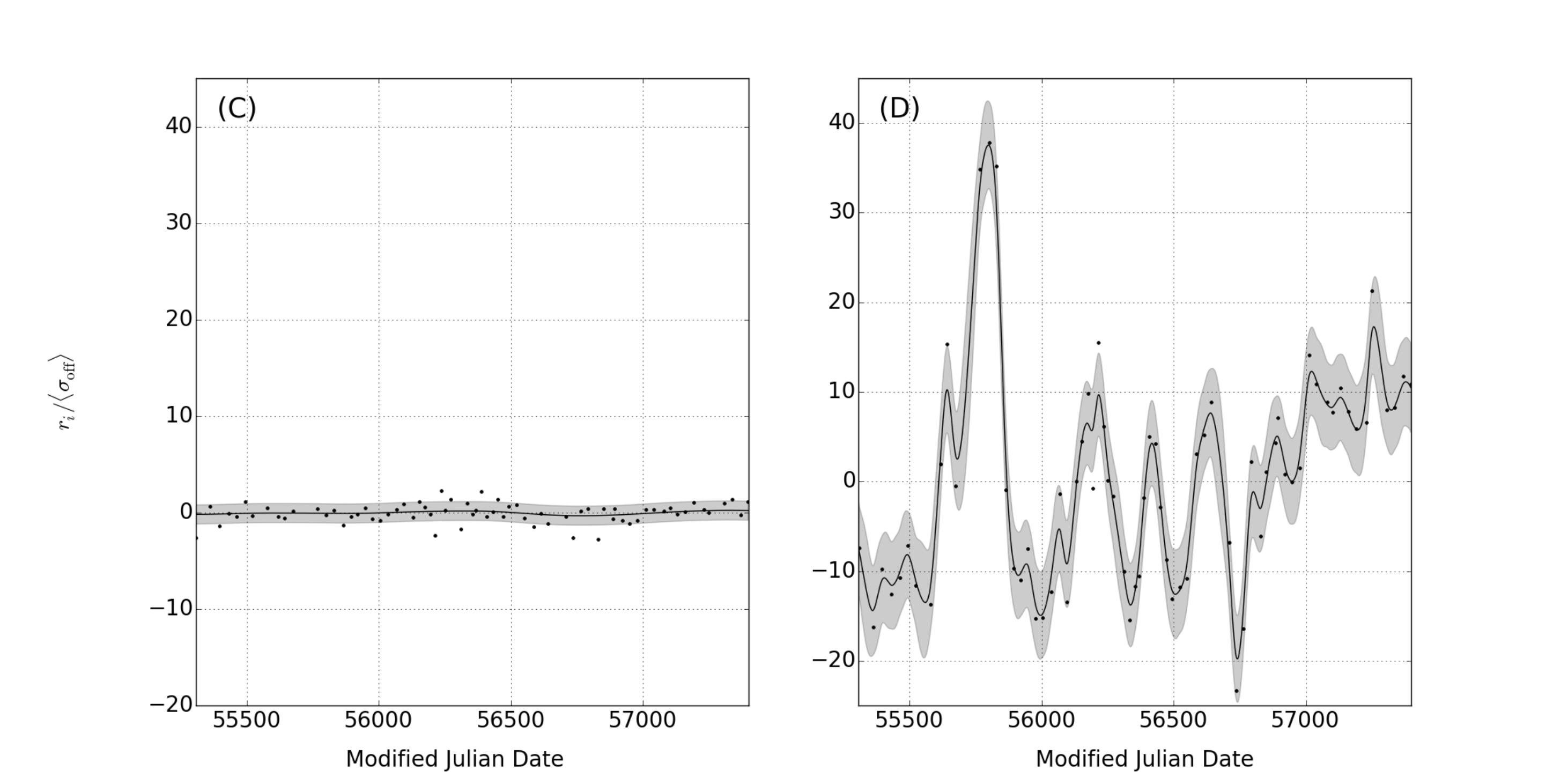}
  \end{tabular}
  \caption{Profile residuals and their GP models. Panel A and Panel
    B show data from example off- and on-pulse phase bins respectively for
    PSR~J1713$+$0747 at an observation frequency of 1500~MHz. Panel
    C and Panel D show data from example off- and on-pulse bins
    respectively for PSR~B1937$+$21 at an observational frequency of
    820~MHz. In each panel, the data points are the profile residuals
    for a single phase bin, the solid black line shows the GP
    mean and the gray area shows the GP standard
    deviation across the phase bin data set. Panel B is an example of a phase
    bin containing predominantly noisy data (with some systematic
    behaviour also embedded within), whereas the data in Panel D show
    a very clearly systematic trend, with relatively little noise. $r_i$ is the
    profile residual of bin $i$,
    $\langle\sigma_{\rm off}\rangle$ is the mean of the off-pulse profile residual standard
    deviation.}
  \label{bin_gp}
\end{figure*}

In \citet{2016MNRAS.456.1374B}, a technique was developed that models
pulse profiles as a function of time, allowing interpolation between
the epochs of observation. For each of the pulse profile phase bins,
we computed a Gaussian process (GP) regression model that best
describes the profile residuals \citep{rasmussen2006gaussian,
  Roberts20110550}. The lengthscale hyperparameter for the GP
regression models was constrained to between 30 and 300 days for every
data set analyzed; we find that this requirement results in the data being well
represented by the models. Full details of the GP regression analysis
can be found in \citet{2016MNRAS.456.1374B}. Examples of this
inference technique are shown in Figure~\ref{bin_gp}.
The individual phase bin models can be combined to produce a \emph{variability map} for each
pulsar. This is an interpolated plot that smoothly maps the evolution
of a pulsar's profile residuals with time. The GP regression technique
can be used to identify subtle long-term trends that are not visible
by eye, as demonstrated in \citet{2016MNRAS.456.1374B}. For each
pulsar discussed in Section \ref{results}, a variability map was
produced for both pre- and post-normalization pulse profiles, so that
the flux density of the observations can be compared with any profile
shape changes seen.

The two pulsars in Figure~\ref{bin_gp} illustrate two different types
of pulse profile variability (as mentioned at the beginning of Section~\ref{analysis}). The systematic nature of the profile
residuals in Panel~D is well modeled by GP regression; the extent
and nature of the profile variability is, therefore, easily captured
by a variability map. In contrast, Panel~B shows that the profile
residuals in the J1713$+$0747 on-pulse phase bin are highly variable
over time, but primarily in a noisy rather than systematic way. As a
consequence, the GP model may infer little or no systematic variability
and simply lie around the mean of the data points
that inform it. The gray band in each Figure~\ref{bin_gp} panel shows the standard
deviation of the model, however, and so provides an alternative
measure of pulse profile variability. In order to show the amount of
noisy variability for a pulsar data set, we also generate a color map
showing the standard deviation of the GP model as a function of pulse
phase and time. For instructional purposes, the color maps for a
stable pulsar data set are shown in the Appendix (Figure~\ref{low_variability}).
\subsection{Quantifying Variability}
\label{quant}
For each pulsar data set analyzed, we computed six metrics to fully
describe the nature of the variability observed in the normalized
pulse profiles.
The metrics are defined as follows:
\\
\begin{enumerate}[label=(\Alph*)]
  \item{Ratio of (i) mean standard deviation of on-pulse phase bins to
    (ii) mean standard deviation of off-pulse phase bins:
    $\langle\sigma_{\rm on}\rangle/\langle\sigma_{\rm off}\rangle$.}
    
    To calculate $\langle\sigma_{\rm on}\rangle$ we found the standard
    deviation of the profile residuals in each on-pulse phase bin
    and then calculated the mean across all epochs. The equivalent
    calculation was done for $\langle\sigma_{\rm off}\rangle$.
  \item{Ratio of (i) maximum standard deviation of on-pulse phase bins to
    (ii) mean standard deviation of off-pulse phase bins:}
    $\sigma_{\rm on, max}/\langle\sigma_{\rm off}\rangle$.

    The maximum standard deviation of the profile residuals in
    individual on-pulse phase bins, provides information regarding any
    variability that may be concentrated over a small section of the
    pulse profile. 
  \item{Ratio of (i) peak systematic variability to (ii) mean standard
    deviation of off-pulse phase bins: $\rm M_{\rm max}/\langle\sigma_{\rm off}\rangle$.}

    $ \rm M_{\rm max}$ is the peak value of the GP model over all on-pulse phase bins.
  \item{Ratio of (i) average systematic variability to (ii) mean standard
        deviation of off-pulse phase bins:
        $\langle\mid \rm M\mid\rangle/\langle\sigma_{\rm off}\rangle$.}
    
    $\langle\mid \rm M\mid\rangle$ is the mean of the absolute value of the GP model for on-pulse phase bins.
  \item{Ratio of (i) noisy variability to (ii) mean standard
        deviation of off-pulse phase bins: $\langle\sigma_{\rm \rm M}\rangle/\langle\sigma_{\rm off}\rangle$.}
    
    $\langle\sigma_{\rm \rm M}\rangle$ is the mean of the standard deviation of the GP model (i.e. the gray
    shaded regions in Figure ~\ref{bin_gp}) across all on-pulse phase
    bins. In pulsars with systematic variability (e.g. Panel~D of
    Figure~\ref{bin_gp}), the standard deviation about the GP model mean
    will be less than the standard deviation of the data themselves.
  \item{Ratio of (i) average systematic variability to (ii) noisy variability: $\langle\mid \rm M \mid\rangle/\langle\sigma_{\rm M}\rangle$.}

    This metric is indicative of the amount of long-term, systematic
    variability in a data set.
    
\end{enumerate}

An on-pulse phase bin is defined as one in which the flux density of the median profile for the data set is more than 3\% of the peak.
An off-pulse phase bin is defined as one in which the flux density of
the median profile for the data set is less than 0.1\% of the
peak. The 3\% and 0.1\% values were chosen empirically to reliably
select only on- and off-pulse phase bins respectively. The gap between
the thresholds exists in order to avoid contamination between the
two. If on- and off-pulse variability is comparable, metrics A and
E will have a value around unity.

\section{Results} 
\label{results}

The results of the pulse profile variability analysis are presented in Table~\ref{var_table}, which is ordered by pulsar right ascension and then by
observing frequency. Only the NANOGrav data sets that consist of 20 or more
observations (after noisy and unreliable profiles are removed) are
featured in the table. This is done to ensure that the GP regression
has sufficient data points to infer an accurate model; 78 data sets
(from a 38-pulsar subset of the 45 pulsars observed in the
NANOGrav 11-year data set)
remain after this requirement. The vast majority of pulsars show
relatively little variability, with the mean standard deviation of
their on-pulse phase bins being less than a factor of two greater than
that of their off-pulse bins (Metric~A of Table~\ref{var_table}). The
three pulsars for which this factor
is greatest (denoted by an asterisk in Table~\ref{var_table}) are PSRs~J1713$+$0747, B1937$+$21 and
J2145$-$0750. We have selected these pulsars for further analysis; below, we discuss the nature and possible causes of the
profile variability for each of them. In addition, we also focus on PSR~J1643$-$1224; after PSR~B1937$+$21, the 820~MHz
data set for this pulsar (denoted by a double asterisk in Table~\ref{var_table}) has the largest average systematic to noisy
variability ratio (Metric~F of Table~\ref{var_table}), which
indicates the presence of long-term variability. PSR~J1643$-$1224 has
also previously demonstrated unusual chromatic timing behavior and
long-term pulse profile shape changes \citep{2016ApJ...828L...1S}.

\subsection{PSR~J1713$+$0747}
\label{1713_results}
Due in part to the high S/N of its pulse profile, PSR~J1713$+$0747 is
one of the most precisely timed pulsars. \citet{2018ApJS..235...37A}
list the standard deviation of the epoch-averaged \emph{timing residuals} (the
differences between observed TOAs and a timing model) for this
pulsar as 116~ns over 11 years of NANOGrav observations. The
high S/N also allows any pulse profile variations to be seen
clearly.

\begin{figure*}[p]
  \sbox0{\begin{tabular}{@{}cc@{}}
      \includegraphics[width=110mm]{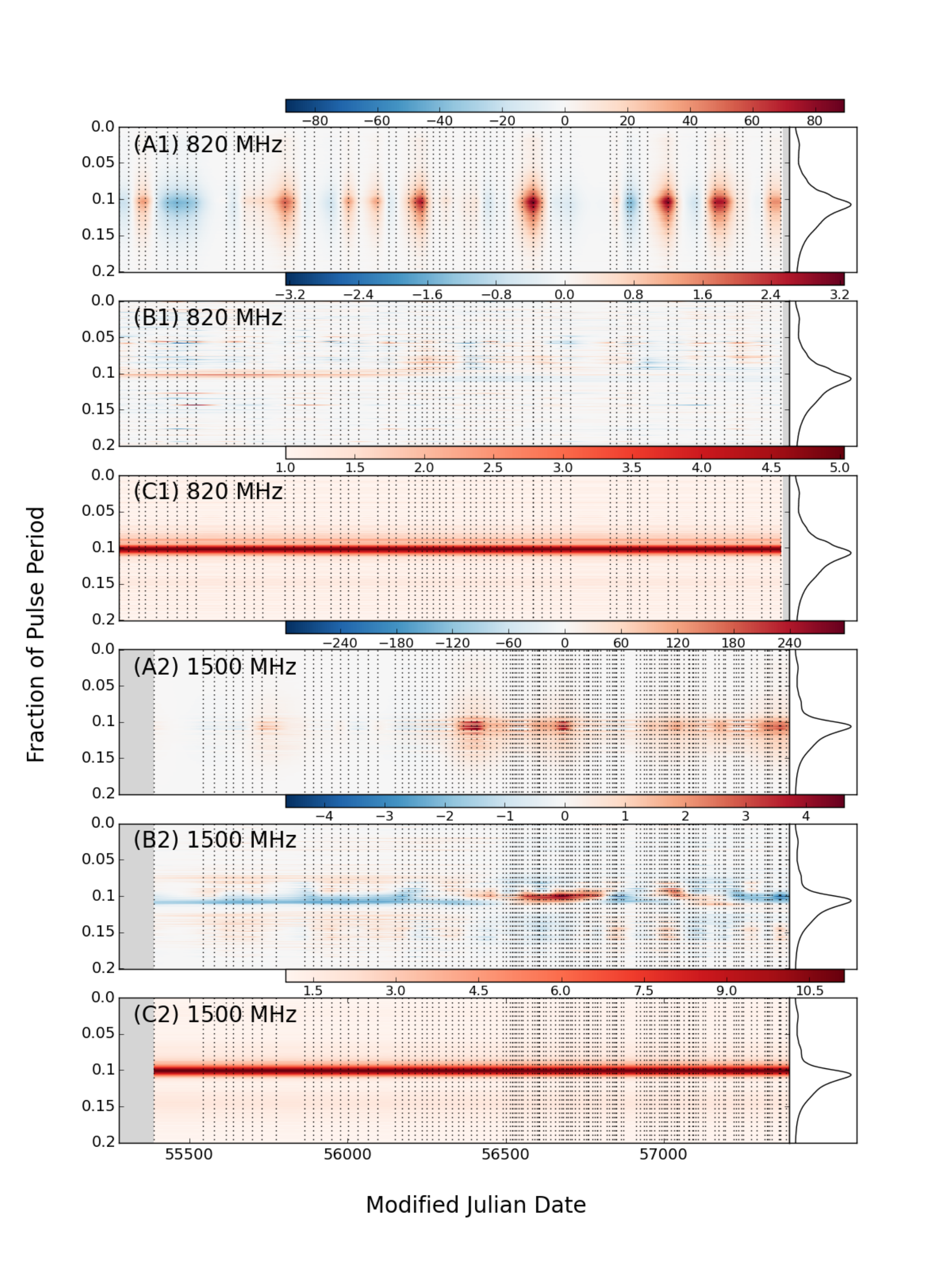} &
      \includegraphics[width=110mm]{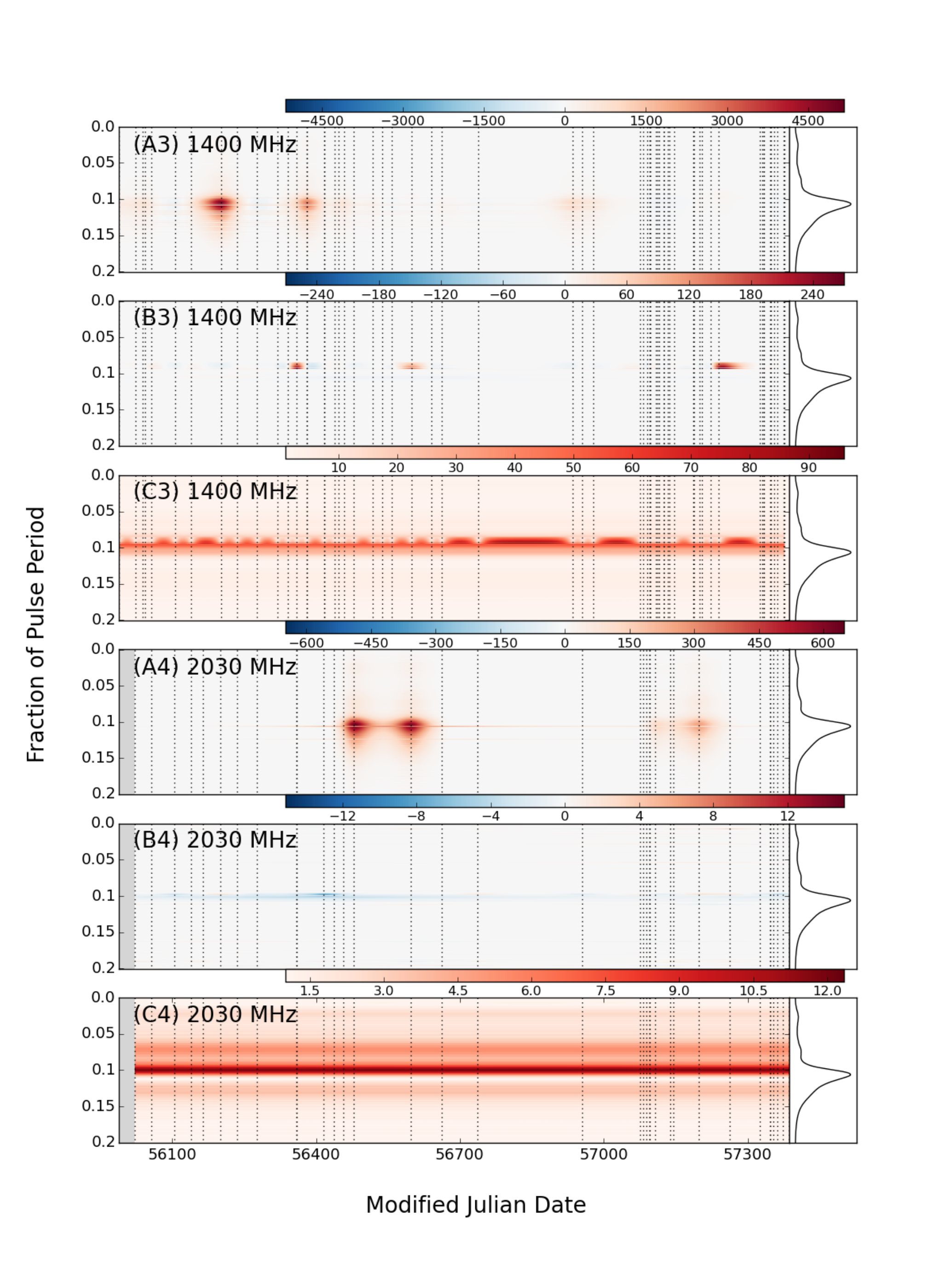}
  \end{tabular}}
  \rotatebox{90}{\begin{minipage}[c][\textwidth][c]{\wd0}
      \usebox0
      \caption{Variability maps for PSR~J1713$+$0747, for central
        observing frequencies of 820~MHz (GBT), 1500~MHz (GBT), 1400~MHz
        (AO) and 2030~MHz (AO). Panel labels prefixed with an A are
        variability maps showing the flux density variations in the
        flux-calibrated, pre-normalized observations. Those prefixed with
        a B show pulse profile shape changes after the observations have
        been normalized. In all of these variability maps, red regions
        indicate where the inferred pulse profile has an excess of flux
        density (normalized or otherwise) compared to the average for the
        data set. Blue indicates where it has a deficit. Panel labels
        prefixed with a C map the standard deviation of the GP model as a
        function of pulse phase and time.  The vertical dotted lines
        indicate the epochs of observation informing the GP models. The
        unit for all panels is the mean of the standard deviation of the
        off-pulse phase bins for the relevant data set. Panel sections
        for which there is no data, are gray. To the right of each
        panel, the average pulse profile for the data set is shown.}     
      \label{1713_vm}
  \end{minipage}}
\end{figure*}

\begin{figure*}[ht]
  \begin{tabular}{@{}cc@{}}
    \includegraphics[width=\textwidth]{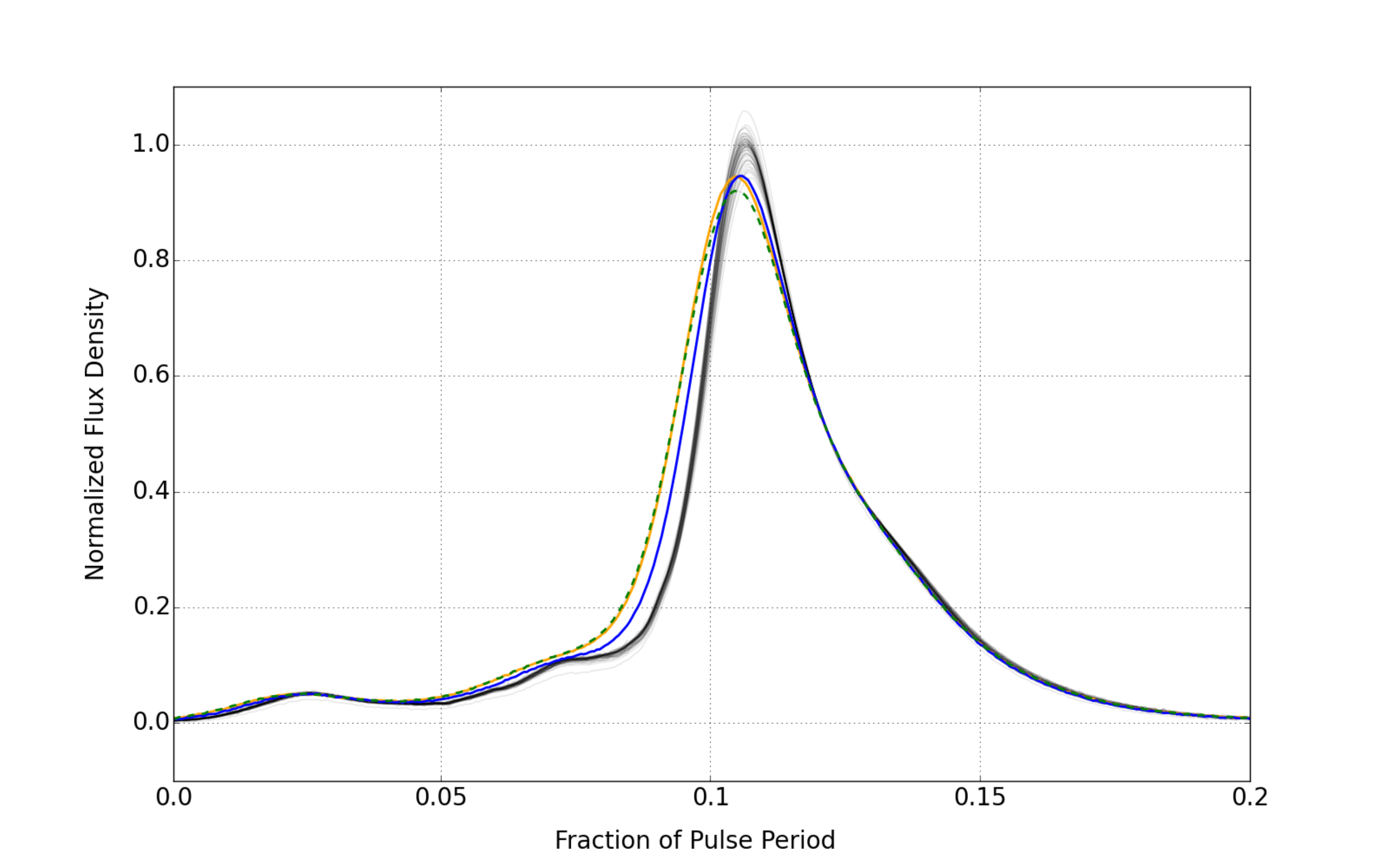}
  \end{tabular}
  \caption{The pulse profile variability seen in PSR~J1713$+$0747 at
    1400~MHz. The orange, blue and green (dashed) profiles were observed on MJDs
    56360, 56598 and 57239 respectively. The gray profiles show the
    other 58 pulse profiles in the data set. See also Figure~\ref{1713_1400_dm}
    and Section~\ref{inaccurate_dm} for a comparision of these three
    noteworthy observations to profile changes caused by inaccurate DMs
    in this data set.}
  \label{1713_extreme}
\end{figure*}

The 1400~MHz AO observations of PSR~J1713$+$0747 display the most
profile variability of all data sets analyzed in this work;
Table~\ref{var_table} shows that this is mostly noisy in
nature (a relatively large value for variability metric E and a relatively
small value for variability metric F).
Despite the average systematic variability of the data set
being low with respect to the noisy variability, the peak of the
systematic variability is high; the GP model is being strongly affected
over short time periods by three observations with anomalous profile
shapes (MJDs 56360, 56598 and 57239). This can be seen in the
variability map in Panel~B3 of Figure~\ref{1713_vm}.
These three profiles
are compared to those typical for the data set in
Figure~\ref{1713_extreme}.
For the observation made on MJD 56360, it is known that during the flux density calibration
procedure, an incorrect pulsed calibration
signal was injected at the epoch of observation. It is not clear whether the
pulse profile shape was affected by this. However, no such calibration issues
exist for the observations made on MJDs 56598 or 57239.
In the ``Discussion'' section,
we compare these three profiles to those expected to be produced by inaccurate DM
measurements. This is shown in Figure~\ref{1713_extreme} and described in Section~\ref{inaccurate_dm}.

\startlongtable
\begin{deluxetable*}{cccccccc}
  \tablecaption{The variability calculated in 78 NANOGrav MSP data sets.\label{var_table}}
  \tablehead{
    \colhead{Pulsar} &
    \colhead{Observing} &
    \colhead{$\langle\sigma_{\rm on}\rangle/\langle\sigma_{\rm off}\rangle$} &
    \colhead{$\sigma_{\rm on,max}/\langle\sigma_{\rm off}\rangle$} &
    \colhead{$\rm M_{\rm max}/\langle\sigma_{\rm off}\rangle$} &
    \colhead{$\langle\mid \rm M\mid\rangle/\langle\sigma_{\rm off}\rangle$} &
    \colhead{$\langle\sigma_{\rm \rm M}\rangle/\langle\sigma_{\rm off}\rangle$} &
    \colhead{$\langle\mid \rm M \mid\rangle/\langle\sigma_{\rm M}\rangle$}\\
    \colhead{Name} & \colhead{Frequency} & \colhead{} & \colhead{} &
    \colhead{} & \colhead{} & \colhead{} & \colhead{}\\
    \colhead{} & \colhead{(MHz)} & \colhead{(A)} & \colhead{(B)} &
    \colhead{(C)} & \colhead{(D)} & \colhead{(E)} & \colhead{(F)}
  }  
  \startdata
  J0023$+$0923 & 430 & 1.13 & 1.97 & 3.58 & 0.12 & 1.10 & 0.11\\
  J0023$+$0923 & 1400 & 1.17 & 2.74 & 3.69 & 0.08 & 1.17 & 0.07\\
  J0030$+$0451 & 430 & 1.02 & 1.76 & 2.81 & 0.06 & 1.02 & 0.06\\
  J0030$+$0451 & 1400 & 1.03 & 1.72 & 1.97 & 0.05 & 1.03 & 0.05\\
  J0340$+$4130 & 820 & 1.03 & 1.70 & 1.95 & 0.08 & 1.02 & 0.08\\
  J0340$+$4130 & 1500 & 1.00\hspace{0.3cm}1.00 &
  1.42\hspace{0.3cm}1.38 & 3.07\hspace{0.3cm}3.44 & 0.09\hspace{0.3cm}0.09 & 0.98\hspace{0.3cm}0.99 & 0.09\hspace{0.3cm}0.09\\
  J0613$-$0200 & 820 & 1.08 & 3.01 & 3.56 & 0.11 & 1.06 & 0.10\\
  J0613$-$0200 & 1500 & 1.04\hspace{0.3cm}1.04 &
  1.47\hspace{0.3cm}1.70 & 3.92\hspace{0.3cm}3.16 & 0.08\hspace{0.3cm}0.08 & 1.03\hspace{0.3cm}1.03 & 0.08\hspace{0.3cm}0.08\\
  J0636$+$5128 & 820 & 1.10 & 1.63  4.16 & 0.14 & 1.07 & 0.13\\
  J0636$+$5128 & 1500 & 1.08\hspace{0.3cm}1.07 &
  1.74\hspace{0.3cm}1.64 & 4.34\hspace{0.3cm}3.54 & 0.10\hspace{0.3cm}0.13 & 1.06\hspace{0.3cm}1.02 & 0.10\hspace{0.3cm}0.13\\
  J0645$+$5158 & 820 & 1.06 & 2.30 & 3.47 & 0.09 & 1.04 & 0.09\\
  J0645$+$5158 & 1500 & 1.06\hspace{0.3cm}1.05 &
  2.09\hspace{0.3cm}2.12 & 5.96\hspace{0.3cm}5.72 & 0.10\hspace{0.3cm}0.12 & 1.03\hspace{0.3cm}1.02 & 0.10\hspace{0.3cm}0.12\\
  J0931$-$1902 & 820 & 1.03 & 1.71 & 4.14 & 0.11 & 1.01 & 0.11\\
  J0931$-$1902 & 1500 & 1.09\hspace{0.3cm}1.05 &
  2.08\hspace{0.3cm}1.87 & 3.73\hspace{0.3cm}3.79 & 0.11\hspace{0.3cm}0.12 & 1.08\hspace{0.3cm}1.02 & 0.10\hspace{0.3cm}0.12\\
  J1012$+$5307 & 820 & 1.09 & 1.77 & 3.57 & 0.12 & 1.07 & 0.11\\
  J1012$+$5307 & 1500 & 1.13\hspace{0.3cm}1.12 &
  2.92\hspace{0.3cm}2.64 & 3.39\hspace{0.3cm}2.56 & 0.06\hspace{0.3cm}0.09 & 1.12\hspace{0.3cm}1.11 & 0.05\hspace{0.3cm}0.08\\
  J1024$-$0719 & 820 & 1.06 & 1.72 & 3.59 & 0.09 & 1.04 & 0.09\\
  J1024$-$0719 & 1500 & 1.05\hspace{0.3cm}1.06 &
  1.71\hspace{0.3cm}1.80 & 3.56\hspace{0.3cm}3.31 & 0.10\hspace{0.3cm}0.08 & 1.03\hspace{0.3cm}1.05 & 0.10\hspace{0.3cm}0.08\\
  J1125$+$7819 & 820 & 1.04 & 1.71 & 4.21 & 0.15 & 0.99 & 0.15\\
  J1455$-$3330 & 820 & 1.10 & 2.04 & 3.44 & 0.11 & 1.09 & 0.10\\
  J1455$-$3330 & 1500 & 1.09\hspace{0.3cm}1.07 &
  3.00\hspace{0.3cm}2.71 & 5.75\hspace{0.3cm}6.23 &
  0.16\hspace{0.3cm}0.15 & 1.04\hspace{0.3cm}1.02 &
  0.15\hspace{0.3cm}0.15\\
  J1600$-$3053 & 820 & 1.05 & 1.40 & 2.36 & 0.09 & 1.04 & 0.09\\
  J1600$-$3053 & 1500 & 1.15\hspace{0.3cm}1.24 &
  2.10\hspace{0.3cm}2.96 & 1.99\hspace{0.3cm}6.31 & 0.08\hspace{0.3cm}0.10 & 1.14\hspace{0.3cm}1.23 & 0.07\hspace{0.3cm}0.08\\
  J1614$-$2230 & 820 & 1.02 & 1.39 & 2.99 & 0.09 & 1.00 & 0.09\\
  J1614$-$2230 & 1500 & 1.01\hspace{0.3cm}1.00 &
  1.29\hspace{0.3cm}1.33 & 2.59\hspace{0.3cm}1.91 & 0.05\hspace{0.3cm}0.06 & 1.00\hspace{0.3cm}1.00 & 0.05\hspace{0.3cm}0.06\\
  J1640$+$2224 & 430 & 1.07 & 1.63 & 2.57 & 0.04 & 1.07 & 0.04\\
  J1640$+$2224 & 1400 & 1.32 & 2.85 & 2.75 & 0.08 & 1.32 & 0.06\\
  \text{*}\text{*}J1643$-$1224 & 820 & 1.14 & 1.75 & 3.40 & 0.29 & 1.05 & 0.28\\
  J1643$-$1224 & 1500 & 1.01\hspace{0.3cm}1.00 &
  1.42\hspace{0.3cm}1.32 & 2.33\hspace{0.3cm}2.26 & 0.09\hspace{0.3cm}0.09 & 1.00\hspace{0.3cm}0.99 & 0.09\hspace{0.3cm}0.09\\
  J1713$+$0747 & 820 & 1.36 & 5.11 & 3.33 & 0.12 & 1.35 & 0.09\\
  \text{*}J1713$+$0747 & 1400 & 8.78 & 52.19 & 269.88 & 0.89 & 8.65 & 0.10\\
  J1713$+$0747 & 1500 & 2.38\hspace{0.3cm}2.21 &
  12.47\hspace{0.3cm}11.54 & 6.85\hspace{0.3cm}6.23 & 0.12\hspace{0.3cm}0.27 & 2.36\hspace{0.3cm}2.18 & 0.05\hspace{0.3cm}0.12\\
  J1713$+$0747 & 2030 & 3.36 & 11.58 & 4.08 & 0.10 & 3.39 & 0.03\\
  J1738$+$0333 & 1400 & 1.34 & 6.00 & 5.88 & 0.14 & 1.29 & 0.11\\
  J1741$+$1351 & 430 & 1.22 & 3.09 & 3.70 & 0.16 & 1.17 & 0.14\\
  J1741$+$1351 & 1400 & 1.11 & 2.37 & 3.52 & 0.10 & 1.10 & 0.09\\
  J1744$-$1134 & 820 & 1.12 & 1.56 & 3.00 & 0.11 & 1.10 & 0.10\\
  J1744$-$1134 & 1500 & 1.38\hspace{0.3cm}1.35 &
  2.26\hspace{0.3cm}2.45 & 2.41\hspace{0.3cm}3.71 & 0.11\hspace{0.3cm}0.11 & 1.37\hspace{0.3cm}1.34 & 0.08\hspace{0.3cm}0.08\\
  J1747$-$4036 & 820 & 1.03 & 1.68 & 2.44 & 0.12 & 1.01 & 0.12\\
  J1747$-$4036 & 1500 & 1.03\hspace{0.3cm}1.03 &
  1.52\hspace{0.3cm}1.62 & 3.05\hspace{0.3cm}2.97 & 0.12\hspace{0.3cm}0.11 & 1.01\hspace{0.3cm}1.01 & 0.12\hspace{0.3cm}0.11\\
  J1832$-$0836 & 820 & 1.02 & 1.55 & 2.69 & 0.09 & 1.01 & 0.09\\
  J1832$-$0836 & 1500 & 1.03\hspace{0.3cm}1.03 &
  1.49\hspace{0.3cm}1.66 & 3.04\hspace{0.3cm}4.91 & 0.09\hspace{0.3cm}0.10 & 1.01\hspace{0.3cm}1.00 & 0.09\hspace{0.3cm}0.10\\
  J1853$+$1303 & 430 & 1.06 & 1.82 & 2.85 & 0.08 & 1.06 & 0.08\\
  J1853$+$1303 & 1400 & 1.07 & 1.66 & 4.46 & 0.10 & 1.05 & 0.10\\
  B1855$+$09 & 430 & 1.03 & 1.46 & 3.61 & 0.08 & 1.02 & 0.08\\
  B1855$+$09 & 1400 & 1.25 & 3.77 & 3.98 & 0.11 & 1.24 & 0.09\\
  J1903$+$0327 & 1400 & 1.10 & 1.75 & 3.19 & 0.21 & 1.05 & 0.20\\
  J1903$+$0327 & 2030 & 1.10 & 2.00 & 4.90 & 0.13 & 1.08 & 0.12\\
  J1909$-$3744 & 820 & 1.59 & 2.95 & 3.65 & 0.24 & 1.54 & 0.16\\
  J1909$-$3744 & 1500 & 1.65\hspace{0.3cm}1.58 &
  3.12\hspace{0.3cm}2.81 & 1.51\hspace{0.3cm}1.74 & 0.04\hspace{0.3cm}0.07 & 1.65\hspace{0.3cm}1.58 & 0.02\hspace{0.3cm}0.04\\
  J1910$+$1256 & 1400 & 1.03 & 1.59 & 2.89 & 0.08 & 1.02 & 0.08\\
  J1910$+$1256 & 2030 & 1.04 & 1.58 & 4.76 & 0.17 & 0.99 & 0.17\\
  J1918$-$0642 & 820 & 1.03 & 1.42 & 1.96 & 0.08 & 1.02 & 0.08\\
  J1918$-$0642 & 1500 & 1.05\hspace{0.3cm}1.05 & 
  1.58\hspace{0.3cm}1.54 & 1.92\hspace{0.3cm}3.04 & 0.05\hspace{0.3cm}0.06 & 1.05\hspace{0.3cm}1.04 & 0.05\hspace{0.3cm}0.06\\
  J1923$+$2515 & 430 & 1.06 & 2.00 & 2.93 & 0.09 & 1.05 & 0.09\\
  J1923$+$2515 & 1400 & 1.09 & 1.95 & 5.11 & 0.10 & 1.07 & 0.09\\
  \text{*}B1937$+$21 & 820 & 5.35 & 12.20 & 37.01 & 2.69 & 3.60 & 0.75\\
  B1937$+$21 & 1400 & 5.09 & 12.86 & 23.42 & 0.91 & 4.82 & 0.19\\
  B1937$+$21 & 1500 & 5.07\hspace{0.3cm}3.36 & 10.85\hspace{0.3cm}7.76
  & 3.82\hspace{0.3cm}3.24 & 0.42\hspace{0.3cm}0.39 & 5.05\hspace{0.3cm}3.33 & 0.08\hspace{0.3cm}0.12\\
  B1937$+$21 & 2030 & 2.93 & 8.30 & 14.32 & 1.31 & 2.36 & 0.56\\
  J1944$+$0907 & 430 & 1.09 & 1.96 & 3.38 & 0.10 & 1.08 & 0.09\\
  J1944$+$0907 & 1400 & 1.12 & 2.06 & 3.91 & 0.13 & 1.09 & 0.12\\
  B1953$+$29 & 430 & 1.10 & 1.91 & 4.22 & 0.22 & 1.03 & 0.21\\
  B1953$+$29 & 1400 & 1.03 & 1.52 & 4.25 & 0.09 & 1.01 & 0.09\\
  J2010$-$1323 & 820 & 1.14 & 2.76 & 3.59 & 0.09 & 1.12 & 0.08\\
  J2010$-$1323 & 1500 & 1.10\hspace{0.3cm}1.10 &
  2.33\hspace{0.3cm}2.42 & 2.80\hspace{0.3cm}2.37 & 0.06\hspace{0.3cm}0.07 & 1.09\hspace{0.3cm}1.10 & 0.06\hspace{0.3cm}0.06\\
  J2017$+$0603 & 1400 & 1.11 & 5.55 & 5.47 & 0.15 & 1.08 & 0.14\\
  J2017$+$0603 & 2030 & 1.16 & 3.46 & 7.79 & 0.17 & 1.13 & 0.15\\
  J2043$+$1711 & 430 & 1.05 & 1.62 & 2.39 & 0.03 & 1.05 & 0.03\\
  J2043$+$1711 & 1400 & 1.04 & 1.74 & 5.11 & 0.10 & 1.02 & 0.10\\
  \text{*}J2145$-$0750 & 820 & 1.76 & 8.62 & 3.79 & 0.31 & 1.72 & 0.18\\
  J2145$-$0750 & 1500 & 1.37\hspace{0.3cm}1.31 &
  3.02\hspace{0.3cm}3.51 & 4.69\hspace{0.3cm}3.82 & 0.22\hspace{0.3cm}0.15 & 1.32\hspace{0.3cm}1.28 & 0.17\hspace{0.3cm}0.12\\
  J2214$+$3000 & 1400 & 1.00 & 1.73 & 5.37 & 0.12 & 0.97 & 0.12\\
  J2302$+$4442 & 820 & 1.02 & 1.57 & 2.34 & 0.07 & 1.01 & 0.07\\
  J2302$+$4442 & 1500 & 1.04\hspace{0.3cm}1.00 &
  1.54\hspace{0.3cm}1.56 & 4.73\hspace{0.3cm}2.66 & 0.10\hspace{0.3cm}0.09 & 1.01\hspace{0.3cm}0.99 & 0.10\hspace{0.3cm}0.09\\
  J2317$+$1439 & 327 & 1.13 & 2.15 & 3.07 & 0.13 & 1.10 & 0.12\\
  J2317$+$1439 & 430 & 1.10 & 1.80 & 2.52 & 0.04 & 1.10 & 0.04\\
  J2317$+$1439 & 1400 & 1.07 & 1.78 & 2.81 & 0.07 & 1.06 & 0.07\\
  \enddata
  \tablecomments{An
    asterisk denotes each of the three data sets with the highest
    values for the ratio of the mean standard deviation of on- to off-pulse phase bins (Metric~A; a
    measurement of the level of profile variability of any
    kind). These data are from the PSRs~J1713$+$0747, B1937$+$21 and
    J2145$-$0750. Highlighted with a double asterisk is the 820~MHz data set for
    PSR~J1643$-$1224, which has the highest ratio of average
    systematic to noisy variability (Metric~F; a measurement of the
    significance of systematic variability) after PSR~B1937$+$21. Each of
    the six variability metrics is described in
    Section~\ref{quant}. The data sets observed with the GBT at
    1500~MHz have two values for each variability metric. The left of
    the pair relates to profiles calibrated by the noise diode, and
    the right to profiles that additionally have full
    Mueller matrix calibration applied
    (see Section~\ref{data}).}
\end{deluxetable*}
\noindent
The observations made at 2030~MHz show a high degree of noisy
variability, particularly in the latter half of the data set; the
ratio of average systematic to noisy variability $\langle\mid \rm M \mid\rangle/\langle\sigma_{\rm M}\rangle$ is the smallest of all data
sets. The ratio of the mean standard deviation of the on- to off-pulse phase bins
$\langle\sigma_{\rm on}\rangle/\langle\sigma_{\rm off}\rangle$ is not
as large as that of the 1400~MHz data set, however, because at
2030~MHz, $\langle\sigma_{\rm off}\rangle$ is much
larger.

All PSR~J1713$+$0747 data sets are dominated by noisy variability, i.e.
profile shape changes largely occur on timescales shorter than the time between
observations, and so the noisy variability far exceeds any systematic
variability modeled by the GP. Some systematic variability is present,
however; the PSR~J1713$+$0747 data set with the highest ratio of average
systematic to noisy variability (Metric~F of Table~\ref{var_table}) is recorded at 1500~MHz.
Panel~B of Figure~\ref{bin_gp} depicts the
systematic behavior of the GP model (embedded in the primarily noisy variability) in an individual phase bin for the data
set. The bin is associated with a pulse period fraction of $\sim$ 0.1 in
Panel~B2 of Figure~\ref{1713_vm}.

As the variability in the PSR~J1713$+$0747 data sets is predominantly
short-term in nature, it is difficult to compare even the pulse profiles that
were observed at similar frequencies, unless they are also observed at
the same time. The 1400~MHz AO and
1500~MHz GBT observations are often made just days apart, but only
simultaneous observations could permit us to observe identical pulse profile
shapes and allow us to confirm the nature of any
profile variability seen, as astrophysical.
\subsection{PSR~B1937$+$21}
\begin{figure*}[p]
  \sbox0{\begin{tabular}{@{}cc@{}}
      \includegraphics[width=110mm]{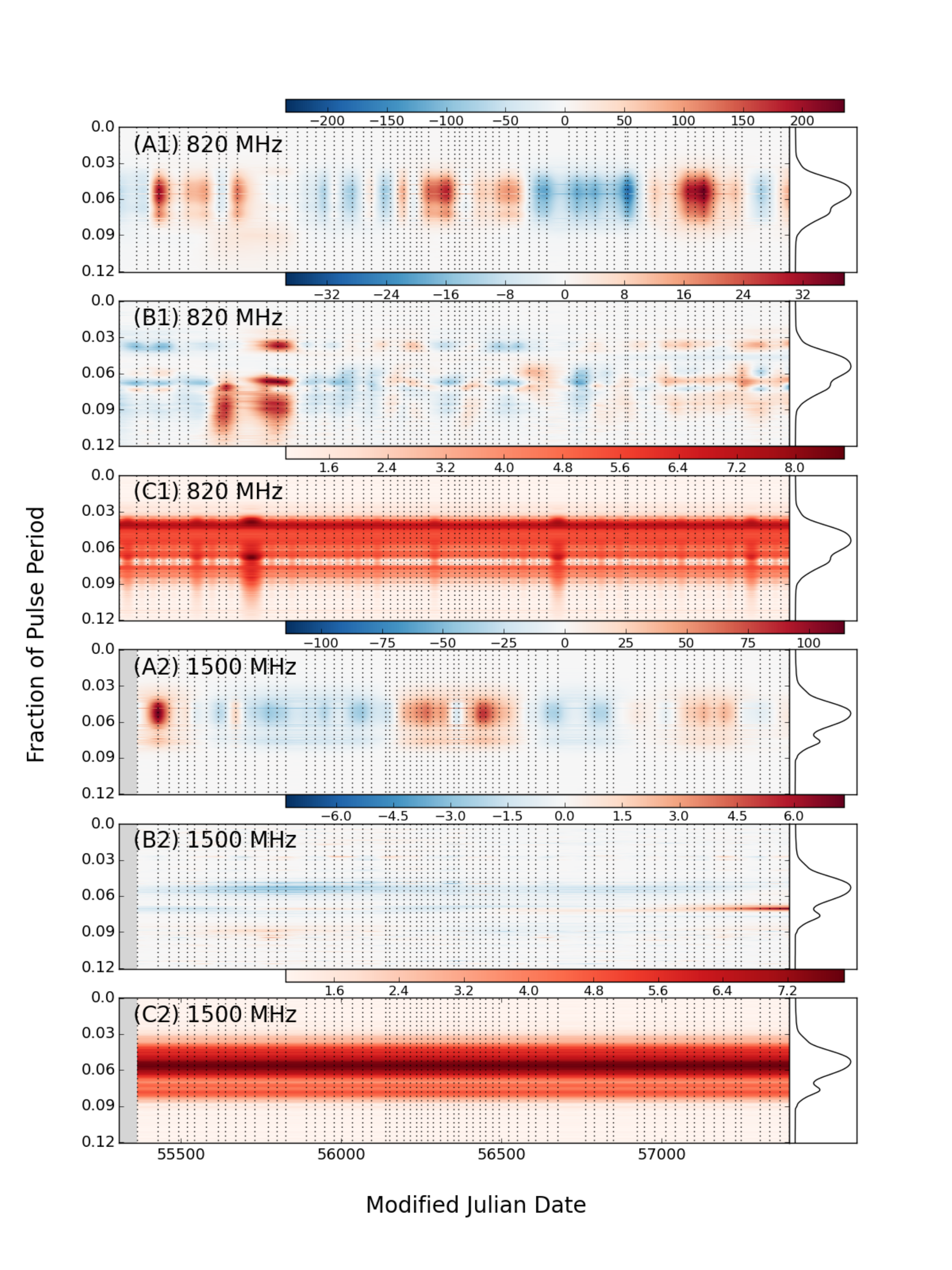} &
      \includegraphics[width=110mm]{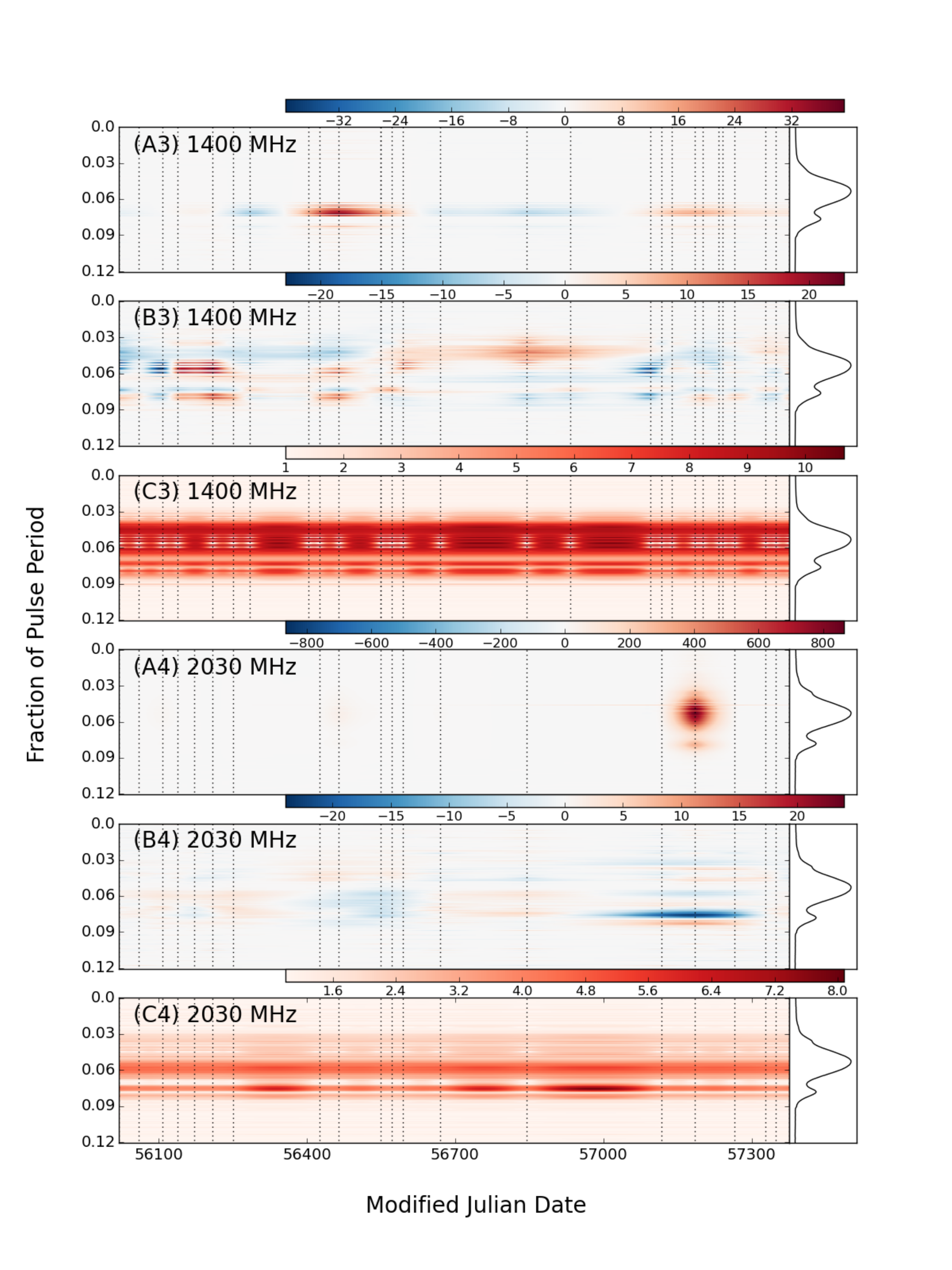}
  \end{tabular}}
  \rotatebox{90}{\begin{minipage}[c][\textwidth][c]{\wd0}
      \usebox0
      \caption{Variability maps for the main pulse of
        PSR~B1937$+$21. Otherwise as Figure~\ref{1713_vm}.}
      \label{1937_mp_vm}
  \end{minipage}}
\end{figure*}

\begin{figure*}[p]
  \sbox0{\begin{tabular}{@{}cc@{}}
      \includegraphics[width=110mm]{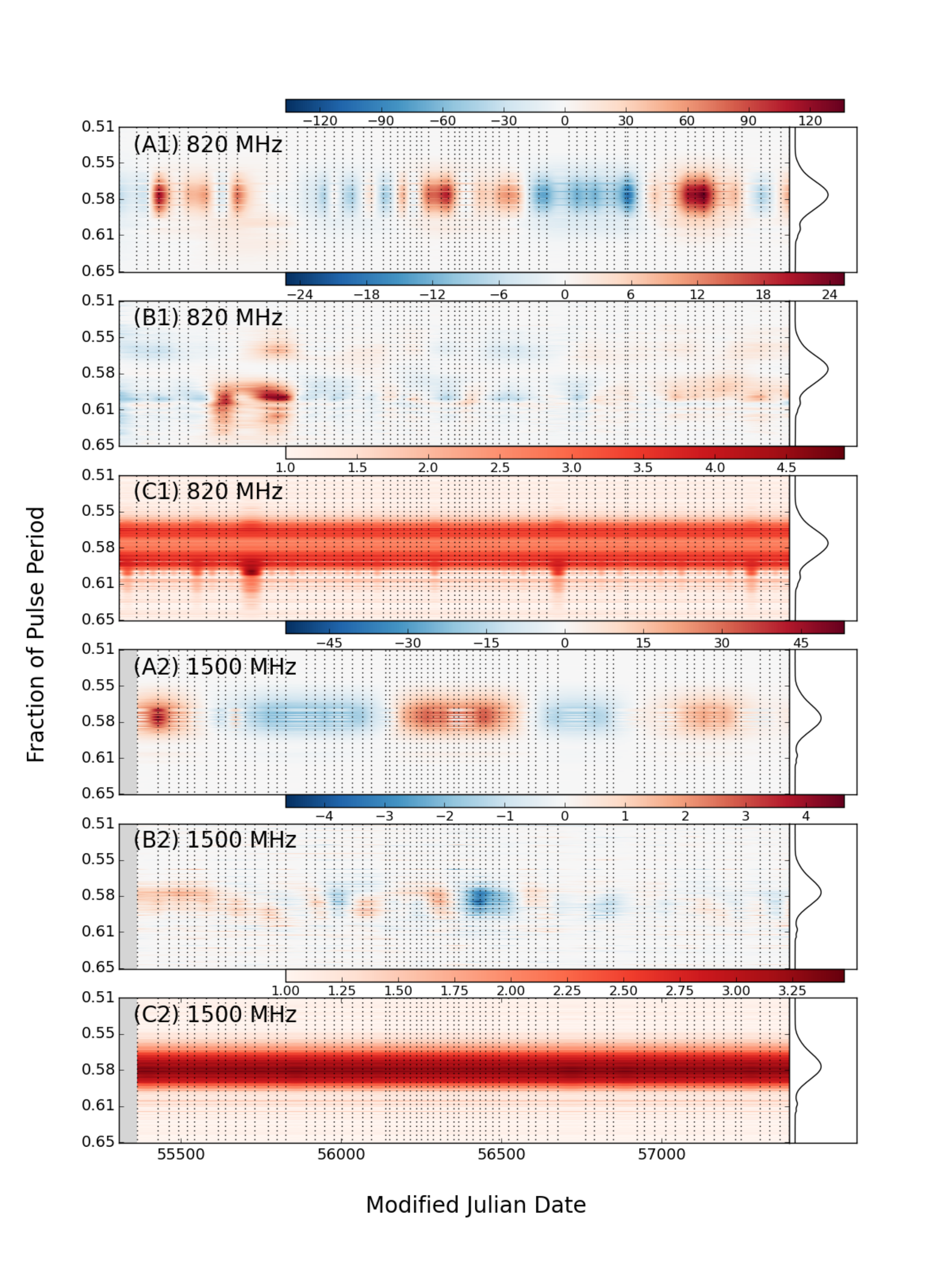} &
      \includegraphics[width=110mm]{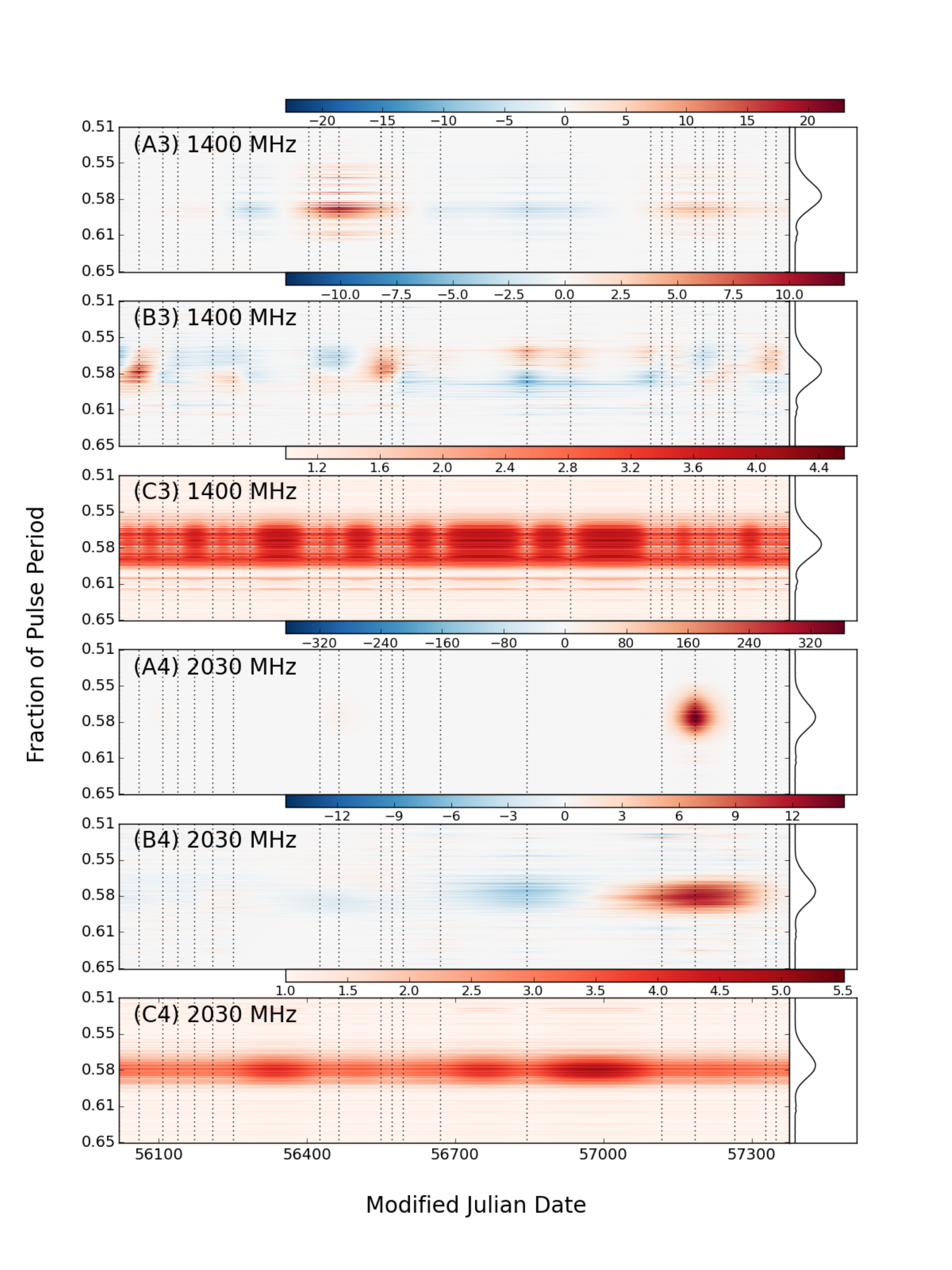}
  \end{tabular}}
  \rotatebox{90}{\begin{minipage}[c][\textwidth][c]{\wd0}
      \usebox0
      \caption{Variability maps for the interpulse of
        PSR~B1937$+$21. Otherwise as Figure~\ref{1713_vm}.}
      \label{1937_ip_vm}
  \end{minipage}}
\end{figure*}

\begin{figure*}[ht]
  \includegraphics[width=\textwidth]{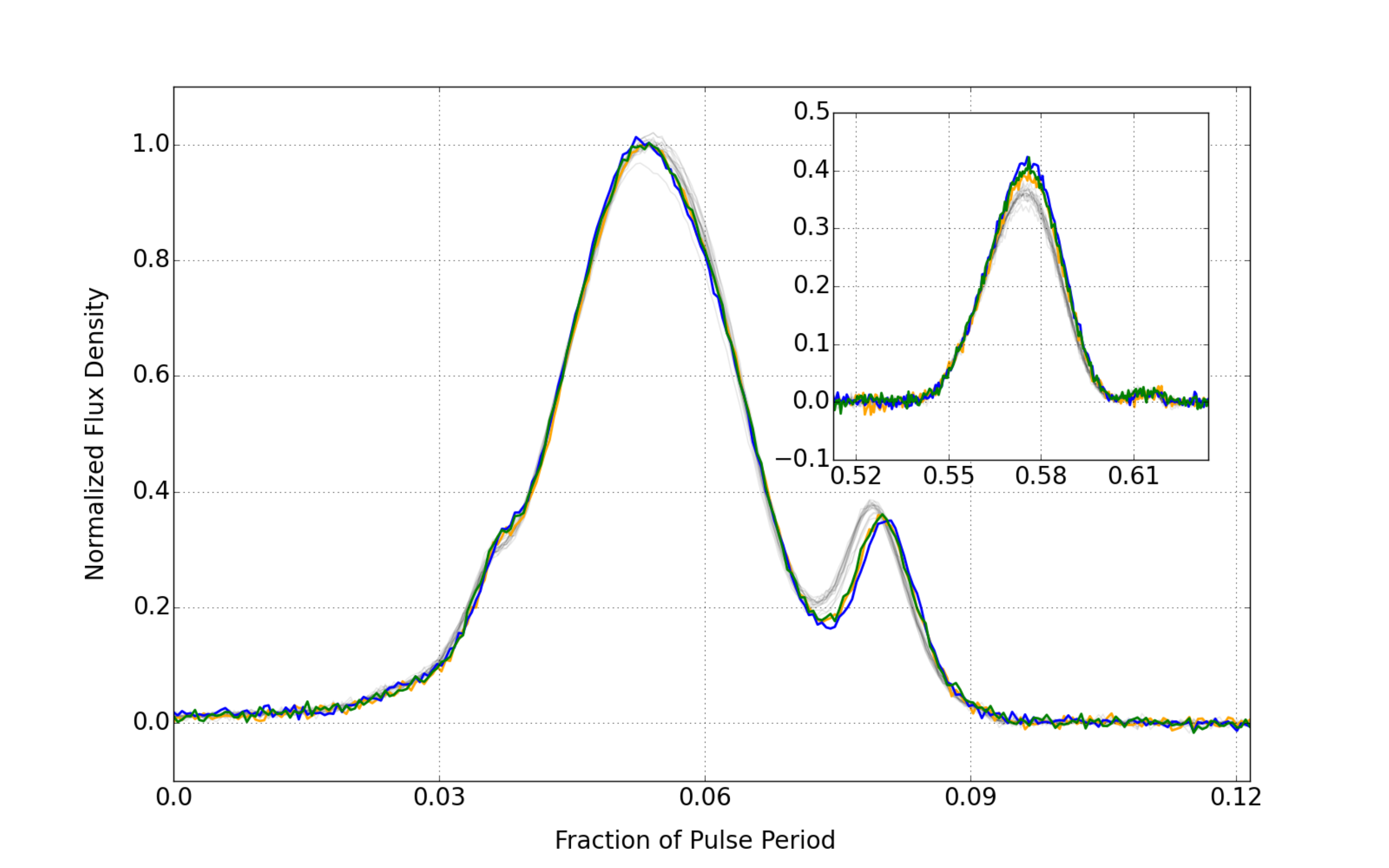}
  \caption{The pulse profile variability of PSR~B1937$+$21 at 2030~MHz. The main panel shows the main pulse, and the inset is the interpulse. The orange, blue and green profiles were
    observed on MJDs 57117, 57185 and 57265 respectively. The gray
    profiles show the other 17 pulse profiles in the data set. See
  also the discussion around polarization calibration in
  Section~\ref{instrumental} and Figure~\ref{1937_55977}.}
  \label{1937_extreme}
\end{figure*}

PSR~B1937$+$21 was the first MSP discovered
\citep{1982Natur.300..615B} and with a rotational frequency of 642~Hz,
remained the most rapidly spinning pulsar known for 24~years after it
was found. This bright, isolated MSP is one of the most precisely
timed pulsars, with the root mean square (rms) value of the white noise
component for the 11-year data set residuals being 109~ns
\citep{2018ApJS..235...37A}. It is also, however, one of
the few MSPs that displays measurable timing noise
\citep{2010ApJ...725.1607S}. Including both the red and white noise
components of the timing residuals, the rms calculated by Arzoumanian et
al. jumps up to 1.5 $\mu$s. Suggested interpretations of the red noise
include intrinsic changes in the spindown rate of the pulsar
\citep{1994ApJ...428..713K}, interstellar propagation effects \citep{1984Natur.307..527A, 1990ARA&A..28..561R, 1994ApJ...428..713K, 1995A&A...296..169C} and the presence of a circumpulsar
asteroid belt \citep{2013ApJ...766....5S}. PSR~B1937$+$21 has also
been seen to exhibit giant pulses; around one in every 10,000
individual pulses has more than 20 times the mean on-pulse flux
density and some pulses have around 300 times this average
\citep{1996ApJ...457L..81C}. This behavior is seen in both the main
and interpulse components, which are separated by approximately half a pulse
period. In our pulse profile analysis,
PSR~B1937$+$21 shows the most systematic variability, present
primarily at 820 and 2030~MHz and in both the main and interpulse
components. This can be seen in Metric~F of Table~\ref{var_table}, where the
ratios of systematic to noisy variability for PSR~B1937$+$21 are much
higher than those for PSR~J1713$+$0747. Panels B1 and B4 of
Figures~\ref{1937_mp_vm} and \ref{1937_ip_vm} also clearly
highlight the systematic evolution of the pulse profile shape over
time. The variability inferred by the GP model at 2030~MHz between $\sim$ MJDs
57000 and 57300 (see Panel~B4 of Figures~\ref{1937_mp_vm} and \ref{1937_ip_vm}) is induced
by three consecutive pulse profiles. They are compared with the rest
of the profiles in the data set in Figure~\ref{1937_extreme}. See
also the discussion around polarization calibration in Section~\ref{instrumental} and Figure~\ref{1937_55977}.

A direct comparison between the profile variability seen by AO at
1400~MHz and by GBT at 1500~MHz is made difficult primarily because
the observations at this frequency have the smallest ratio of
systematic to noisy variability of the PSR~B1937$+$21 data sets
(Metric~F of Table~\ref{var_table}). Therefore, much of the
variability is noisy, but only longer timescale systematic trends can be directly
compared, as the observations are generally not made on the same
days. Some systematic structure appears in the 1500~MHz GBT observation (Panel~B2 of
Figures~\ref{1937_mp_vm} and \ref{1937_ip_vm}), but the units of these
panels show that the systematic variability is weak, with the GP model
reaching levels only a few times higher than the levels of off-pulse
noise. At such levels, the GP model can be substantially influenced by the behavior
of even one or two pulse profiles. Additionally, the GBT and AO data sets
analyzed here span different dates; much of the systematic variability
in the 1500~MHz GBT observations occurs around MJD 56000, which is
before the 1400~MHz AO data were recorded by PUPPI. Furthermore, the
relatively sparse sampling of the AO observations also inhibits direct
comparison with the GBT GP variability models.


\subsection{PSR~J2145$-$0750}
\begin{figure*}[ht]
  \centering
  \begin{tabular}{@{}cc@{}}
    \includegraphics[width=\textwidth]{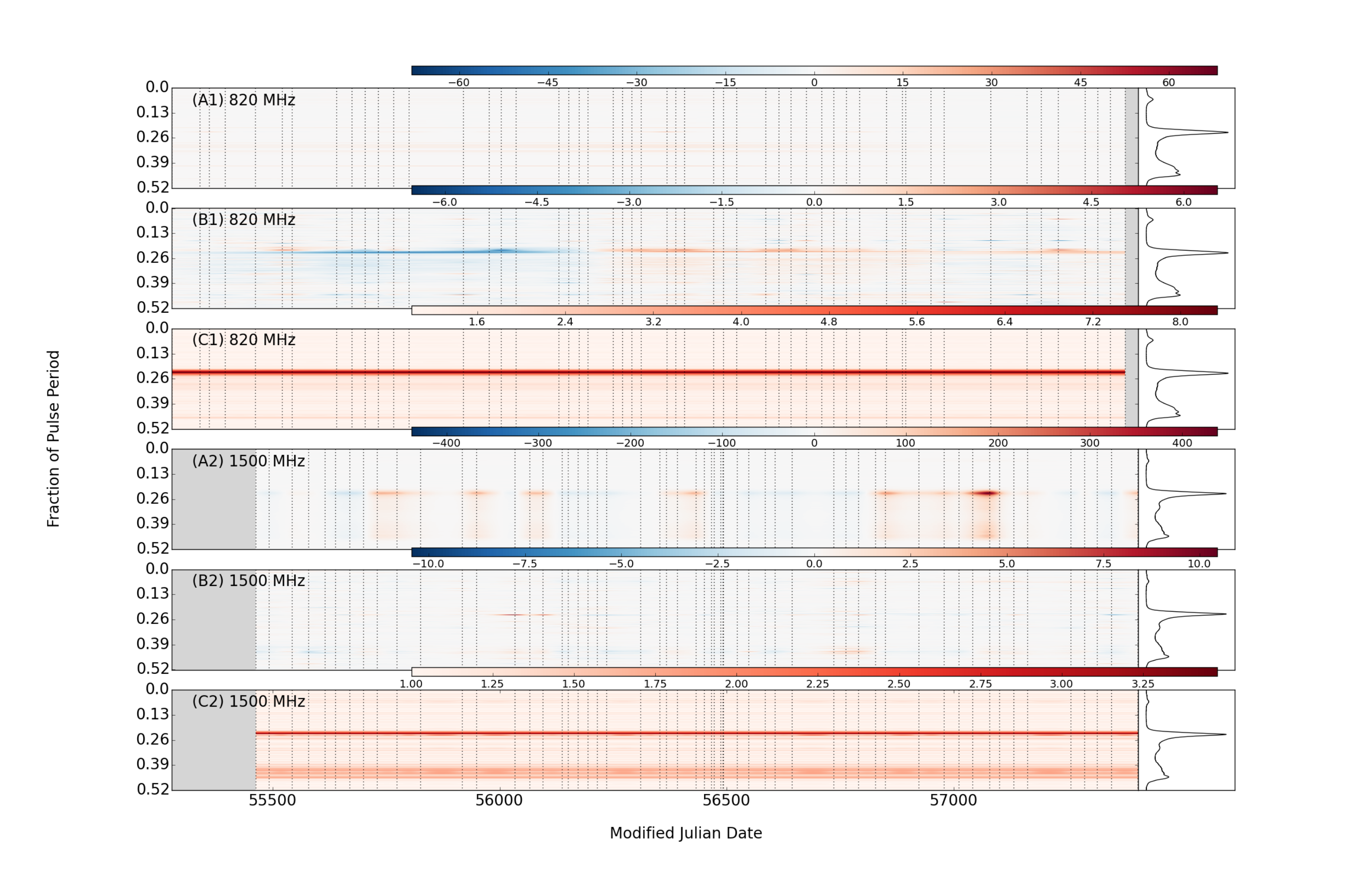}
  \end{tabular}
  \caption{Variability maps for PSR~J2145$-$0750. Otherwise as Figure~\ref{1713_vm}.}
  \label{2145_vm}
\end{figure*}

PSR~J2145$-$0750 has the third highest variability levels (by the metric in Metric~A of
Table~\ref{var_table}) of the pulsars in our analysis. It is
found to have a mean standard deviation of on-pulse phase bins that is
a factor of 1.76 larger than that of the off-pulse phase bins (at
820 MHz). As with PSR~J1713$+$0747, most of the variability is noisy;
Panel~B1 of Figure~\ref{2145_vm} shows a long-timescale change in the
pulse profile shape, but the magnitude of this change is small
compared to the standard deviation of the data (Panel~C1).

\subsection{PSR~J1643$-$1224}
\label{1643_results}
\begin{figure*}[ht]
  \centering
  \begin{tabular}{@{}cc@{}}
    \includegraphics[width=\textwidth]{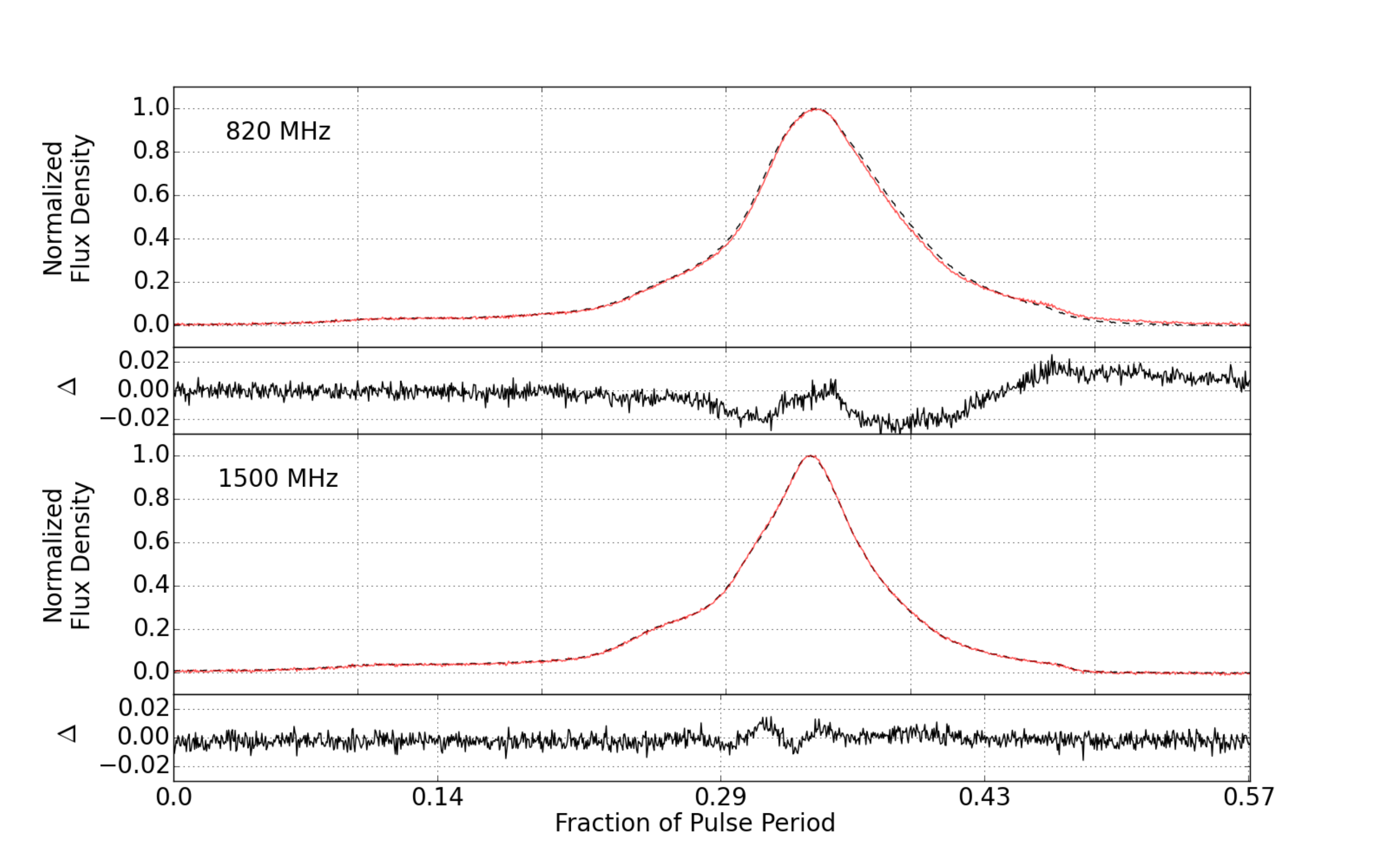}
  \end{tabular}
  \caption{The combined pulse profiles of PSR~J1643$-$1224 before
    (red) and after (black dashed) MJD 57074 (2015 February 21). The
    narrower panels show the red profile minus the black dotted profile
    ($\Delta$). All profiles are normalized to the peak.}
  \label{1643_comp}
\end{figure*}

\begin{figure*}[ht]
  \centering
  \begin{tabular}{@{}cc@{}}
    \includegraphics[width=\textwidth]{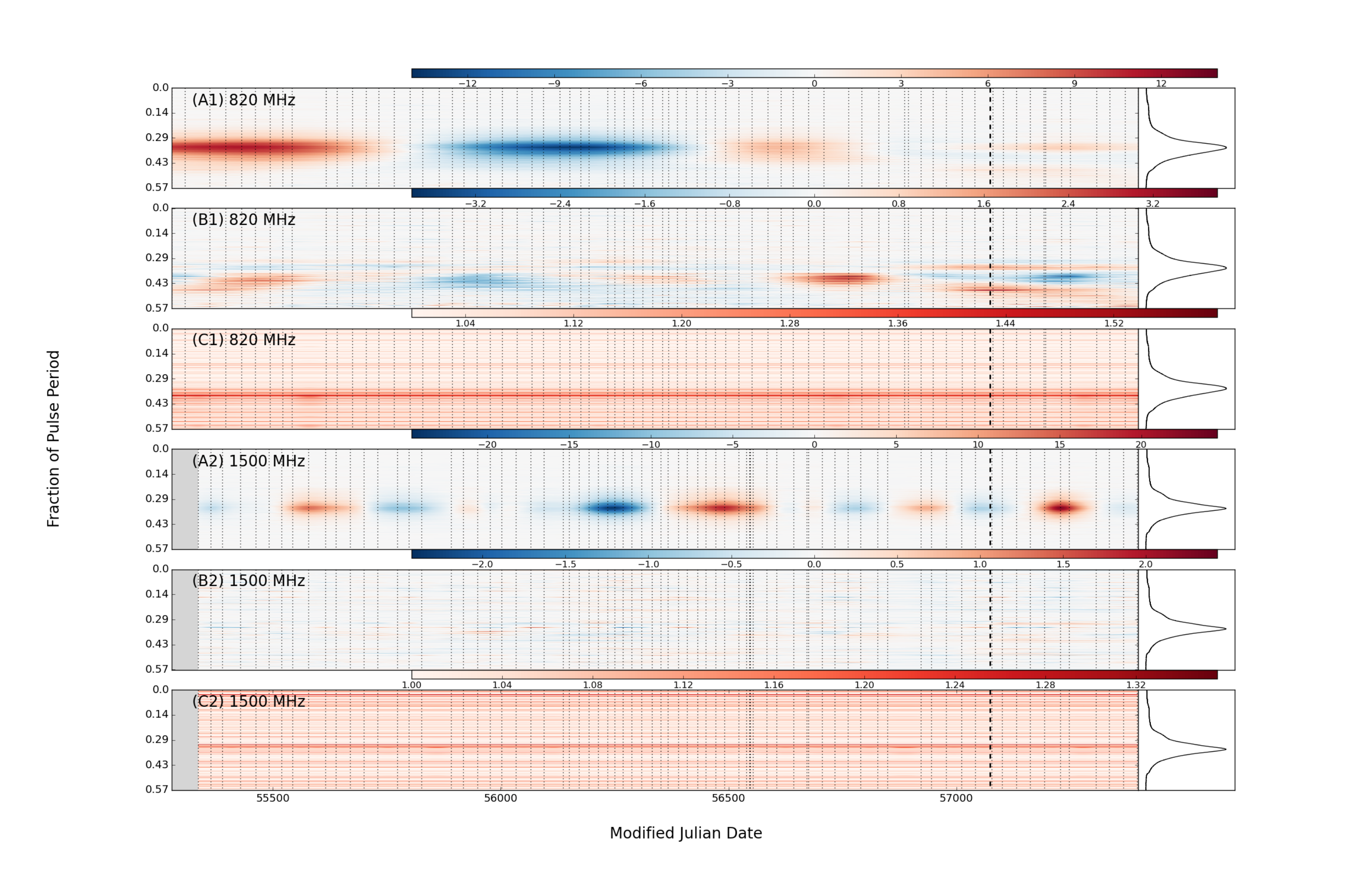}
  \end{tabular}
  \caption{Variability maps for PSR~J1643$-$1224. Otherwise as Figure~\ref{1713_vm}.}
  \label{1643_vm}
\end{figure*}

PSR~J1643$-$1224 has been observed since 2003 as part of the Parkes
Pulsar Timing Array project \citep{2013PASA...30...17M} at 700, 1400
and 3100~MHz. \citet{2016ApJ...828L...1S} noticed TOA perturbations,
which they attributed to unmodeled changes in pulse shape, observed to
occur around MJD~57074 (2015 February 21). These timing perturbations
are most significant at 3100~MHz, and Shannon et al. only show profile
changes at that frequency. Figure~\ref{1643_comp} provides a
comparison of the pulse profile shapes of PSR~J1643$-$1224 before and
after 2015 February 21, as observed by the GBT. The upper panels show
a significant difference at 820~MHz, but little change at
1500~MHz. \citet{2016ApJ...828L...1S} compare the shape variations of
PSR~J1643$-$1224 to those observed in PSR~J0738$-$4042
\citep{2011MNRAS.415..251K} and also point out that the new components
are unpolarized in both pulsars. A new component is seen to appear in
PSR~J0738$-$4042 after a drifting feature is observed to move
centrally over a span of $\sim$ 100 days
\citep{2014ApJ...780L..31B}. The variability map in Panel~B1 of Figure~\ref{1643_vm} shows
that changes in the profile shape appear to be occurring across the
data set, and not just abruptly after MJD~57074. In particular, red
colored drifting features can be seen at the beginning and end of the
data set. The drifting at the end of the data, in which an emission feature moves
away from the center of the pulse profile over a few hundred days, is
more clearly shown in Figure~\ref{1643_drift}.
  
\section{Discussion}
\label{disc}
We have used a new profile alignment technique, GP regression and
multiple metrics to characterize the pulse
profile evolution of 78 NANOGrav MSP data sets.
All pulsars show flux density variations due to DISS and RISS. After
flux density levels are normalized, the differences between the
constant average model (for a particular data set) and the observed
profiles for most of the pulsars is consistent with being due to
additive white Gaussian noise; for the vast majority of pulsars, the
mean standard deviation of their on-pulse phase bins is less than a
factor of two greater than that of their off-pulse bins.
The three pulsars for which this factor
is greatest are PSRs~J1713$+$0747, B1937$+$21 and
J2145$-$0750. Additionally, PSR~J1643$-$1224 shows significant
long-term variability, which has been previously identified.
\subsection{Profile and Timing Variability in J1643$-$1224}
\begin{figure*}[ht]
  \centering
  \begin{tabular}{@{}cc@{}}
    \includegraphics[width=\textwidth]{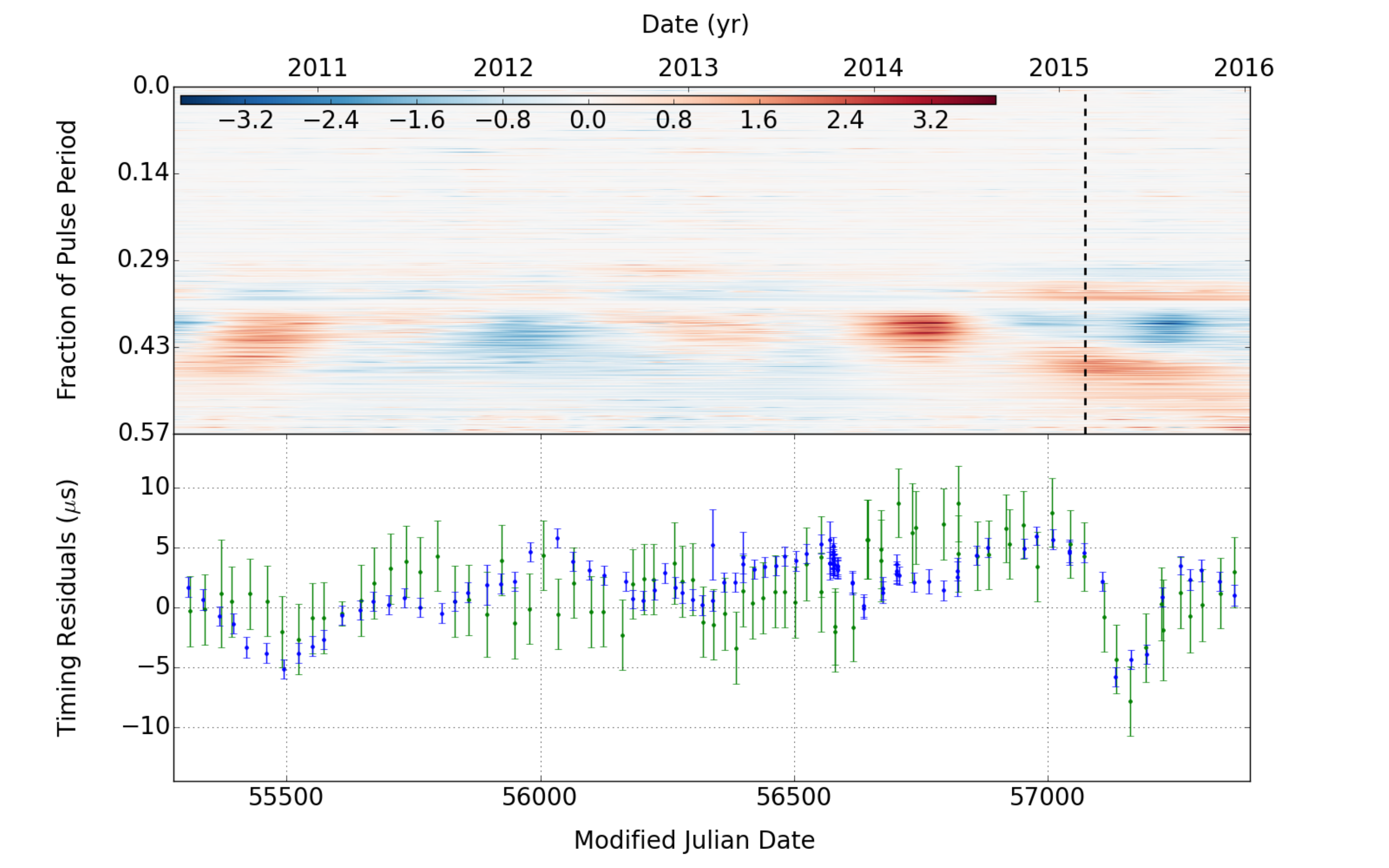}
  \end{tabular}
  \caption{The profile and timing residuals of the J1643$-$1224
    820~MHz data set observed at the GBT. The top panel is a variability map showing pulse
    profile shape changes after the observations have been normalized.
  Red regions indicate where the inferred pulse profile has an excess
  of flux density compared to the average for the data set. Blue
  indicates where it has a deficit. The unit for the
  variability map is the mean of the standard deviation of the
  off-pulse phase bins for the data set. This panel depicts the same
  data as Panel~B1 of Figure~\ref{1643_vm}, with the addition of a
  solid vertical line, which denotes 2015 February 21 (MJD 57074),
  the date around which timing and profile changes were seen by \citet{2016ApJ...828L...1S}. The bottom panel shows the TOA residuals for PSR~J1643$-$1224 \citep{2018ApJS..235...37A} at 820~MHz (green) and 1500~MHz (blue).}
  \label{1643_profile_timing}
\end{figure*}

\begin{figure*}[ht]
  \centering
  \begin{tabular}{@{}cc@{}}
    \includegraphics[width=\textwidth]{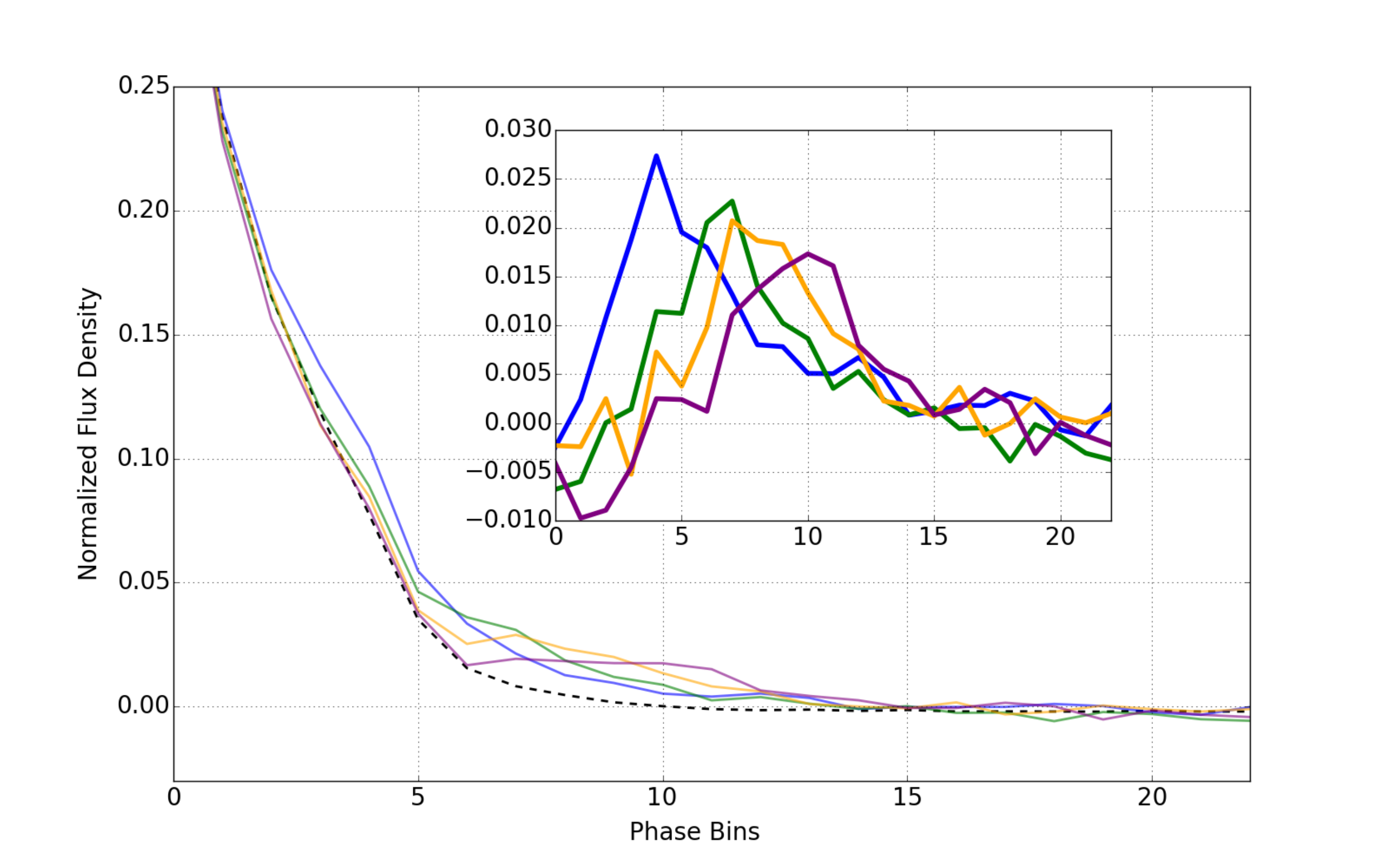}
  \end{tabular}
  \caption{The main plot shows the trailing edge of the pulse profile
    for PSR~J1643$-$1224 observed at 820~MHz. The 2048 phase bins that span the pulse period have been resampled to 64 in order to increase the profile S/N. The average profile for the data set is shown by a black dashed line. The four solid lines are the profiles as observed on MJDs~57102 (blue), 57307 (green), 57336 (orange) and 57369 (purple). The inset shows the difference between each colored profile in the main plot and the average pulse profile (observation minus average). The phase location of the maximum deviation can be seen to drift from left to right with time.}
  \label{1643_drift}
\end{figure*}

As mentioned in Section~\ref{1643_results}, around 2015 February 21 (MJD 57074), \citet{2016ApJ...828L...1S}
observe both TOA and pulse profile variations in PSR J1643$-$1224. The
largest TOA perturbations are at 3100~MHz. They report a change that leaves
permanent excess power in the leading edge of the 3100~MHz and
1400~MHz profiles. Their 700~MHz observations show the least amount of
timing variation around this date. TOA perturbations that begin around MJD~57074 can also be
seen in NANOGrav data at both 820~MHz and 1500~MHz (Figure~\ref{1643_profile_timing}).
As seen in column~3 of Table~\ref{var_table}, however, the
PSR~J1643$-$1224 1500~MHz pulse profiles are the most stable of all the
data sets analyzed in this work. This suggests that the pulse profile
changes are not the cause of the TOA disruptions.
At 820~MHz, we see more obvious pulse profile changes, but they occur
across the whole data set, rather than abruptly around MJD~57074, as
seen by Shannon et al. at 3100~MHz. The 820~MHz
changes seem to occur primarily at the trailing edge of the profile,
which is, again, in contrast to the 3100~MHz Parkes data, in which
Shannon et al. see a significant excess signal in the leading edge 
after $\sim$ MJD~57074, occurring concurrently with a significant change in the
timing residuals. At around the same time, the NANOGrav 820~MHz GBT data
show a profile feature drifting away from the central peak over a span
of a few hundred days (Figure~\ref{1643_drift}). A similar phenomenon occurred in the Crab pulsar in 1997
\citep{2000ApJ...543..740B}, which has been attributed to
refraction and multiple imaging at the edge of a plasma cloud in the
outer region of the Crab Nebula \citep{2011MNRAS.410..499G}.
Only profiles at 3100~MHz are shown in the Shannon et al. paper, and
so a direct comparison of Parkes and GBT PSR~J1643$-$1224 pulse
profiles at similar frequencies around the time of the TOA disturbance
has yet to been done.

Whenever considering the pulse phase at which profile variability occurs,
it should be understood that different methods for alignment can show the variability to occur at
different parts of the pulse profile.
\subsection{The Effect of Pulse Profile Shape Changes on TOAs}

A TOA is determined by a technique that matches the pulse
profile from an individual observation, with a static pulse profile
model \citep{Taylor117, 2006ApJ...642.1004V}. Any evolution of the pulse profile with time,
therefore, will affect the TOA that is produced by this template
matching procedure. As discussed in Section~\ref{intro}, in the NANOGrav
11-year data set each frequency subband is used to produce TOAs
via the template matching procedure. It should be stressed,
therefore, that there are phenomena that could cause profile changes in the
frequency-integrated pulse profiles, but have little effect on the
profiles of individual frequency channels and, therefore, on the
NANOGrav TOAs.

In order to assess the magnitude of changes in TOA that result from
the frequency-integrated pulse profile shape variability that we have
seen, we employ the template matching procedure using the
\emph{PYPULSE} software package \citep{2017ascl.soft06011L}.
The \emph{fitPulse} function performs the template matching
procedure described in \citet{Taylor117}; any two pulse
profiles are cross-correlated in order to calculate a difference in
TOA between them.

Before the template matching analysis was carried out, the relative
alignment of the pulse profiles was performed.
We have employed a new, objective pulse profile alignment technique
that maximizes the number of pulse phase bins
that are in agreement between profiles (see Section~\ref{align_scale}
for details).
The nature of the alignment technique is such that we are
  insensitive to phase shifts caused by astrophysical processes such
  as timing noise.
In this paradigm, the definition of the \emph{fiducial point} (a
reference point for timing measurements) becomes the phase at which
the pulse profile
has the least variability; without knowledge of the physical processes
involved in the profile shape changes, we assert that this is
a reasonable thing to do. In some cases, the profile shape change is
quite dramatic and consequently has a dramatic effect on the TOA, as
calculated in
Table~\ref{template_table}.
The mean and standard deviations in the table may be dominated by such
outliers and be skewed
as a consequence. The metric of $\left<|\Delta\rm {TOA}|\right>$ and
standard timing residuals are
difficult to compare; we are not using traditional pulsar timing.
Instead we are instead essentially assuming that the pulsar is a
perfect rotator and have defined the fiducial
point as the most stable phase of the pulsar.
Additionally, as discussed in Section~\ref{align_scale}, we are only aligning in
single bin increments (with 2048
bin resolution) as it is sufficient for the profile profile
variability
analysis that is the focus of this work. Not aligning to fractions of
a phase bin may also inflate the values in Table~\ref{template_table}.

Table~\ref{template_table} shows the TOA changes
($\Delta\rm {TOA}$) induced by the pulse profile shape changes seen
in PSRs J1643$-$1224, J1713$+$0747, B1937$+$21 and J2145$-$0750. The
template model used is the average profile of all normalized
observations that survived the analysis in Section~\ref{analysis}. A
value of $\Delta\rm {TOA}$ was calculated for each
observation. Template matching the average model with itself produces
a $\Delta\rm {TOA}$ value of zero by definition.

\begin{deluxetable*}{cccccc}
  \tablecaption{The changes in TOA induced by pulse profile variation
    for PSRs J1643$-$1224, J1713$+$0747, B1937$+$21 and
    J2145$-$0750.}
  \tablehead{\colhead{Pulsar} & \colhead{Observing Frequency (MHz)} & \colhead{$\left<|\Delta\rm {TOA}|\right>$ ($\mu$s)} & \colhead{$\sigma\Delta\rm {TOA}$ ($\mu$s)} & \colhead{Max. $\Delta\rm {TOA}$ ($\mu$s)} & \colhead{$\left<\sigma\rm {TOA}\right>$ (ns)}}
  \startdata
  J1643$-$1244 & 820 & 0.75 & 0.93 & 5.00 & 3.44\\
  J1643$-$1244 & 1500 & 0.39\hspace{0.3cm}0.44 & 0.34\hspace{0.3cm}0.40 & 1.65\hspace{0.3cm}1.86 & 3.70\hspace{0.3cm}3.52\\
  J1713$+$0747 & 1400 & 1.37 & 2.36 & 12.81 & 2.50\\
  J1713$+$0747 & 2030 & 1.27 & 1.45 & 7.40 & 4.90\\
  J1713$+$0747 & 1500 & 0.89\hspace{0.3cm}0.88 & 0.84\hspace{0.3cm}0.84 & 4.94\hspace{0.3cm}3.92 & 2.92\hspace{0.3cm}2.90\\
  J1713$+$0747 & 820 & 0.92 & 0.92 & 5.00 & 4.21\\
  B1937$+$21 & 820 & 0.21 & 0.15 & 0.74 & 0.74\\
  B1937$+$21 & 1500 & 0.23\hspace{0.3cm}0.17 & 0.25\hspace{0.3cm}0.14 & 2.08\hspace{0.3cm}0.58 & 0.74\hspace{0.3cm}0.64\\
  B1937$+$21 & 1400 & 0.16 & 0.11 & 0.43 & 0.84\\
  B1937$+$21 & 2030 & 0.22 & 0.12 & 0.48 & 1.88\\
  J2145$-$0750 & 820 & 3.26 & 2.61 & 10.10 & 15.26\\
  J2145$-$0750 & 1500 & 1.96\hspace{0.3cm}1.92 & 1.33\hspace{0.3cm}1.22 & 5.42\hspace{0.3cm}4.12 & 15.18\hspace{0.3cm}16.10\\
  \enddata
  \tablecomments{$\left<|\Delta\rm {TOA}|\right>$ is the mean of
    the absolute TOA change induced by the pulse profile shape
    changes. $\sigma\Delta\rm {TOA}$ is the standard deviation of
    the $\Delta\rm {TOA}$ distribution. Max. $\Delta\rm {TOA}$
    is the largest TOA changed induced in the data set, and
    $\left<\sigma\rm {TOA}\right>$ is the mean uncertainty in the
    TOA calculations. The data sets observed at 1500~MHz have two
    values for each variability metric. The left of the pair relates
    to profiles that were polarization calibrated only by a local noise diode, and the right to
    profiles that were additionally polarization calibrated using the full Mueller matrix (see
    Section~\ref{analysis}).\label{template_table}}
\end{deluxetable*}
In the 12 data sets analyzed, the average value of the magnitude
of $\Delta\rm {TOA}$ induced by the changing pulse profile
shape is typically around three orders of magnitude larger than the
average 1$\sigma$ uncertainty of the TOA measurements
$\left<\sigma\rm {TOA}\right>$.
\\

In general, the potential causes of the pulse profile variability are
scintillation, inaccurate DM, scatter broadening,
instrumental and interference issues, jitter or other emission changes
intrinsic to the pulsar. We discuss each possibility in detail in the
following.
\subsection{Diffractive Interstellar Scintillation (DISS)}
\begin{figure*}
  \begin{center}
    \includegraphics[width=170mm]{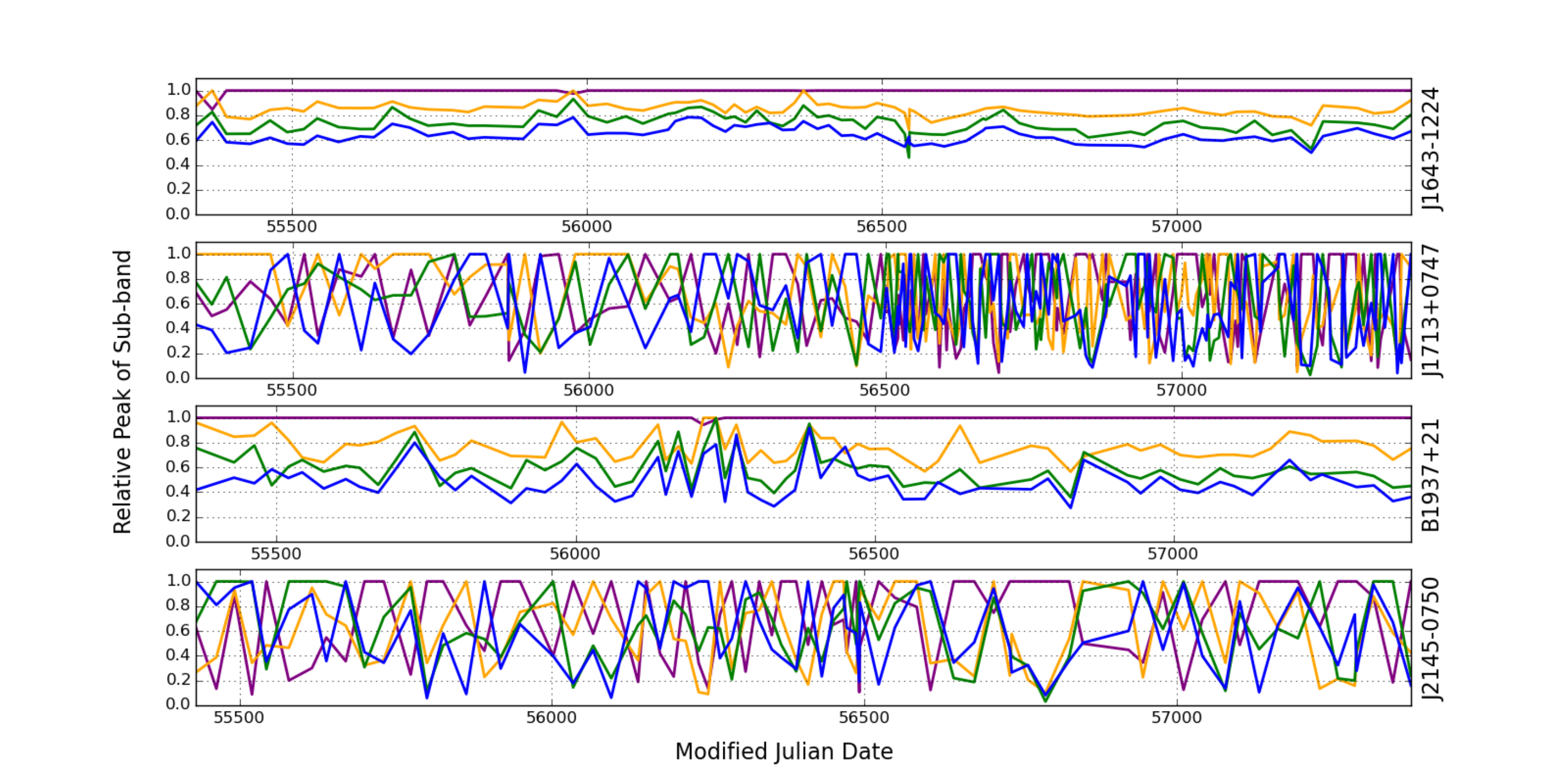}
    \caption{The relative brightness of frequency subbands within the
      observing band at 1500~MHz. The $\sim$ 800~MHz bandwidths are
      divided into four $\sim$ 200~MHz bandwidths. From lowest to highest
      frequency, the subbands are represented by purple, orange, green
      and blue. From top to bottom, the panels show the relative
      subband brightness for PSRs~J1643$-$1224, J1713$+$0747, B1937$+$21 and
      J2145$-$0750.}
    \label{rel_flux}
  \end{center}
\end{figure*}

DISS is the frequency-dependent modulation of pulsar flux density.
If a pulse profile is a
composite of a wide range of equally weighted frequency channels (as it is in this
analysis), scintillation will necessarily lead to pulse profile
changes, providing that (i) the profile evolves with frequency across
the observing band, (ii) the scintillation bandwidth is not
much smaller than the observing bandwidth and (iii) the
scintillation timescale is not much smaller than the timescale of
the observation. In each data set for PSRs~J1643$-$1224, J1713$+$0747, B1937$+$21
and J2145$-$0750, scintillation is occurring to differing degrees,
affecting the relative flux in different parts of the observing band
(see Figure~\ref{rel_flux}). There is also some pulse profile shape
evolution across the observing band for PSRs~J1713$+$0747, B1937$+$21
and J2145$-$0750. Therefore, scintillation plays
at least some part in the pulse profile variability seen in the
analysis of these three pulsars. Relatively little pulse profile shape evolution is seen
across the observing band for PSR~J1643$-$1224.

\cite{2016ApJ...818..166L} showed that the average scintillation
bandwidth for PSR~B1937$+$21 at 1500~MHz is around 2.8~MHz, which is close to the
resolution limit. \citet{2013MNRAS.429.2161K} give a value of
1.2~MHz at a reference frequency of 1500~MHz. The fractional
uncertainty in the scintillation bandwidth is $1/\sqrt{N_{iss}}$, where
\begin{equation}
  N_{iss} \approx \left(1+\zeta \frac{\Delta \nu}{\Delta
    \nu_{d}}\right)\left(1+\zeta \frac{T}{\Delta t_{d}}\right),
  \label{frac_uncertainty}
\end{equation}
and where $\zeta$ is an empirically determined coefficient ($\zeta \approx$ 0.1-0.2), $\Delta \nu$ is
the receiver bandwidth, $\Delta \nu_{d}$ is the scintillation bandwidth,
$T$ is the integration time of the observation and $\Delta t_{d}$ is the
scintillation timescale \citep{1990ApJ...349..245C}.
Using $\Delta \nu_{d} =$ 1.2~MHz and $\Delta t_{d} =$ 327~s from
\citet{2013MNRAS.429.2161K} and $\zeta =$ 0.2, the fractional
uncertainty in scintillation bandwidth for a 30 minute, 800~MHz
bandwidth observation is $\sim$ 6\%
For the 1400 and 2030 MHz centered observations in
this analysis, the observing bandwidth is 800~MHz, for 1500~MHz it is
700~MHz and for the 820~MHz centered observations it is 200~MHz. As
these bandwidths are so much larger than the scintillation
bandwidths for PSR~B1937$+$21, we expect to see scintillation effects largely
averaging out across the observing band.
Figure~\ref{rel_flux} shows
that although the scintillation observed in PSR~B1937$+$21 is much
less than that observed in PSRs~J1713$+$0747 and J2145$-$0750, the
relative weighting of different parts of the observing band does
change with time.

Levin et al.\ were unable to calculate the scintillation
bandwidth for PSR~J1643$-$1224, limited by the frequency resolution of
their observations. \citet{2013MNRAS.429.2161K} give a value of 22~kHz
at a reference frequency of 1500~MHz. Using
Equation~\ref{frac_uncertainty}, setting $\Delta \nu_{d}$ to 22~kHz and $\Delta t_{d}$ to 582~s from
Keith et al.\ and setting $\zeta$ to 0.2, gives a
scintillation bandwidth fractional uncertainty for a 30 minute, 800~MHz bandwidth observation of
$\sim$ 1\%. Again, these observing bandwidths are so much
larger than the scintillation bandwidth, we expect to see
scintillation effects averaging out across the observing band. Despite
this, the top panel of Figure~\ref{rel_flux} indicates that scintillation is causing some
changes in the relative flux density across the observing band for PSR~J1643$-$1224.

The average scintillation bandwidth reported at 1500~MHz by
Levin et al.\ is 21.1~MHz for PSR J1713$+$0747 and
47.8~MHz for PSR~J2145$-$0750. The second to top and bottom panels in
Figure~\ref{rel_flux} indicate clear scintillation for
PSRs~J1713$+$0747 and J2145$-$0750 respectively.

\begin{figure*}
  \begin{center}
    \includegraphics[width=145mm]{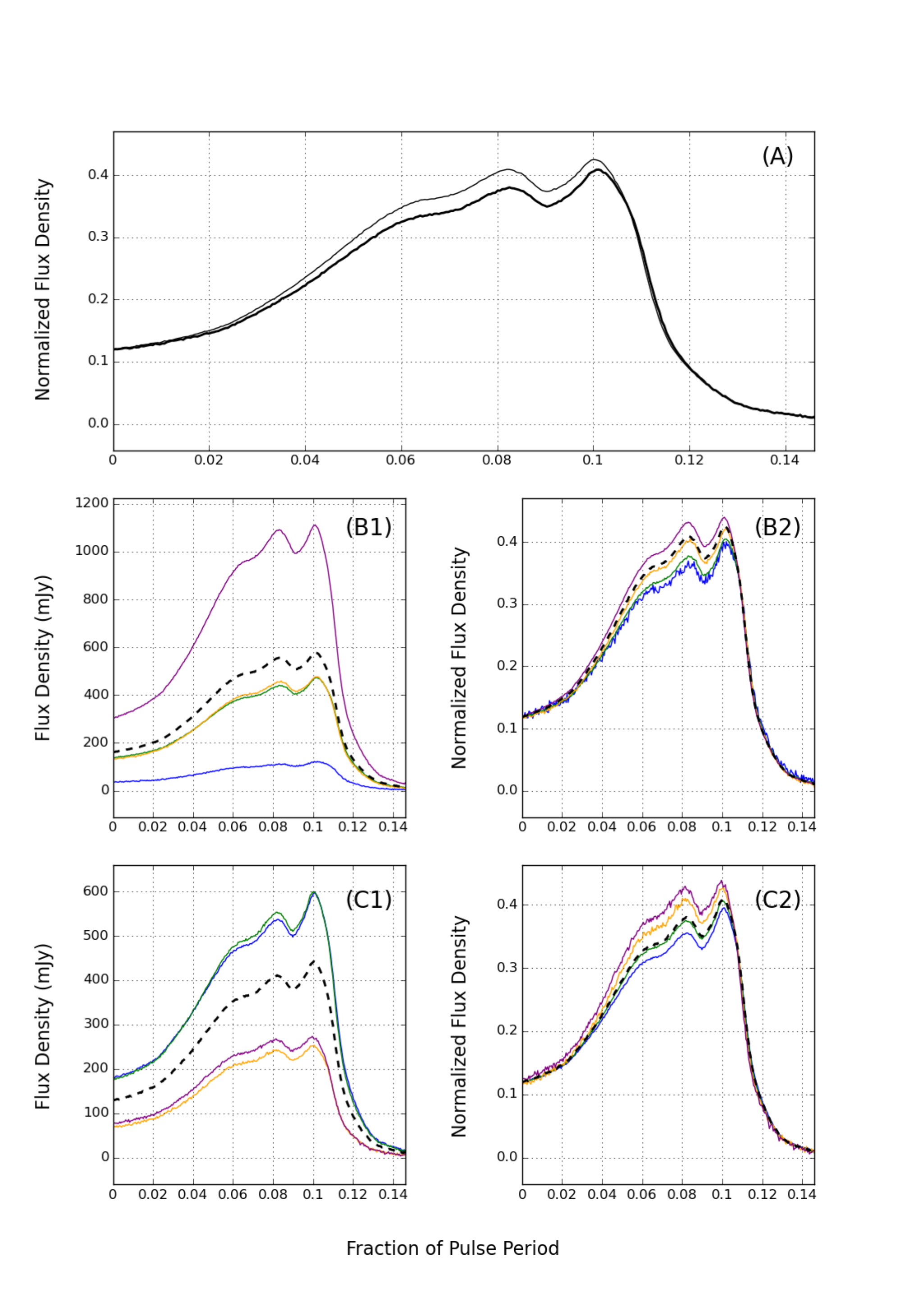}
    \caption{Two contrasting pulse profiles of PSR~J2145$-$0750 at
      820~MHz. Panel A shows two pulse profiles aligned and normalized
      by their peaks. The peak is not shown; all panels focus only on a subsection of the profile. This is done to allow shape changes to be seen
      clearly. The thin line is the pulse profile as observed on
      MJD~55361 and the thick line on MJD 56792. Panel B1 shows the
      MJD~55361 observation split into four frequency subbands, each
      spanning 50~MHz. From lowest to highest frequency, the subbands
      are represented by purple, orange, green and blue profiles. The
      black dashed line shows the frequency-integrated profile. In
      Panel B2, the subband and frequency-integrated profiles have
      also been normalized to, and aligned by, the peak. Panel C1 and C2
      show the same as B1 and B2 respectively, but for the observation
      made on MJD 56792.}
    \label{2145_820_sub}
  \end{center}
\end{figure*}

Figure~\ref{2145_820_sub} illustrates the nature of typical pulse
profile variations that we see in the PSR J2145$-$0750 data set
at 820~MHz. The figure shows that when divided into four $\sim$ 50~MHz
frequency bands, the pulse profile shapes of the subbands
are largely stable between MJDs 55361 and 56792 (see Panels~B2
and C2) and the relative flux densities are not (see Panels~B1 and
C1). Between these two observation dates, the relative weighting of
parts of the observing band has been changed by scintillation. As the
different parts of the band have different profile shapes, a modification of
the frequency-integrated pulse profile necessarily results. The pulse profile changes
that are seen in PSR~J2145$-$0750 are, therefore, consistent with the effects of
scintillation.

\begin{figure*}[p]
  \sbox0{\begin{tabular}{@{}cc@{}}
      \includegraphics[width=110mm]{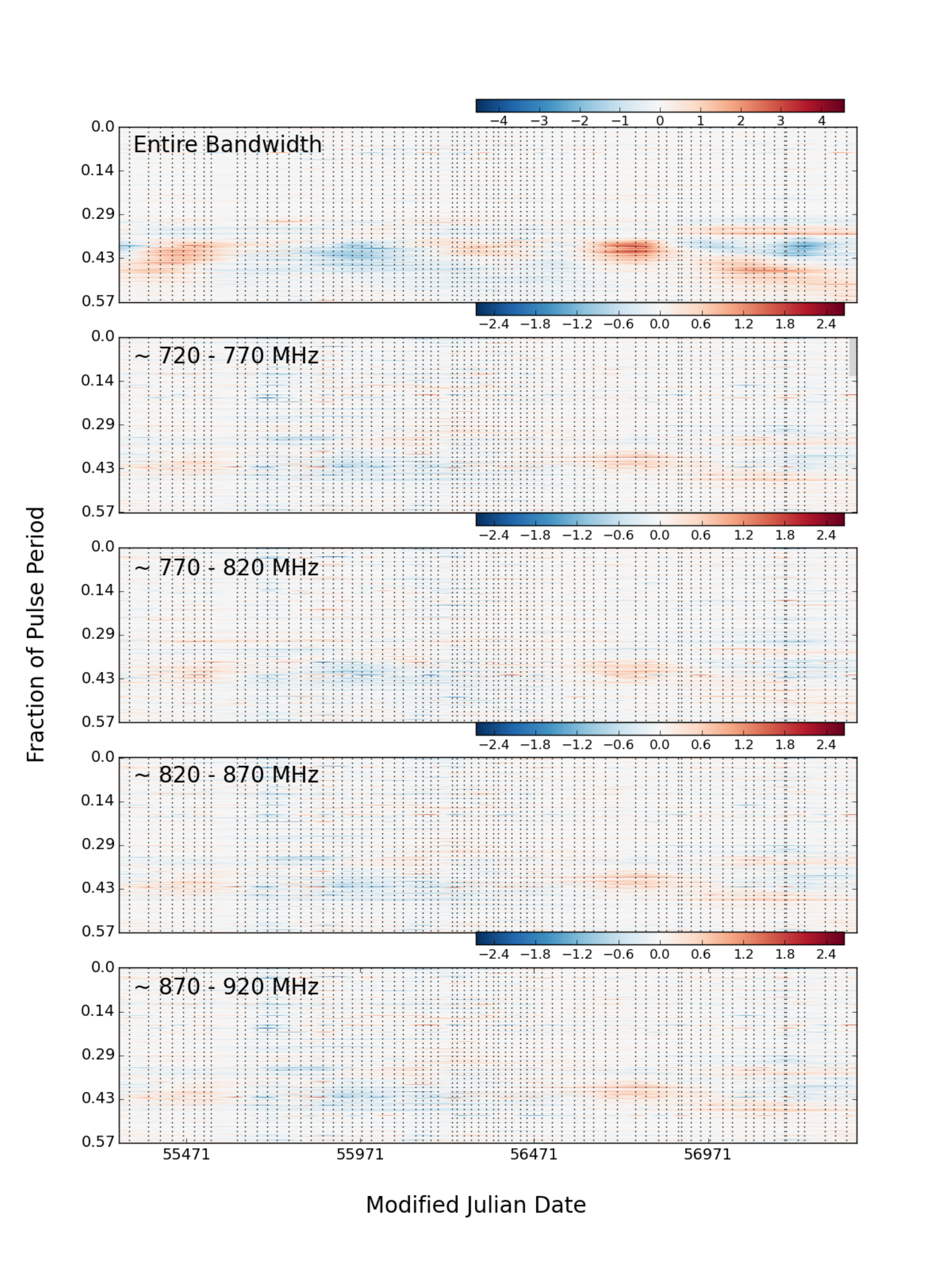} &
      \includegraphics[width=110mm]{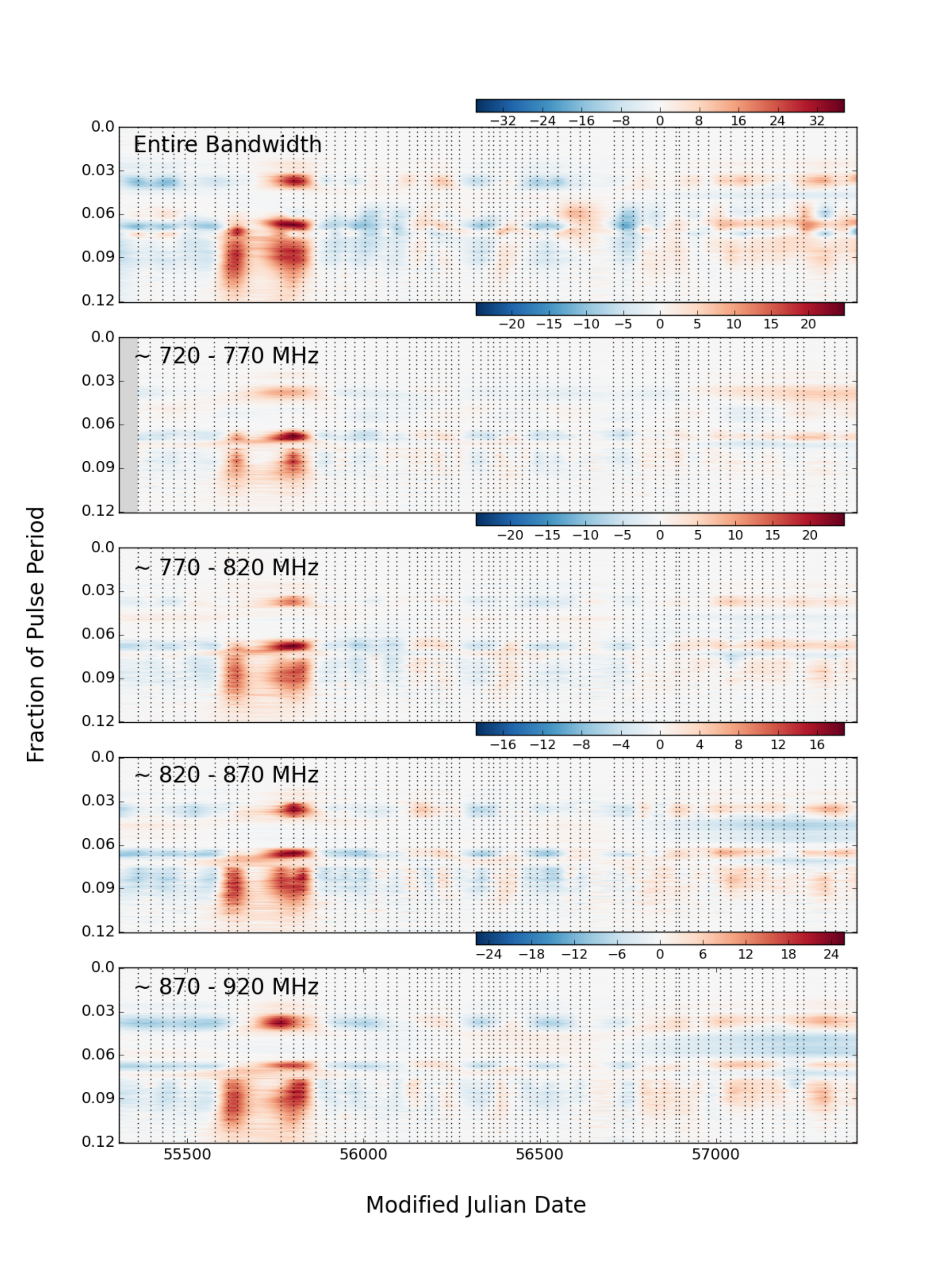}
  \end{tabular}}
  \rotatebox{90}{\begin{minipage}[c][\textwidth][c]{\wd0}
      \usebox0
      \caption{Variability maps for GUPPI data
        sets broken down into subbands of observational frequency. The left panel
        panel shows observations of PSR~J1643$-$1224 with a central frequency of
        $\sim$ 820~MHz. The right panel shows the same for PSR~B1937$+$21. The unit for all
        panels is the mean of the standard deviation of the off-pulse
        phase bins for the relevant subband data set. Otherwise as Figure~\ref{1713_vm}.}
    \end{minipage}}
    \label{by_freq}
\end{figure*}

\begin{figure}[ht]
  \centering
  \begin{tabular}{@{}cc@{}}
    \includegraphics[width=\columnwidth]{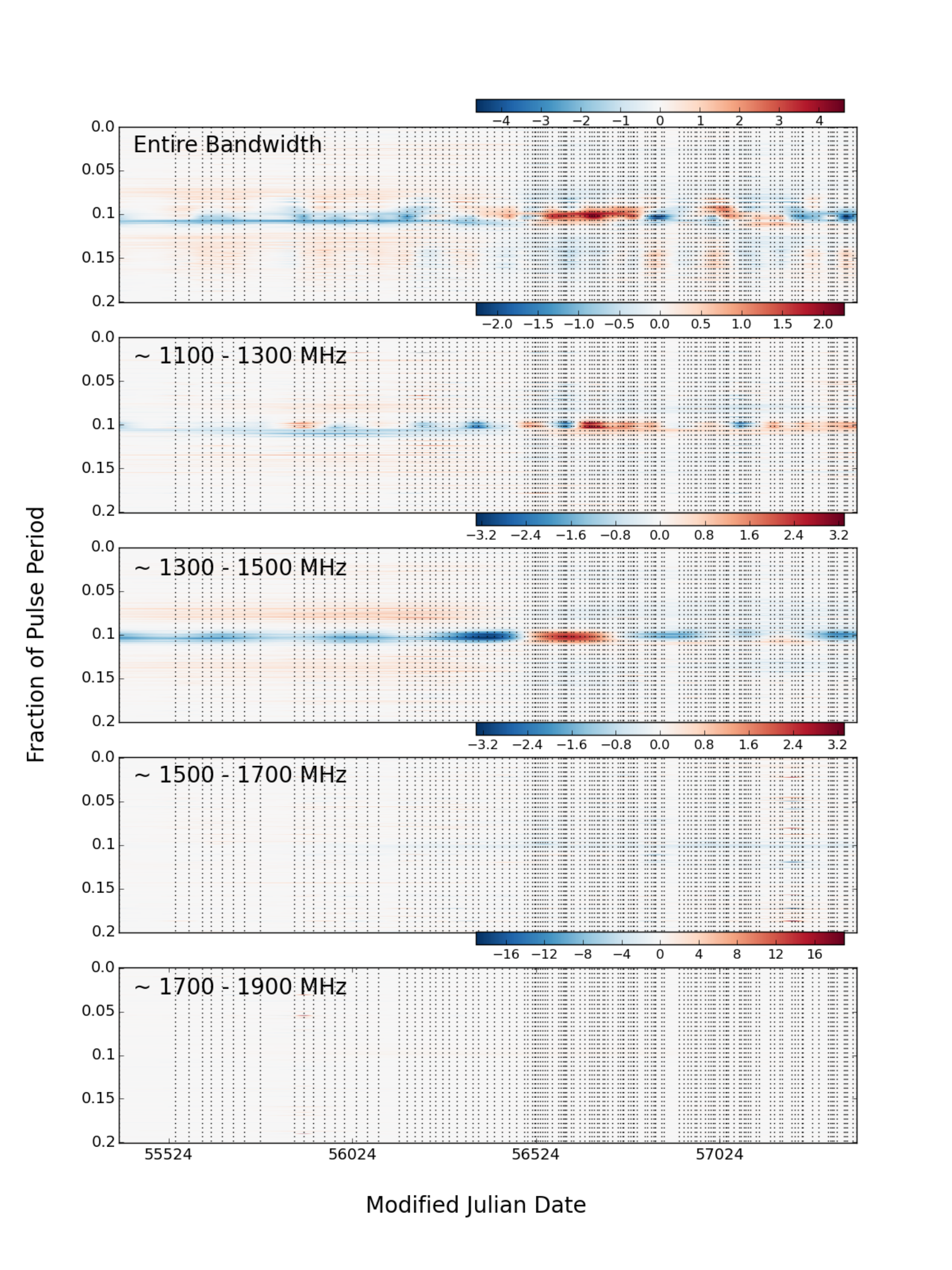}
  \end{tabular}
  \caption{Variability maps for PSR~J1713$+$0747 GUPPI observations at
    a central frequency of $\sim$ 1500~MHz, broken down into subbands
    of observational frequency. The unit for all panels is the mean of
    the standard deviation of the off-pulse phase bins for the
    relevant subband data set. Otherwise as Figure~\ref{1713_vm}.}
  \label{1713_by_freq}
\end{figure}

For PSRs~J1643$-$1224, J1713$+$0747 and B1937$+$21, some
variability is not consistent with the effects of scintillation. In
Figure~\ref{by_freq}, the pulse profile
variations centered at 820~MHz are seen to correlate across the
observing band for both PSR~J1643$-$1224 and PSR~B1937$+$21. We do not expect
such effects to be the result of scintillation.
Figure~\ref{1713_by_freq} shows that the systematic pulse
profile variability of PSR~J1713$+$0747 at 1500~MHz is seen in only two of the four
frequency subbands.

\begin{figure*}
  \begin{center}
    \includegraphics[width=145mm]{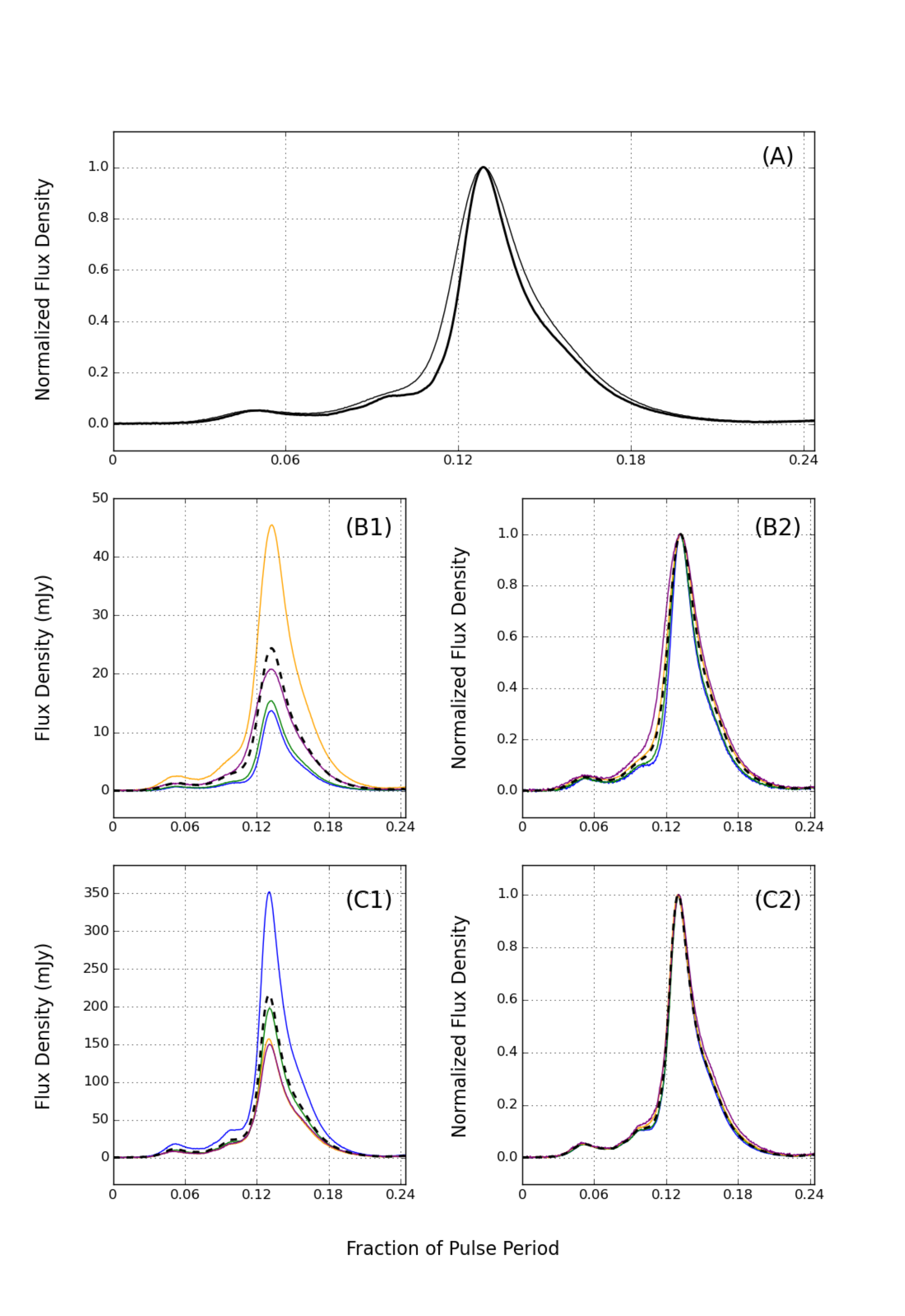}
    \caption{Two contrasting pulse profiles of PSR~J1713$+$0747 at
      1400~MHz. Panel A shows two pulse profiles aligned and normalized
      by their peaks. This is done to allow shape changes to be seen
      clearly. The thin line is the pulse profile as observed on MJD
      56360 and the thick line on MJD 57076. Panel B1 shows the MJD
      56360 observation split into four frequency subbands, each
      spanning 200~MHz. From lowest to highest frequency,
      the subbands are represented by purple, orange, green and blue
      profiles. The black dotted line shows the frequency-integrated
      profile. In Panel B2, the subband and frequency-integrated profiles have also been
      normalized to and aligned by the peak. Panel C1 and C2 show the
      same as B1 and B2 respectively, but for the observation made on
      MJD 57076. We see from Panels~B2 and C2 that the evolution of
      the pulse profile across the observing band, is different for
      the two observations.}
    \label{1713_1400_sub}
  \end{center}
\end{figure*}

Furthermore, for some observations, we see that the evolution of the
pulse profile across the observing band, is different
than for others. An example in PSR~J1713$+$0747 can be seen by comparing panels B2 and C2
in Figure~\ref{1713_1400_sub}. It is not clear how such differences
could be caused by scintillation.

\begin{figure*}
  \begin{center}
    \includegraphics[width=\textwidth]{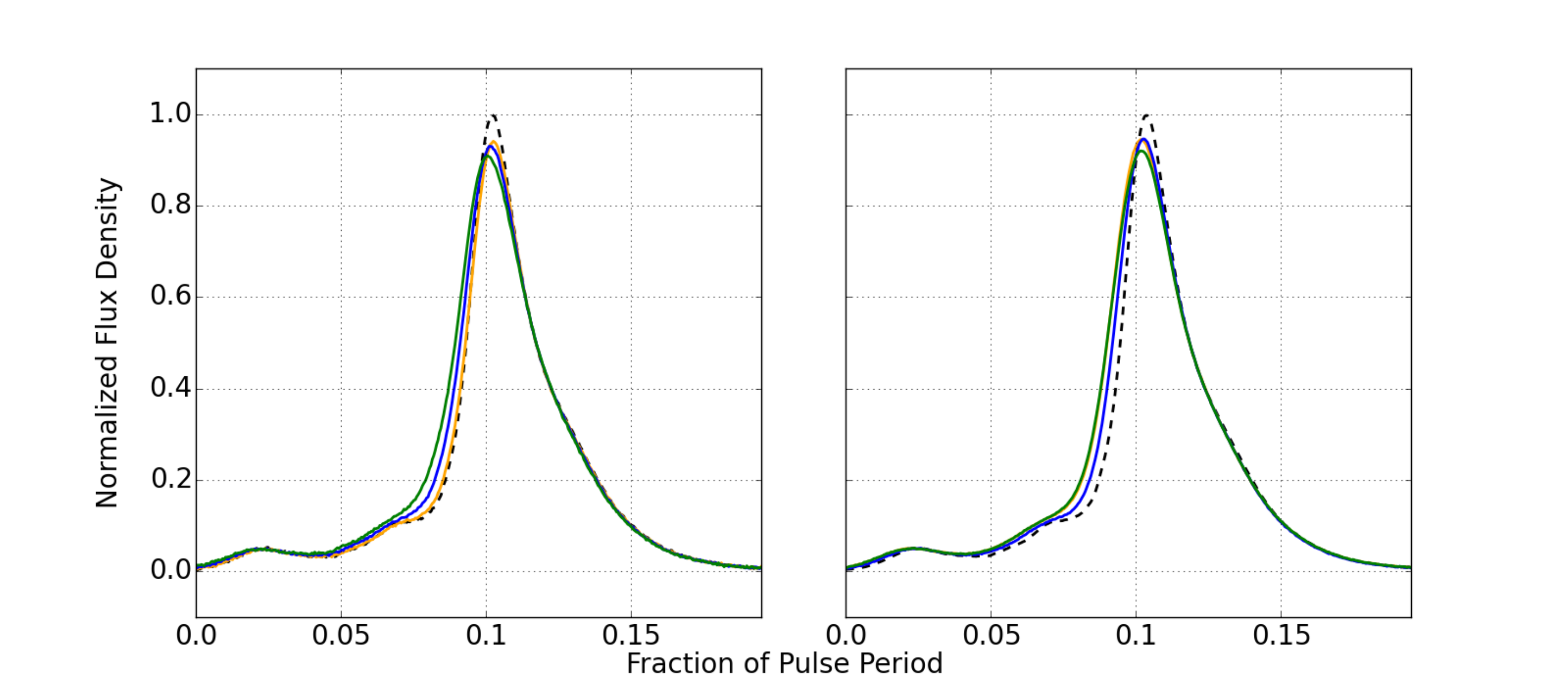}
    \caption{How the frequency-integrated pulse profiles of
      PSR~J1713$+$0747 change with DM at 1400~MHz. Left panel: The
      black dashed profile was dedispersed at a DM of
      15.990~pc~cm$^{-3}$. The orange, blue and green profiles were
      dedispersed with 0.04, 0.06 and 0.08~pc~cm$^{-3}$ added
      respectively. Right panel: Three observations in the data set
      showing large deviations from the average pulse profile. The
      black dashed profile is the average for the data set. The orange,
      blue and green profiles were recorded on MJDs 56360, 56598 and
      57239 respectively.}
    \label{1713_1400_dm}
  \end{center}
\end{figure*}

\subsection{Inaccurate DM}
\label{inaccurate_dm}
As an electromagnetic signal travels though the IISM, its interaction
with free electrons produces a frequency-dependent time delay that scales
as $\nu^{-2}$ where $\nu$ is the signal frequency. The magnitude of
this delay is proportional to the integrated column density of
electrons along the path of the signal, which is known as the DM. If we fail to correct for such frequency
dependent time delays, an integrated pulse profile that is created by
summing a signal detected across a range of observing frequencies,
will necessarily appear smeared out when compared to the intrinsic
pulse shape. Although correcting for such signal dispersion is
routine, DMs are well known to vary with epoch both systematically and
stochastically \citep{2013MNRAS.429.2161K,2016ApJ...821...66L,2017ApJ...841..125J}
due primarily to a changing line of sight. NANOGrav measures the value of DM
at nearly every observing epoch \citep{2015ApJ...813...65T}, but an
inaccurate DM value can lead to a modified pulse profile.

Because NANOGrav calculates TOAs for all frequency channels, a further
complication is added to the determination of DM. Pulse shapes vary
with frequency, but only a single standard template is used in the
template-matching procedure. This produces small systematic
frequency-dependent perturbations in the TOAs in addition to the
$\nu^{-2}$ offsets due to dispersion. To compensate for this, an
additional timing delay is added to all timing models, where
\begin{equation}
  \Delta t_{\rm {FD}} = \sum c_{i}\mathrm{log}\left(\frac{\nu}{\mathrm{1GHz}}\right)^{i}
\end{equation}
and the coefficients $c_{i}$ are fit parameters in the timing
  model \citep{2015ApJ...813...65T}. When finding the best-fit timing
model parameters for a pulsar, DM and $\Delta t_{\rm {FD}}$ are somewhat
covariant, and so the best-fit DM value can change significantly,
dependent on whether $\Delta t_{\rm {FD}}$ is included in the timing model.

For the purposes of creating the frequency-integrated pulse profiles
employed in this variability analysis, we have calculated the best-fit
DM parameters without the inclusion of the $\Delta t_{\rm {FD}}$ parameters
that are necessary for TOA determination in individual frequency
channels. This minimizes smearing when generating the
frequency-integrated pulse profiles.

\citet{2017ApJ...841..125J} report that PSR~J1713$+$0747 has a DM of
$\sim$ 16~pc cm$^{-3}$, which is typically seen to vary on the order of
$10^{-4}$~pc~cm$^{-3}$ on approximately yearly timescales. A 1400~MHz observation that has a DM
inaccuracy of a few $10^{-4}$~pc~cm$^{-3}$ would only introduce a
delay across an 800~MHz bandwidth of a few tenths of a microsecond. A
single phase bin in our analysis of PSR~J1713$+$0747 covers an order
of magnitude more time than this (2.23 $\mu$s). Figure~\ref{1713_1400_dm}
demonstrates that to produce some of the most modified pulse profiles
in the 1400~MHz data set, the DM would have to be incorrect by the order of
$10^{-2}$~pc~cm$^{-3}$, which is around a hundred times larger than
the DM variations that we observe for this pulsar.

\begin{figure*}
  \begin{center}
    \includegraphics[width=\textwidth]{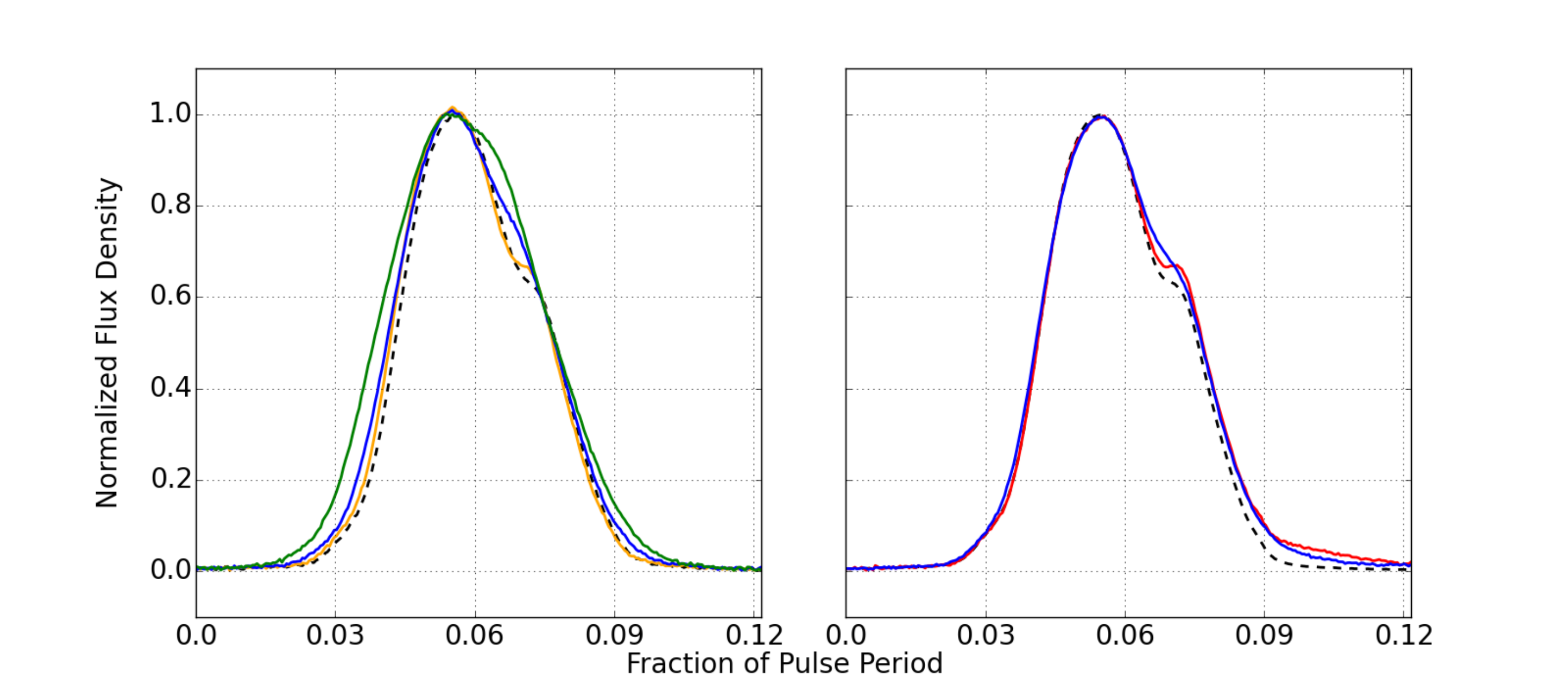}
    \caption{How the frequency-integrated pulse profiles of
      PSR~B1937$+$21 change with DM at 820~MHz. Left panel: The black
      dashed profile was dedispersed at a DM of
      71.025~pc~cm$^{-3}$. The orange, blue and green profiles were
      dedispersed with 0.005, 0.01 and 0.015~pc~cm$^{-3}$ subtracted
      respectively. Right panel: Two observations in the data set
      showing large deviations from the average pulse profile. The
      black dashed profile is the average for the data set. The red
      profile was recorded on MJD 55641 and the blue on MJD 55765.}
    \label{1937_820_dm}
  \end{center}
\end{figure*}

PSR~B1937$+$21 is calculated to have a DM of $\sim$ 71~pc~cm$^{-3}$, which is
typically seen to vary on the order of $10^{-3}$~pc~cm$^{-3}$ on
approximately yearly timescales. A
1400~MHz observation that has a DM inaccuracy of a few
$10^{-3}$~pc~cm$^{-3}$ would introduce a delay across an 800~MHz
bandwidth of a few microseconds. This is the equivalent of a few
PSR~B1937$+$21 phase bins (each spanning 0.76 $\mu$s).  Figure~\ref{1937_820_dm} shows that to
produce some of the profile variations seen in the 820~MHz data
set, the DM would have to change by around $10^{-3} -
10^{-2}$~pc~cm$^{-3}$, which is comparable to the typical DM
fluctuations seen in this pulsar.

A DM of 62.4~pc~cm$^{-3}$ with approximately yearly fluctuations of around $10^{-3}$~pc~cm$^{-3}$ is reported for
PSR~J1643$-$1224 by Jones et al.; an incorrect DM value
of this magnitude would introduce a
delay at 820~MHz across an 200~MHz
bandwidth of a few microseconds. This is the equivalent of one or two
PSR~J1643$-$1224 phase bins (each spanning 2.26 $\mu$s). However, the
phase drifts that are seen in Figure~\ref{1643_drift} are not
suggestive of profile changes induced by incorrect DM measurements, as
the modifications in each observation are localized in relatively
narrow regions of pulse phase; a more
smeared effect would be expected from incorrect DM values.

PSR~J2145$-$0750 has a DM of 9~pc cm$^{-3}$ with typical variations on the
order of $10^{-3}$~pc~cm$^{-3}$ occurring on approximately yearly timescales as reported by Jones et al.; inaccuracies on this scale would
introduce a delay of a few microseconds across the 200~MHz bandwidth
at 820~MHz. This is only a fraction of a J2145$-$0750 phase bin which
spans $\sim$ 8 $\mu$s. Additionally, even in the most deviant pulse
profiles in the data set, some sharp features remain, which would be
smeared out when a DM inaccuracy (of the magnitude needed to replicate
the profile changes) exists.

Based on these calculations, an inaccurate (but realistic) DM value used to dedisperse
the pulsar signal when producing a frequency-integrated pulse profile
could produce shape changes in PSRs~B1937$+$21 and J1643$-$1224, but is unlikely to
produce those observed in PSR~J1713$+$0747 or J2145$-$0750.
\subsection{Temporal Broadening from Scattering}
Electromagnetic waves traveling through the IISM are scattered and
follow different paths to the observer. This can, therefore, lead to
the broadening of an observed pulse profile; an intrinsically narrow
pulse will broaden due to scattering, producing an exponential decay
of the pulse with a characteristic timescale $\tau$ known as the
scattering timescale. Scatter broadening is a frequency-dependent
effect, with $\tau \propto \nu^{-4.4}$ for a thin screen scattering
model \citep{1991ApJ...376..123C}.

\begin{figure*}
  \begin{center}
    \includegraphics[width=\textwidth]{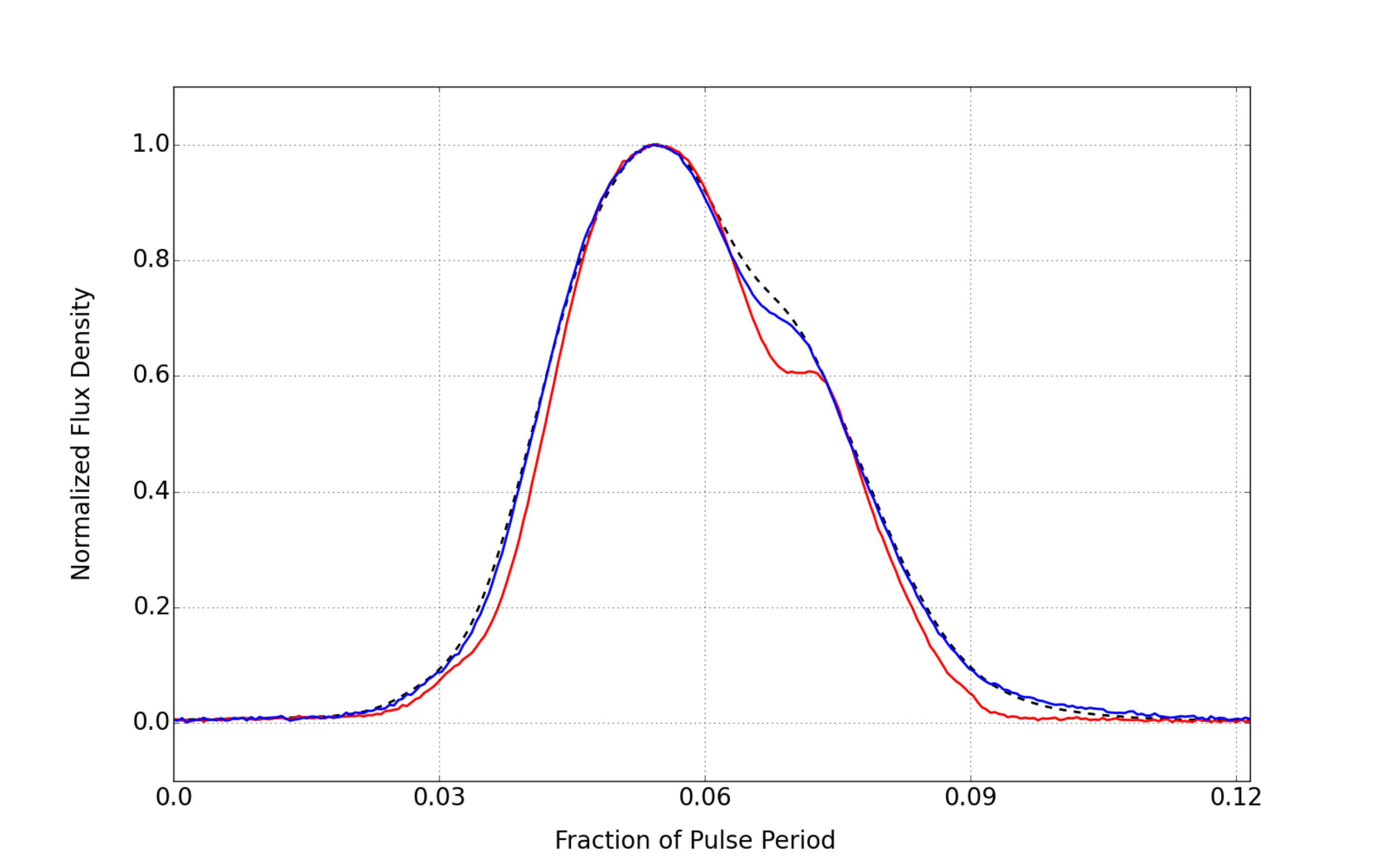}
    \caption{A scatter broadening simulation for PSR~B1937$+$21. The red line is the
        1400~MHz AO pulse profile observed at MJD~55361, and the blue
        profile was observed on MJD~55828. The black dashed line is the result of a convolution of
        the red profile and a one-sided exponential function, in order to
        simulate the effects of scatter broadening. The $\tau$ value
        of the exponential function is 9.1~$\mu$s.}
    \label{1937_820_scat}
  \end{center}
\end{figure*}

\citet{2016ApJ...818..166L} determine average scattering timescales
via the measurement of scintillation bandwidths $\Delta\nu$ in the dynamic spectra
of the observations, using the relationship
\begin{equation}
  2\pi\Delta\nu\tau\sim1,
\end{equation}
\citep{1998ApJ...507..846C}. For PSR~B1937$+$21, Levin et al.\ calculate that at
1500~MHz, the average $\tau$ is around 44~ns. This is close to the
limit imposed by the
frequency resolution of their
observations. For some epochs, therefore, no scintillation bandwidth could
be measured, meaning only a lower limit for $\tau$ on the order of
tens of nanoseconds could be inferred. At 820~MHz, these values
translate to scattering timescales on the order of a microsecond or more. \citet{1990ApJ...349..245C} measure the scattering
timescale at 430~MHz for the main pulse of PSR~B1937$+$21 to
be 25$\pm$2~$\mu$s and 30$\pm$2~$\mu$s for the interpulse. Assuming a thin screen scattering model, this translates to a
$\tau$ value of a few microseconds at 820~MHz. At 327~MHz,
\citet{2006ApJ...645..303R} measure 120~$\mu$s with an rms variation
of 20~$\mu$s.  With the thin screen scattering assumption, 120~$\mu$s
at 327~MHz translates to approximately 3~$\mu$s at 820~MHz.
\citet{2013MNRAS.429.2161K} give a scintillation bandwidth of 1.2~MHz
for PSR~B1937$+$21, which translates to a scattering timescale of
approximately 1.8~$\mu$s at 820~MHz. 
We simulate the effects of thin screen scatter broadening by
convolving a pulse profile with a one-sided exponentially decaying
function.
We use a one-sided exponential function as an approximation to the
pulse broadening function caused by interstellar scattering. Actual
pulse broadening functions are more rounded at the origin due to the
finite thickness of a scattering screen and can have more slowly
decaying tails if there is a wide range of scattering length scales,
as with a Kolmogorov medium.
In
Figure~\ref{1937_820_scat}, such a simulation of scatter broadening for
  PSR~B1937$+$21 shows that if the nature and magnitude of the pulse
profile shape changes we see in the 820~MHz data set were produced by
thin screen scatter broadening, then $\tau$ would have to be on the order of
microseconds, which is consistent with the findings of Levin et al.,
Cordes et al., Ramachandran et al.\ and Keith et al.\ This translates
to a scintillation bandwidth less than 1~MHz. Additionally, the
strongly correlated variability seen between
the main pulse and the interpulse of PSR~B1937+21 (most clearly
illustrated at 820~MHz in Panel~B1 of
Figure~\ref{1937_mp_vm} and Figure~\ref{1937_ip_vm}), is consistent
with what would be expected from a scatter broadened signal (or one
modified by propagation effects in general). However, a similar effect could
also result from global changes in the pulsar magnetosphere, and so
intrinsic variability cannot be ruled out on this basis.

For PSRs~J1713$+$0747 and J2145$-$0750,
\citet{2016ApJ...818..166L} find that $\tau$ is on the order of
ns at 1500~MHz. These scattering timescales are much too small for scatter
broadening to significantly affect the pulse profile shapes.

The nature of the pulse profile shape changes observed in PSR~J1643$-$1224
cannot be well replicated simply by the convolution of a one-sided
decaying exponential function; the phase range over which the profile
is modified is usually relatively narrow and is also seen to drift
with time. However, IISM structure that is close to the line of sight,
could permit such transient profile components via the deflection of radio
waves back to the observer. Such behavior has been seen previously in
other pulsars \citep{2000ApJ...543..740B,2018MNRAS.476.2704M}.

\subsection{Instrumental Issues \& Radio-Frequency Interference}
\label{instrumental}

\begin{figure*}[ht]
  \begin{tabular}{@{}cc@{}}
    \includegraphics[width=\textwidth]{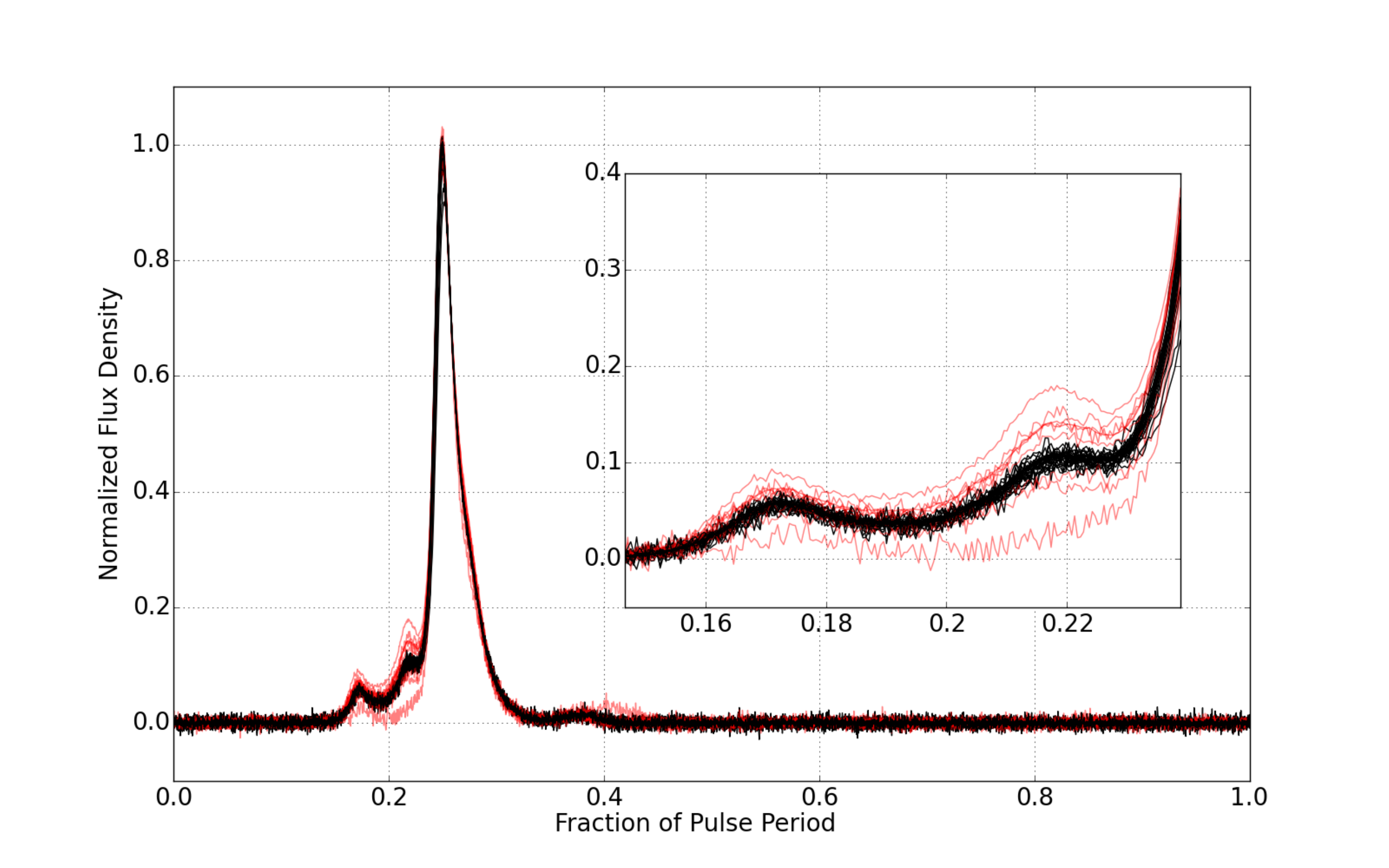}
  \end{tabular}
  \caption{All 36 pulse profiles of PSR~J1713$+$0747 observed at
    2030~MHz included in the variability analysis. The red profiles
    were observed on or between MJDs~57083 and 57263. All other
    profiles are black.}
  \label{1713_bad}
\end{figure*}

\begin{figure*}
  \begin{center}
    \includegraphics[width=150mm]{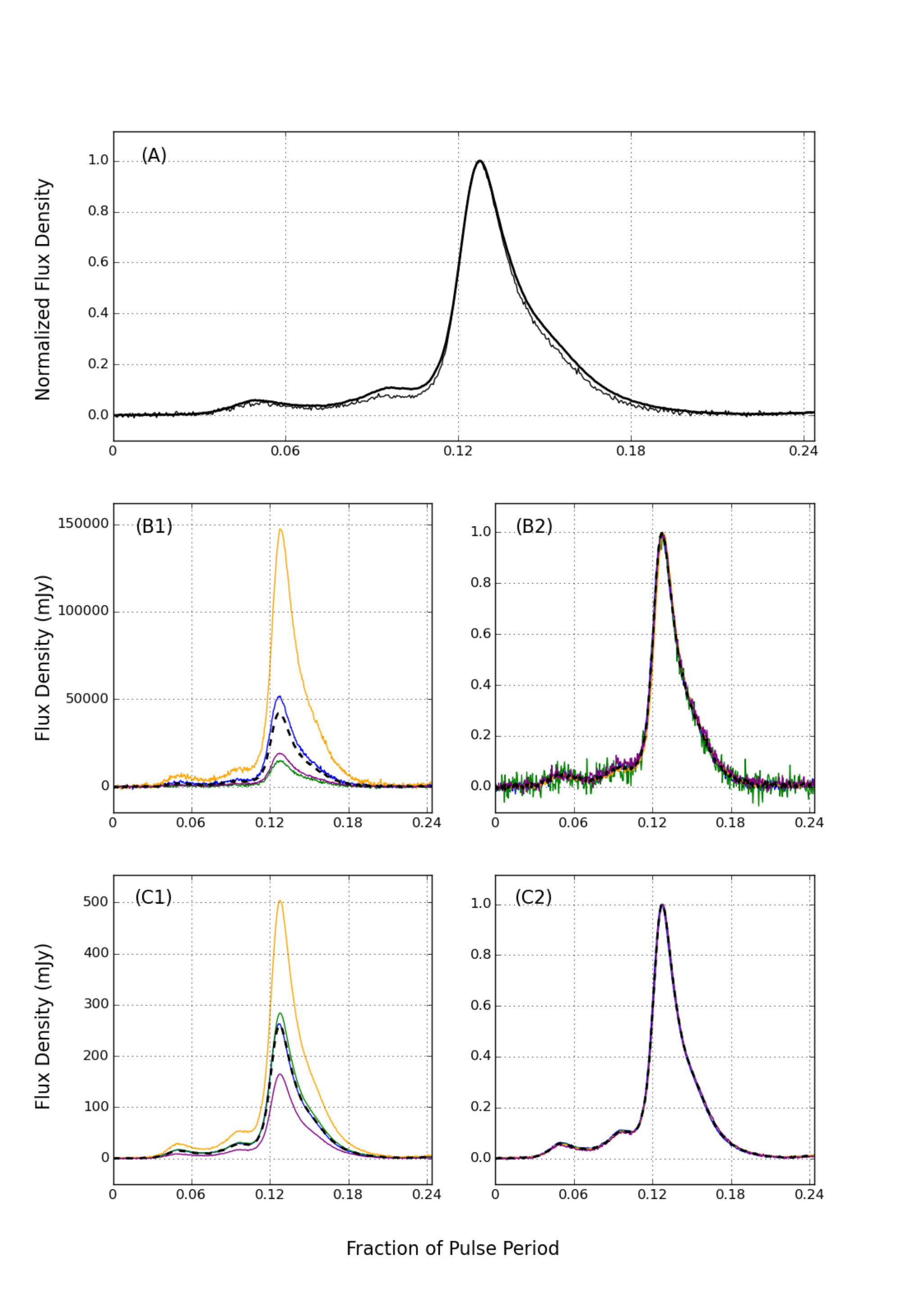}
    \caption{Two contrasting pulse profiles of PSR~J1713$+$0747 at
      2030~MHz. The thin line in Panel~A and Panels~B1 and B2 show
      the observation made on MJD 57108. The thick line and Panels~C1
      and C2 show the observation made on MJD 57375. The plots are
      as Figure~\ref{1713_1400_sub} otherwise. A comparison of
      Panels~B2 and C2 shows that shape of the frequency-integrated profiles
      and all of the contributing subbands is different for the two
      observations. This is most noticable at a pulse phase fraction
      $\sim$ 0.10. The absolute flux in Panel~B1 is recorded as
      much larger than expected.}
    \label{1713_2030_sub}
  \end{center}
\end{figure*}

The pulse profile shape changes of PSR~J1713$+$0747 at 2030~MHz seem
to mainly occur approximately between the MJDs of 57083 (2015 March 2)
and 57263 (2015 August 29), as shown in Figure~\ref{1713_bad}.
During
this time there are various observations in which the shape of the
frequency-integrated profiles and all of the contributing subbands is
modified with respect to the average profile shape of the data
set. Additionally, during this period there is a large fraction of
observations in which the absolute fluxes are recorded as much larger
than expected. The S/N of these observations suggests that the high
flux density is due to miscalibration rather than a very bright
signal. Both of these phenomena are shown on MJD~57108 in
Figure~\ref{1713_2030_sub}. The high concentration of pulse profile
changes during this time period, along with their nature, may suggest an
non-astrophysical cause.
This hypothesis is bolstered by the fact that in the 2030~MHz
observations of PSR~B1937$+$21, the most significant pulse profile
variations also occur within this date range, as can be seen in Panel
B4 of Figure~\ref{1937_mp_vm} and also in Figure~\ref{1937_extreme}. When
analyzing all pulsar observations that appeared in the 2030~MHz data
set of PSR~B1937$+$21 before any profiles were removed due to low S/N,
three out of the four highest flux density observations fell within
this MJD~57083--57263 range. The absolute fluxes of these observations
are very large, with profile peaks around 27, 44 and 65~Jy, yet all
have comparatively low S/N, as was the case for
PSR~J1713$+$0747.
This span of time coincides with an era of particularly high levels of
RFI around 2000~MHz at AO. The RFI was
eventually mitigated by a new filter installed by the observatory in
October 2015. As part of the processing, extra RFI removal was carried
out for all PUPPI 2030~MHz data before MJD 57300. Much of the
frequency band had to be removed from many of these observations, but
residual effects of the RFI may well remain and be responsible for the
AO profile changes at 2030~MHz.

\begin{figure*}[ht]
  \centering
  \begin{tabular}{@{}cc@{}}
    \includegraphics[width=\textwidth]{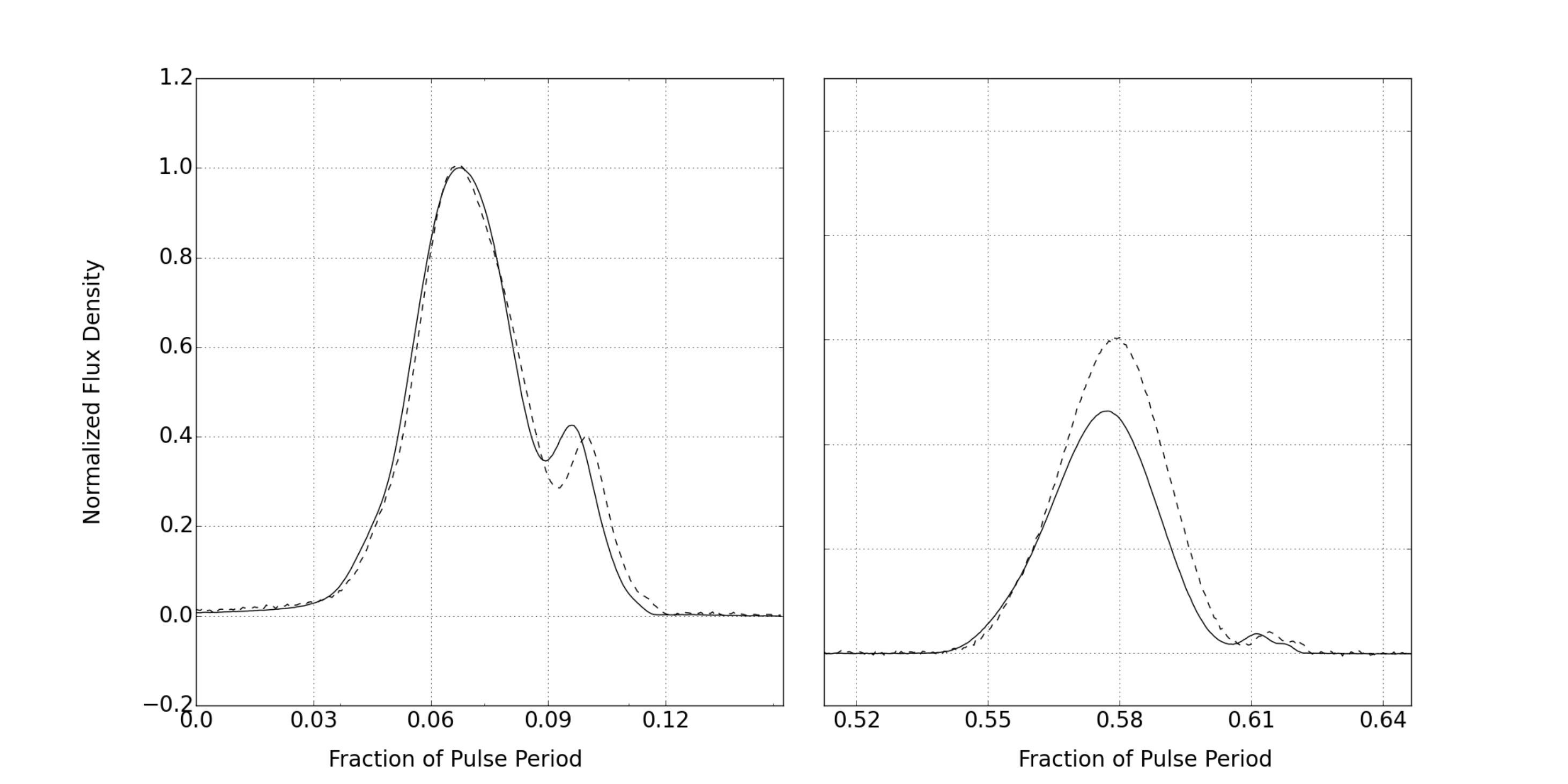}
  \end{tabular}
  \caption{Pulse profile deviations from PSR~B1937$+$21 1500~MHz GBT observations that seem to result from incorrect
    polarization calibration. The solid line shows the average profile for the data
    set. The dashed profile is an MJD 55977 observation which has had
    noise diode polarization calibration applied. This discrepancy
    is not present when full Mueller matrix calibration is employed.}
  \label{1937_55977}
\end{figure*}

Further information regarding the cause of these profile changes is provided when
comparing the 2030~MHz PSR~B1937$+$21 profiles occurring in the MJD 57083-57263 range (discussed above; Figure~\ref{1937_extreme}) with Figure~\ref{1937_55977}.
As discussed in Section~\ref{data}, only the 1500~MHz GUPPI data
has undergone two parallel methods of polarization calibration: using
a noise diode and the more sophisticated full Mueller matrix
calibration. This data set, therefore, gives us an opportunity to see
how the different calibration techniques affect the resulting pulse
profiles (see also \citealp{2018ApJ...862...47G}). Only relatively subtle changes are produced by the different
polarization calibration methods for most observations. However, there
are some observation days that show large pulse profile variability
when calibrated using only the noise diode. One such day is
MJD~55977. Figure~\ref{1937_55977} shows the pulse profile modifications that
take place in the PSR~B1937$+$21 1500~MHz noise diode calibrated
observations made on that
day. These deviations from the average all but disappear when full
Mueller matrix calibration is applied. The same phenomenon is seen on
the same day for PSR~J1713$+$0747.

The PUPPI
2030~MHz profiles that fall in the problematic MJD 57083-57263 range and are highlighted in Figure~\ref{1937_extreme}, are very similar in nature to the GUPPI 1500~MHz
MJD~55977 profile that was polarization calibrated using only a noise diode, both in the main pulse and the interpulse. It
is likely, therefore, that the 2030~MHz PSR~B1937$+$21 PUPPI profiles highlighted
in Figure~\ref{1937_extreme} are also the result of incorrect
polarization calibration.
Extrapolating further, the PSR~J1713$+$0747
2030~MHz PUPPI profile changes that also occur in the same problematic
date range as the PSR~B1937$+$21 2030~MHz PUPPI observations may also
be due to incorrect polarization calibration. The polarimetric
calibration of some NANOGrav MSPs is addressed in detail in
\citet{2018ApJ...862...47G}. As discussed in Section~\ref{data},
Gentile et al.\ have
performed full Mueller matrix polarization calibration for the PUPPI
data. This is done using a method called Measurement Equation Template
Matching \citep{2013ApJS..204...13V}, a technique that uses pulsars with known polarization
profiles to act as \emph{standard sources} in order to generate
polarimetric responses for any epoch of observation. Unfortunately,
the standard sources used by Gentile et al.\ were PSRs~J1713$+$0747 and
B1937$+$21. The polarization profiles for these two pulsars are,
therefore, assumed to be unchanging and so are not calculated for
each observation. In general, Gentile et al.\ find that the
polarimetric responses of AO's 1400 and 2030~MHz receivers vary
significantly with time.

In general, it is possible that some pulse profile shape changes are the
result of flux and polarization calibration issues. 
As discussed in Section~\ref{1713_results}, the flux density calibration
procedure was not undertaken correctly for the 1400~MHz AO observation
of PSR~J1713+0747 made on MJD 56360; an incorrect pulsed calibration
signal was injected at the epoch of observation. It is not clear
whether the change in pulse profile shape was affected by this, as pulse profiles
with similar shapes were also seen in the data set, for which no such calibration issues
were seen (MJDs 56598 and 57239).

The profile shape changes of PSR~J1643$-$1224 have now been
observed by both the GBT and the Parkes Radio Telescope and,
therefore, an instrumental cause can be ruled out.

\subsection{Jitter}
Pulsars are known to exhibit stochastic, broadband, single-pulse
variations that are intrinsic to the pulsar emission process and
affect the shape of the integrated pulse profile. This phenomenon is
known as \emph{jitter} and contributes noise to the
TOAs. \citet{1985ApJS...59..343C} showed that on timescales ranging
from one pulse period to integrations of up to an hour, TOA variations
exceed what is expected from radiometer noise alone in long-period
pulsars. Studies of MSPs show similar findings
\citep[e.g.][]{2014MNRAS.443.1463S}, and this is generally true for
NANOGrav MSPs \citep{2016ApJ...819..155L}.

As jitter is expected to be uncorrelated from one pulse period to the
next, it should not be responsible for any systematic profile changes
such as those seen in PSRs~B1937$+$21 and J1643$-$1224 at 820~MHz. Using the AO 1400~MHz receiver,
\citet{2012ApJ...761...64S} studied the impact that jitter has on the
timing stability of PSR~J1713$+$0747. They predict that for a
30~minute observation (comprising $\sim 10^{5.6}$ pulses), jitter will
produce a scatter $\sigma_{\rm {J}}$ in the arrival times of $\sim$ 40~ns. Similarly, \citet{2016ApJ...819..155L} calculate $\sigma_{\rm {J}}$ for
pulsars in the NANOGrav nine-year data set and find values for
PSR~J1713$+$0747 that range from 39~ns in the AO 1400~MHz data to
91~ns in the GBT 820~MHz data. They find $\sigma_{\rm {J}}$ for PSR~B1937$+$21
to be between 5.7~ns (AO 1400~MHz) and 32~ns (GBT 820~MHz). For
PSR~J2145$-$0750, $\sigma_{\rm {J}}$ is calculated to be 89~ns at 820~MHz and
120~ns at 1500~MHz using GBT data. These values are much smaller than
the changes in TOA induced by the observed changes in pulse profile,
as seen in Table~\ref{template_table}. This indicates that pulse
jitter is not the dominant source of the profile changes we observe in
PSRs~J1713$+$0747, B1937$+$21 and J2145$-$0750. Furthermore, many other
pulsars observed by the NANOGrav collaboration show evidence of more
jitter noise but less profile shape variability.

\citet{2016ApJ...819..155L} calculate $\sigma_{\rm {J}}$ for
PSR~J1643$-$1224 as 162 and 219~ns at 820 and 1500~MHz
respectively. This is the same order of magnitude as the changes in
TOA induced by the observed changes in pulse profile
(Table~\ref{template_table}). However, the drifting and systematic
nature of the profile changes in the data set is not indicative of
jitter, which is uncorrelated in time.
\subsection{Other Pulsar Emission Changes}
Some pulse profile variability observed in PSRs J2145$-$0750 and
B1937$+$21 is consistent with effects of the propagation of a radio
signal through the IISM; scintillation and scatter broadening
respectively. Some profile modulations in PSRs~J1713$+$0747 and
B1937$+$21 may also be the product of improper polarization
calibration. Other pulse profile shape changes elude a comprehensive explanation and so emission
changes intrinsic to the pulsar (besides jitter) cannot be ruled out.

As described above, the changes in PSR~J1643$-$1224 profile do not seem to be
characteristic of modulations induced by propagation effects,
inaccurate DMs, jitter or instrumental issues. As pointed out by \citet{2016ApJ...828L...1S},
the drifting nature of pulse profile disturbances is reminiscent of
that seen in PSR~J0738$-$4042; a pulsar displaying simultaneous changes in
emission and rotation, which were assessed to be intrinsic to the
neutron star \citep{2014ApJ...780L..31B}.

We also note here that PSR~B1937$+$21 is known to emit giant
pulses \citep{1996ApJ...457L..81C}. The longitude at which the giant pulses are seen to occur is
not consistent with the pulse profile shape changes that we
see. Additionally, the profile variability in PSR~B1937$+$21 occurs on
timescales of hundreds of days; no such timescale is known for giant
pulse activity.

In general, there are few obvious correlations between the profile shape changes
and pulsar flux density (as seen by comparing the A and B
prefixed panels in the variability maps). The notable exception is the
period between MJDs~57083 and 57263 at 2030~MHz in PSRs~J1713$+$0747,
B1937$+$21, as discussed in Section~\ref{instrumental}.

\begin{figure*}[ht]
  \centering
  \begin{tabular}{@{}cc@{}}
    \includegraphics[width=\textwidth]{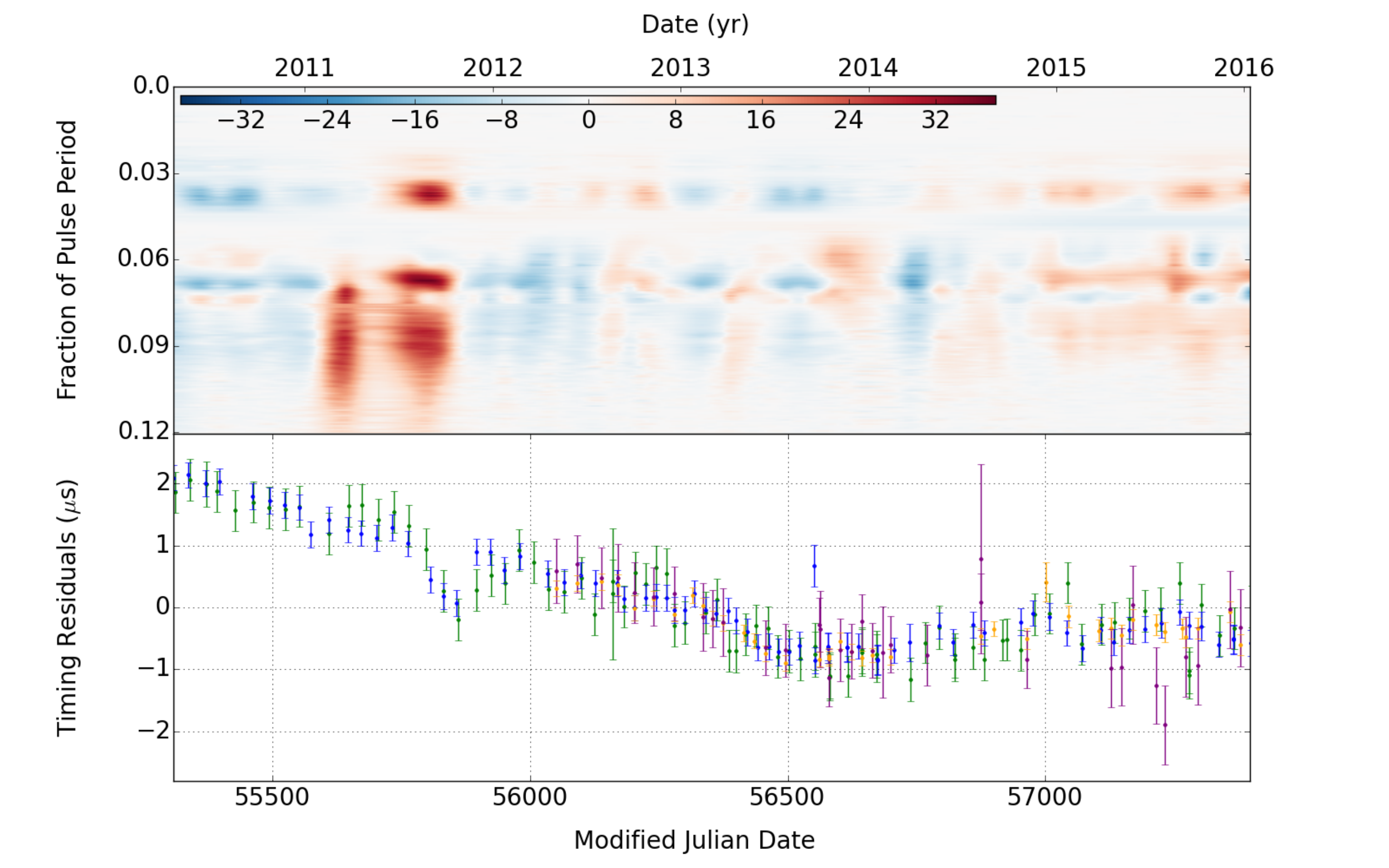}
  \end{tabular}
  \caption{The 820~MHz profile residuals and the timing residuals of
    B1937$+$21, observed at the GBT.
    The top panel is a variability map showing pulse
    profile shape changes after the observations have been normalized
    and depicts the same data as Panel~B1 of Figure~\ref{1937_mp_vm}.
    Red regions indicate where the inferred pulse profile has an excess
    of flux density compared to the average for the data set. Blue
    indicates where it has a deficit. The unit for the
    variability map is the mean of the standard deviation of the
    off-pulse phase bins for the data set. The bottom panel shows
    the TOA residuals for PSR~B1937$-$21 at 820~MHz (green), 1400~MHz
    (orange), 1500~MHz (blue) and 2030~MHz (purple).}
  \label{1937_profile_timing}
\end{figure*}

Other links between the profile variability and the rotational behavior
of a pulsar may provide further clues regarding the source of any
variability. Figure~\ref{1937_profile_timing} shows the behavior of both profile
and timing residuals for PSR~B1937$+$21. The profile residuals shown
are at an observing frequency of 820~MHz (the
data set displaying the most systematic variability).
A more detailed analysis of any relationship between
the emission and rotational properties of these pulsars will be left
to future work.\\

Whatever the cause of unmodeled pulse profile changes, they are
detrimental to the template matching technique of TOA determination
and, therefore, to pulsar timing. For PSRs~J1643$-$1224, J1713$+$0747, B1937$+$21
and J2145$-$0750 the TOA inaccuracies induced due to some changes in pulse
profile are on the order of hundreds of nanoseconds to microseconds. The
frequency-integrated pulse profile changes that we have focused on may
not translate to profile changes in the narrow individual frequency channels
that the NANOGrav collaboration uses to produces its TOAs however;
pulse profiles that result from the combination of a relatively wide
band of frequency channels are far more sensitive to shape changes
induced by the effects of signal propagation.
It is also important that highly aberrant pulse profiles that appear
in a data set have their corresponding TOAs removed in order to ensure
that the most accurate timing models are
produced.

When looking at figures that show the phase location
of profile variability, we must be cognizant of the fact that
different methods for alignment will show the variability to occur at
different parts of the pulse profile. Different alignment methods,
however, should largely be in agreement regarding the amount of
variability that is contained within a data set, even if they disagree
on the phases at which it occurs. A priori, we expect the magnitude of
profile variability to be most around the profile peak if we assume
that the amount of variability will be proportional to the profile
intensity at the phase at which it occurs.

When quantifying the variability seen in pulse profiles, we have
discounted the number of pulsar rotations that contribute to the
observations. With its very short period of 1.56~ms, the
PSR~B1937$+$21 data sets typically have around 10$^{6}$ rotations per
observation. Conversely, PSR~J2145$-$0750 has the longest period of
all pulsars analyzed in this work at 16.05~ms. Consequently, the data
sets for this pulsar have only around 10$^{5}$ rotations per
observation. All else being equal, 10 times more pulses contributing
to an integrated pulse profile would increase the S/N by approximately
$\sqrt{10}$ and would also decrease the pulse profile variability due
to jitter, thereby decreasing the variability
measured. Falling in between these extremes, PSRs~J1643$-$1224 and
J1713$+$0747 have pulse periods of 4.62 and 4.57~ms respectively and
so rotate approximately 3-4$\times$10$^{5}$ times per observation.

In future work, MSP pulse profile variability information could lead to the
mitigation of timing aberrations caused by the unmodeled pulse
profile changes we observe. For example, more NANOGrav data are
currently undergoing full Mueller matrix polarization calibration; we
have shown that this process can correct pulse profile shape
distortions that may result from imperfect calibration when using only
a local
noise diode. In the case of a pulsar in which pulse profile
variability is primarily due to temporal broadening from scattering,
we can apply techniques such as cyclic spectroscopy to recover the
intrinsic pulse profiles
\citep{2011MNRAS.416.2821D,2013ApJ...779...99W} from the effects of interstellar scattering. In these two
examples, as the differences between the shape of the observed
profiles
and the timing template are reduced, so too are the timing
residuals. If profile shape changes are entirely due to DISS, then
the
consequences for timing can be minimized by calculating the TOAs for
relatively narrow frequency subbands, as is already done by NANOGrav.

To create the smooth, continuous variability maps seen throughout this
paper, we have inferred the behavior of the flux density for each
phase bin (and, therefore, of the pulse profile as a whole) between
observations using GP regression. For pulsars that show systematic
variability, such modeling techniques would also permit the
extrapolation of pulse profiles shapes. A predicted profile shape
could then be used as a dynamic template for the TOA
calculation. Using an accurate template shape (if one can be
calculated) will necessarily also improve the accuracy of the TOA
recorded. For pulsars with more erratic shape changes and less
systematic variability, such extrapolations will be difficult to
make. However, throughout this analysis, we have also used a new pulse
profile alignment technique which maximizes the number of pulse phase
bins that are in agreement (see Section~\ref{align_scale} for
details). As a result, only the stable parts of the pulse profiles are
used in their alignment. Using only these stable phase bins in the
template matching procedure could potentially result in reduced timing
residuals for some pulsar data sets.

The question of how the
variability measured in this work will impact the predicted timeline
for nanohertz gravitational wave detection is a difficult
one. Relatively little research has been done on long-term pulse
profile variability in MSPs. The physical origin of much of the
emerging profile variability is uncertain, can be different for each
data set analyzed and must be a mixture of multiple effects to varying
degrees. Mitigation of profile variability will require further
investigation and, therefore, it is not clear how soon we will be able
to accommodate such profile changes in a pulsar timing model. As
evidence for MSP profile variability grows, so too will the voicing of
suggestions that precision pulsar timing should not be done using the
standard template matching techniques, but instead, using other
techniques that are more accommodating to such variability, e.g. the
profile domain pulsar timing analysis of
\citet{2015MNRAS.447.2159L}. Such discussions make the analyses in
this paper more interesting and relevant.

\section{Conclusions}

The primary aim of this work was to analyze the long-term pulse
profile behavior in the 11-year data set employed by the NANOGrav
collaboration to search for nanohertz frequency gravitational waves;
significant profile variability is detrimental to the effort if
overlooked.

PSRs~J1713$+$0747, B1937$+$21 and J2145$-$0750 show the highest levels
of variability of the pulsars analyzed, with PSR~B1937$+$21 showing
significant long-timescale trends. These pulsars are also three
of the brightest observed by the NANOGrav collaboration. This is not
entirely surprising as any pulse profile shape changes are more easily
classified as such in bright pulsars, and also, the variability metric is in
units of the rms levels of the off-pulse regions, which will be
relatively small in such pulsars. Despite this, some
of the profile changes seen in these pulsars are of a magnitude that
means they would also be visible in pulsars with a much lower
S/N; the method used for detecting long-term variability has been
shown to be able to do so down to a level that is comparable in magnitude to the rms of the observation noise
\citep{2016MNRAS.456.1374B}.  Systematic variability is also observed
in the PSR~J1643$-$1224 data, which has been identified previously in
observations by the Parkes radio telescope. The cause is not yet clear
beyond being astrophysical in nature. The variability seen in PSR~J2145$-$0750 is consistent
with scintillation effects. Some of the profile modification seen in
PSRs~J1713$+$0747 and B1937$+$21 is likely due to improper
polarization calibration and in the 2030~MHz AO observations, RFI is suspected to
have strongly influenced some of the changes. However, some variability in B1937$+$21 also
seems consistent with scatter broadening, while some profile changes
in PSRs~J1713$+$0747 are due to effects of scintillation.

In the future, the impact of pulse profile variability on precision
timing can be minimized by techniques such as full Mueller matrix
polarization calibration, cyclic spectroscopy, the employment of
dynamic templates in the template matching procedure and timing to the
most stable parts of a pulse profile.

\section{Acknowledgements}

The NANOGrav project is supported by NSF Physics Frontiers Center
award PHYS-1430284. P.R.B. is supported by Track I award
OIA-1458952. The Green Bank Observatory is a facility of the National Science Foundation
operated under cooperative agreement by Associated Universities,
Inc. The Arecibo Observatory is operated by SRI International under a
cooperative agreement with the NSF (AST-11000968), and in alliance
with the Ana G. Méndez-Universidad Metropolitana, and the Universities
Space Research Association. Pulsar research at the University of
British Columbia is supported by an NSERC Discovery Grant and by the
Canadian Institute for Advanced Research. J.A.E. was partially
supported by NASA through Einstein Fellowship
grant PF4-150120. W.W.Z. is supported by the CAS
Pioneer Hundred Talents Program and the Strategic
Priority Research Program of the Chinese Academy of Sciences Grant
No. XDB23000000. Portions of this work performed at NRL were supported
by the Chief of Naval Research.

\appendix
\section{A Variability Map Demonstrating a Stable Pulse Profile Shape}
\begin{figure*}
  \begin{center}
    \includegraphics[width=170mm]{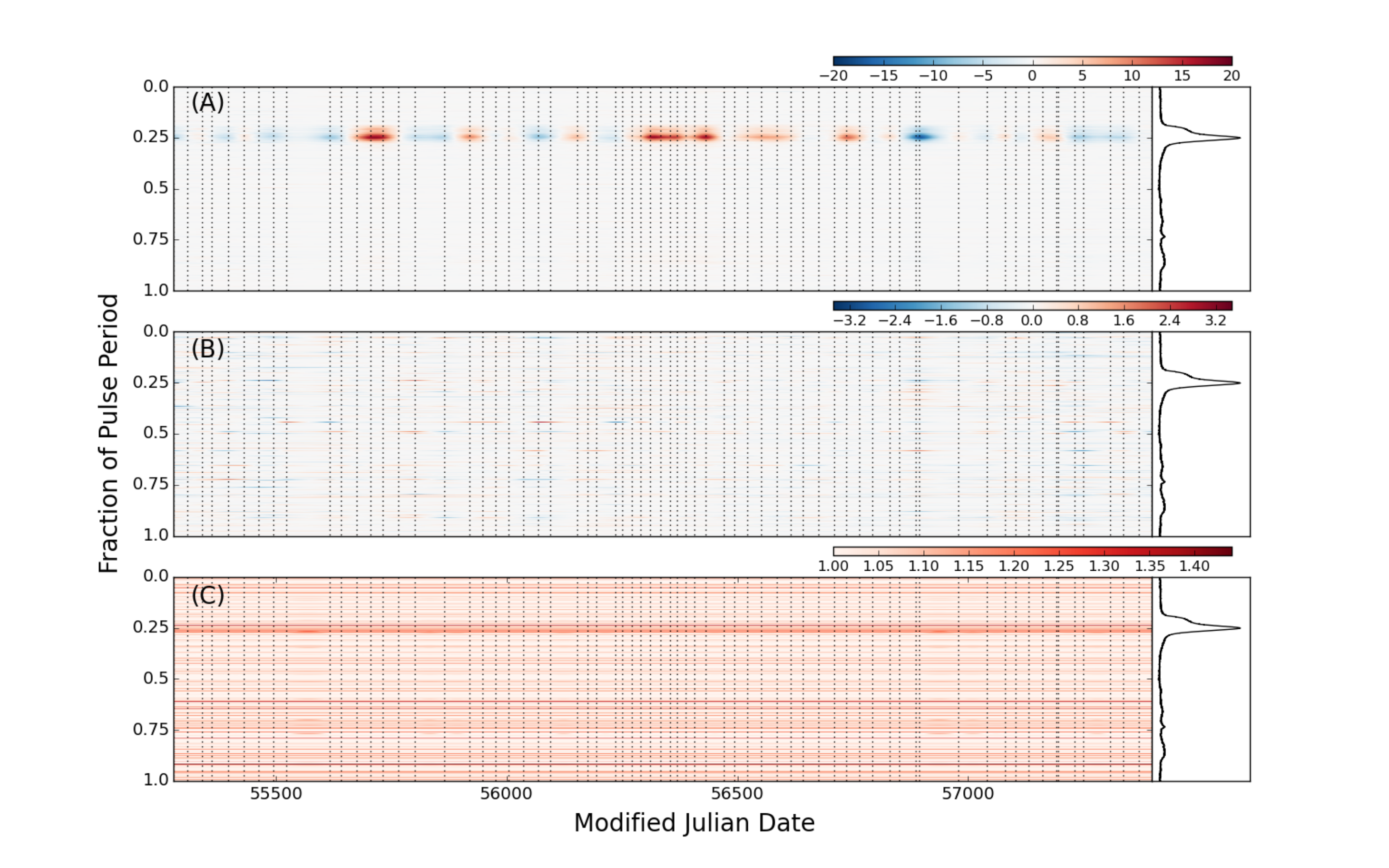}
    \caption{A variability map for a data set in which the pulse
      profile shape is relatively stable. Panel A shows the flux density
      variations in the flux-calibrated, pre-normalized observations
      of PSR~J1614$-$2230 at 820~MHz. The flux density variations are
      due to refractive and diffractive scintillation. Red regions
      indicate where the inferred pulse profile has an excess of flux
      density compared to the average for the data set. Blue indicates
      where it has a deficit. Most variability is seen around the peak
      of the profile at a pulse period fraction of $\sim$ 0.25. Panel
      B shows any pulse profile shape changes after the observations
      have been normalized. As any variability in this data set is consistent with
      additive white Gaussian noise, the variability map is almost entirely devoid of
      color; in each phase bin, the GP model lies around zero, i.e. the mean of the data points
      which inform it. Panel C shows the standard deviation of the
      inferred GP model as a function of pulse phase and time. The
      unit for all panels is the mean of the standard deviation of the
      off-pulse phase bins for the data set. In Panel~C the value is close to
      unity across the whole pulse profile; the variance of the data
      is approximately the same whether looking at on- or off-pulse
      phase bins. The vertical dotted lines indicate the epochs of
      observation informing the GP models. To the right of each panel, the average
      pulse profile for the data set is shown.}
    \label{low_variability}
  \end{center}
\end{figure*}

For illustrative purposes, we present in Figure~\ref{low_variability}, a variability map
for a pulsar that has a stable pulse profile.

\bibliographystyle{aasjournal}
\bibliography{pulse_profile_variability}

\begin{thebibliography}{}
\expandafter\ifx\csname natexlab\endcsname\relax\def\natexlab#1{#1}\fi

\bibitem[{{Armstrong}(1984)}]{1984Natur.307..527A}
{Armstrong}, J.~W. 1984, \nat, 307, 527

\bibitem[{{Arzoumanian} {et~al.}(2015){Arzoumanian}, {Brazier},
  {Burke-Spolaor}, {Chamberlin}, {Chatterjee}, {Christy}, {Cordes}, {Cornish},
  {Crowter}, {Demorest}, {Dolch}, {Ellis}, {Ferdman}, {Garver-Daniels},
  {Gonzalez}, {Jenet}, {Jones}, {Jones}, {Kaspi}, {Koop}, {Lam}, {Lazio},
  {Levin}, {Lommen}, {Lorimer}, {Luo}, {Lynch}, {Madison}, {McLaughlin},
  {McWilliams}, {Nice}, {Palliyaguru}, {Pennucci}, {Ransom}, {Siemens},
  {Stairs}, {Stinebring}, {Stovall}, {Swiggum}, {Vallisneri}, {van Haasteren},
  {Wang}, \& {Zhu}}]{2015ApJ...813...65T}
{Arzoumanian}, Z., {Brazier}, A., {Burke-Spolaor}, S., {et~al.} 2015, \apj,
  813, 65

\bibitem[{{Arzoumanian} {et~al.}(2018){Arzoumanian}, {Brazier},
  {Burke-Spolaor}, {Chamberlin}, {Chatterjee}, {Christy}, {Cordes}, {Cornish},
  {Crawford}, {Thankful Cromartie}, {Crowter}, {DeCesar}, {Demorest}, {Dolch},
  {Ellis}, {Ferdman}, {Ferrara}, {Fonseca}, {Garver-Daniels}, {Gentile},
  {Halmrast}, {Huerta}, {Jenet}, {Jessup}, {Jones}, {Jones}, {Kaplan}, {Lam},
  {Lazio}, {Levin}, {Lommen}, {Lorimer}, {Luo}, {Lynch}, {Madison}, {Matthews},
  {McLaughlin}, {McWilliams}, {Mingarelli}, {Ng}, {Nice}, {Pennucci}, {Ransom},
  {Ray}, {Siemens}, {Simon}, {Spiewak}, {Stairs}, {Stinebring}, {Stovall},
  {Swiggum}, {Taylor}, {Vallisneri}, {van Haasteren}, {Vigeland}, {Zhu}, \&
  {The NANOGrav Collaboration}}]{2018ApJS..235...37A}
---. 2018, \apjs, 235, 37

\bibitem[{{Backer}(1970{\natexlab{a}})}]{1970Natur.228...42B}
{Backer}, D.~C. 1970{\natexlab{a}}, \nat, 228, 42

\bibitem[{{Backer}(1970{\natexlab{b}})}]{1970Natur.228.1297B}
---. 1970{\natexlab{b}}, \nat, 228, 1297

\bibitem[{{Backer} {et~al.}(1982){Backer}, {Kulkarni}, {Heiles}, {Davis}, \&
  {Goss}}]{1982Natur.300..615B}
{Backer}, D.~C., {Kulkarni}, S.~R., {Heiles}, C., {Davis}, M.~M., \& {Goss},
  W.~M. 1982, \nat, 300, 615

\bibitem[{{Backer} {et~al.}(2000){Backer}, {Wong}, \&
  {Valanju}}]{2000ApJ...543..740B}
{Backer}, D.~C., {Wong}, T., \& {Valanju}, J. 2000, \apj, 543, 740

\bibitem[{{Brook} {et~al.}(2014){Brook}, {Karastergiou}, {Buchner}, {Roberts},
  {Keith}, {Johnston}, \& {Shannon}}]{2014ApJ...780L..31B}
{Brook}, P.~R., {Karastergiou}, A., {Buchner}, S., {et~al.} 2014, \apjl, 780,
  L31

\bibitem[{{Brook} {et~al.}(2016){Brook}, {Karastergiou}, {Johnston}, {Kerr},
  {Shannon}, \& {Roberts}}]{2016MNRAS.456.1374B}
{Brook}, P.~R., {Karastergiou}, A., {Johnston}, S., {et~al.} 2016, \mnras, 456,
  1374

\bibitem[{{Camilo} {et~al.}(2012){Camilo}, {Ransom}, {Chatterjee}, {Johnston},
  \& {Demorest}}]{2012ApJ...746...63C}
{Camilo}, F., {Ransom}, S.~M., {Chatterjee}, S., {Johnston}, S., \& {Demorest},
  P. 2012, \apj, 746, 63

\bibitem[{{Cognard} {et~al.}(1995){Cognard}, {Bourgois}, {Lestrade}, {Biraud},
  {Aubry}, {Darchy}, \& {Drouhin}}]{1995A&A...296..169C}
{Cognard}, I., {Bourgois}, G., {Lestrade}, J.-F., {et~al.} 1995, \aap, 296, 169

\bibitem[{{Cognard} {et~al.}(1996){Cognard}, {Shrauner}, {Taylor}, \&
  {Thorsett}}]{1996ApJ...457L..81C}
{Cognard}, I., {Shrauner}, J.~A., {Taylor}, J.~H., \& {Thorsett}, S.~E. 1996,
  \apjl, 457, L81

\bibitem[{{Cordes} \& {Downs}(1985)}]{1985ApJS...59..343C}
{Cordes}, J.~M., \& {Downs}, G.~S. 1985, \apjs, 59, 343

\bibitem[{{Cordes} \& {Lazio}(1991)}]{1991ApJ...376..123C}
{Cordes}, J.~M., \& {Lazio}, T.~J. 1991, \apj, 376, 123

\bibitem[{{Cordes} \& {Rickett}(1998)}]{1998ApJ...507..846C}
{Cordes}, J.~M., \& {Rickett}, B.~J. 1998, \apj, 507, 846

\bibitem[{{Cordes} {et~al.}(1990){Cordes}, {Wolszczan}, {Dewey}, {Blaskiewicz},
  \& {Stinebring}}]{1990ApJ...349..245C}
{Cordes}, J.~M., {Wolszczan}, A., {Dewey}, R.~J., {Blaskiewicz}, M., \&
  {Stinebring}, D.~R. 1990, \apj, 349, 245

\bibitem[{{Demorest}(2011)}]{2011MNRAS.416.2821D}
{Demorest}, P.~B. 2011, \mnras, 416, 2821

\bibitem[{{DuPlain} {et~al.}(2008){DuPlain}, {Ransom}, {Demorest}, {Brandt},
  {Ford}, \& {Shelton}}]{2008SPIE.7019E..1DD}
{DuPlain}, R., {Ransom}, S., {Demorest}, P., {et~al.} 2008, in \procspie, Vol.
  7019, Advanced Software and Control for Astronomy II, 70191D

\bibitem[{{Ford} {et~al.}(2010){Ford}, {Demorest}, \&
  {Ransom}}]{2010SPIE.7740E..0AF}
{Ford}, J.~M., {Demorest}, P., \& {Ransom}, S. 2010, in \procspie, Vol. 7740,
  Software and Cyberinfrastructure for Astronomy, 77400A

\bibitem[{{Gentile} {et~al.}(2018){Gentile}, {McLaughlin}, {Demorest},
  {Stairs}, {Arzoumanian}, {Crowter}, {Dolch}, {DeCesar}, {Ellis}, {Ferdman},
  {Ferrara}, {Fonseca}, {Gonzalez}, {Jones}, {Jones}, {Lam}, {Levin},
  {Lorimer}, {Lynch}, {Ng}, {Nice}, {Pennucci}, {Ransom}, {Ray}, {Spiewak},
  {Stovall}, {Swiggum}, \& {Zhu}}]{2018ApJ...862...47G}
{Gentile}, P.~A., {McLaughlin}, M.~A., {Demorest}, P.~B., {et~al.} 2018, \apj,
  862, 47

\bibitem[{{Graham Smith} {et~al.}(2011){Graham Smith}, {Lyne}, \&
  {Jordan}}]{2011MNRAS.410..499G}
{Graham Smith}, F., {Lyne}, A.~G., \& {Jordan}, C. 2011, \mnras, 410, 499

\bibitem[{{Heiles} {et~al.}(2001){Heiles}, {Perillat}, {Nolan}, {Lorimer},
  {Bhat}, {Ghosh}, {Lewis}, {O'Neil}, {Salter}, \&
  {Stanimirovic}}]{2001PASP..113.1274H}
{Heiles}, C., {Perillat}, P., {Nolan}, M., {et~al.} 2001, \pasp, 113, 1274

\bibitem[{{Helfand} {et~al.}(1975){Helfand}, {Manchester}, \&
  {Taylor}}]{1975ApJ...198..661H}
{Helfand}, D.~J., {Manchester}, R.~N., \& {Taylor}, J.~H. 1975, \apj, 198, 661

\bibitem[{{Hobbs}(2013)}]{2013CQGra..30v4007H}
{Hobbs}, G. 2013, Classical and Quantum Gravity, 30, 224007

\bibitem[{{Hotan} {et~al.}(2005){Hotan}, {Bailes}, \&
  {Ord}}]{2005ApJ...624..906H}
{Hotan}, A.~W., {Bailes}, M., \& {Ord}, S.~M. 2005, \apj, 624, 906

\bibitem[{{Jones} {et~al.}(2017){Jones}, {McLaughlin}, {Lam}, {Cordes},
  {Levin}, {Chatterjee}, {Arzoumanian}, {Crowter}, {Demorest}, {Dolch},
  {Ellis}, {Ferdman}, {Fonseca}, {Gonzalez}, {Jones}, {Lazio}, {Nice},
  {Pennucci}, {Ransom}, {Stinebring}, {Stairs}, {Stovall}, {Swiggum}, \&
  {Zhu}}]{2017ApJ...841..125J}
{Jones}, M.~L., {McLaughlin}, M.~A., {Lam}, M.~T., {et~al.} 2017, \apj, 841,
  125

\bibitem[{{Karastergiou} {et~al.}(2011){Karastergiou}, {Roberts}, {Johnston},
  {Lee}, {Weltevrede}, \& {Kramer}}]{2011MNRAS.415..251K}
{Karastergiou}, A., {Roberts}, S.~J., {Johnston}, S., {et~al.} 2011, \mnras,
  415, 251

\bibitem[{{Kaspi} {et~al.}(1994){Kaspi}, {Taylor}, \&
  {Ryba}}]{1994ApJ...428..713K}
{Kaspi}, V.~M., {Taylor}, J.~H., \& {Ryba}, M.~F. 1994, \apj, 428, 713

\bibitem[{{Keith} {et~al.}(2013){Keith}, {Coles}, {Shannon}, {Hobbs},
  {Manchester}, {Bailes}, {Bhat}, {Burke-Spolaor}, {Champion}, {Chaudhary},
  {Hotan}, {Khoo}, {Kocz}, {Os{\l}owski}, {Ravi}, {Reynolds}, {Sarkissian},
  {van Straten}, \& {Yardley}}]{2013MNRAS.429.2161K}
{Keith}, M.~J., {Coles}, W., {Shannon}, R.~M., {et~al.} 2013, \mnras, 429, 2161

\bibitem[{{Kramer}(1998)}]{1998ApJ...509..856K}
{Kramer}, M. 1998, \apj, 509, 856

\bibitem[{{Kramer} \& {Champion}(2013)}]{2013CQGra..30v4009K}
{Kramer}, M., \& {Champion}, D.~J. 2013, Classical and Quantum Gravity, 30,
  224009

\bibitem[{{Kramer} {et~al.}(2006){Kramer}, {Lyne}, {O'Brien}, {Jordan}, \&
  {Lorimer}}]{2006Sci...312..549K}
{Kramer}, M., {Lyne}, A.~G., {O'Brien}, J.~T., {Jordan}, C.~A., \& {Lorimer},
  D.~R. 2006, Science, 312, 549

\bibitem[{{Lam}(2017)}]{2017ascl.soft06011L}
{Lam}, M.~T. 2017, PyPulse, Astrophysics Source Code Library, , , ascl:1706.011

\bibitem[{{Lam} {et~al.}(2016{\natexlab{a}}){Lam}, {Cordes}, {Chatterjee},
  {Jones}, {McLaughlin}, \& {Armstrong}}]{2016ApJ...821...66L}
{Lam}, M.~T., {Cordes}, J.~M., {Chatterjee}, S., {et~al.} 2016{\natexlab{a}},
  \apj, 821, 66

\bibitem[{{Lam} {et~al.}(2016{\natexlab{b}}){Lam}, {Cordes}, {Chatterjee},
  {Arzoumanian}, {Crowter}, {Demorest}, {Dolch}, {Ellis}, {Fonseca},
  {Gonzalez}, {Jones}, {Jones}, {Levin}, {Madison}, {McLaughlin}, {Nice},
  {Pennucci}, {Ransom}, {Siemens}, {Stairs}, {Stovall}, {Swiggum}, \&
  {Zhu}}]{2016ApJ...819..155L}
---. 2016{\natexlab{b}}, \apj, 819, 155

\bibitem[{{Lentati} {et~al.}(2015){Lentati}, {Alexander}, \&
  {Hobson}}]{2015MNRAS.447.2159L}
{Lentati}, L., {Alexander}, P., \& {Hobson}, M.~P. 2015, \mnras, 447, 2159

\bibitem[{{Levin} {et~al.}(2016){Levin}, {McLaughlin}, {Jones}, {Cordes},
  {Stinebring}, {Chatterjee}, {Dolch}, {Lam}, {Lazio}, {Palliyaguru},
  {Arzoumanian}, {Crowter}, {Demorest}, {Ellis}, {Ferdman}, {Fonseca},
  {Gonzalez}, {Jones}, {Nice}, {Pennucci}, {Ransom}, {Stairs}, {Stovall},
  {Swiggum}, \& {Zhu}}]{2016ApJ...818..166L}
{Levin}, L., {McLaughlin}, M.~A., {Jones}, G., {et~al.} 2016, \apj, 818, 166

\bibitem[{{Lorimer} {et~al.}(2012){Lorimer}, {Lyne}, {McLaughlin}, {Kramer},
  {Pavlov}, \& {Chang}}]{2012ApJ...758..141L}
{Lorimer}, D.~R., {Lyne}, A.~G., {McLaughlin}, M.~A., {et~al.} 2012, \apj, 758,
  141

\bibitem[{{Lyne} {et~al.}(2010){Lyne}, {Hobbs}, {Kramer}, {Stairs}, \&
  {Stappers}}]{2010Sci...329..408L}
{Lyne}, A., {Hobbs}, G., {Kramer}, M., {Stairs}, I., \& {Stappers}, B. 2010,
  Science, 329, 408

\bibitem[{{Lyne} {et~al.}(1971){Lyne}, {Smith}, \&
  {Graham}}]{1971MNRAS.153..337L}
{Lyne}, A.~G., {Smith}, F.~G., \& {Graham}, D.~A. 1971, \mnras, 153, 337

\bibitem[{{Lyne} {et~al.}(2017){Lyne}, {Stappers}, {Freire}, {Hessels},
  {Kaspi}, {Allen}, {Bogdanov}, {Brazier}, {Camilo}, {Cardoso}, {Chatterjee},
  {Cordes}, {Crawford}, {Deneva}, {Ferdman}, {Jenet}, {Knispel}, {Lazarus},
  {van Leeuwen}, {Lynch}, {Madsen}, {McLaughlin}, {Parent}, {Patel}, {Ransom},
  {Scholz}, {Seymour}, {Siemens}, {Spitler}, {Stairs}, {Stovall}, {Swiggum},
  {Wharton}, \& {Zhu}}]{2017ApJ...834...72L}
{Lyne}, A.~G., {Stappers}, B.~W., {Freire}, P.~C.~C., {et~al.} 2017, \apj, 834,
  72

\bibitem[{{Manchester} {et~al.}(2013){Manchester}, {Hobbs}, {Bailes}, {Coles},
  {van Straten}, {Keith}, {Shannon}, {Bhat}, {Brown}, {Burke-Spolaor},
  {Champion}, {Chaudhary}, {Edwards}, {Hampson}, {Hotan}, {Jameson}, {Jenet},
  {Kesteven}, {Khoo}, {Kocz}, {Maciesiak}, {Oslowski}, {Ravi}, {Reynolds},
  {Sarkissian}, {Verbiest}, {Wen}, {Wilson}, {Yardley}, {Yan}, \&
  {You}}]{2013PASA...30...17M}
{Manchester}, R.~N., {Hobbs}, G., {Bailes}, M., {et~al.} 2013, \pasa, 30, e017

\bibitem[{{McLaughlin}(2013)}]{2013CQGra..30v4008M}
{McLaughlin}, M.~A. 2013, Classical and Quantum Gravity, 30, 224008

\bibitem[{{Michilli} {et~al.}(2018){Michilli}, {Hessels}, {Donner},
  {Grie{\ss}meier}, {Serylak}, {Shaw}, {Stappers}, {Verbiest}, {Deller},
  {Driessen}, {Stinebring}, {Bondonneau}, {Geyer}, {Hoeft}, {Karastergiou},
  {Kramer}, {Os{\l}owski}, {Pilia}, {Sanidas}, \&
  {Weltevrede}}]{2018MNRAS.476.2704M}
{Michilli}, D., {Hessels}, J.~W.~T., {Donner}, J.~Y., {et~al.} 2018, \mnras,
  476, 2704

\bibitem[{{Pennucci} {et~al.}(2014){Pennucci}, {Demorest}, \&
  {Ransom}}]{2014ApJ...790...93P}
{Pennucci}, T.~T., {Demorest}, P.~B., \& {Ransom}, S.~M. 2014, \apj, 790, 93

\bibitem[{{Ramachandran} {et~al.}(2006){Ramachandran}, {Demorest}, {Backer},
  {Cognard}, \& {Lommen}}]{2006ApJ...645..303R}
{Ramachandran}, R., {Demorest}, P., {Backer}, D.~C., {Cognard}, I., \&
  {Lommen}, A. 2006, \apj, 645, 303

\bibitem[{Rasmussen \& Williams(2006)}]{rasmussen2006gaussian}
Rasmussen, C.~E., \& Williams, C.~K.~I. 2006, Gaussian Processes for Machine
  Learning (Cambridge, MA: MIT Press)

\bibitem[{{Rathnasree} \& {Rankin}(1995)}]{1995ApJ...452..814R}
{Rathnasree}, N., \& {Rankin}, J.~M. 1995, \apj, 452, 814

\bibitem[{{Rickett}(1990)}]{1990ARA&A..28..561R}
{Rickett}, B.~J. 1990, \araa, 28, 561

\bibitem[{Roberts {et~al.}(2012)Roberts, Osborne, Ebden, Reece, Gibson, \&
  Aigrain}]{Roberts20110550}
Roberts, S., Osborne, M., Ebden, M., {et~al.} 2012, Philosophical Transactions
  of the Royal Society of London A: Mathematical, Physical and Engineering
  Sciences, 371, 1984

\bibitem[{{Shannon} \& {Cordes}(2010)}]{2010ApJ...725.1607S}
{Shannon}, R.~M., \& {Cordes}, J.~M. 2010, \apj, 725, 1607

\bibitem[{{Shannon} \& {Cordes}(2012)}]{2012ApJ...761...64S}
---. 2012, \apj, 761, 64

\bibitem[{{Shannon} {et~al.}(2013){Shannon}, {Cordes}, {Metcalfe}, {Lazio},
  {Cognard}, {Desvignes}, {Janssen}, {Jessner}, {Kramer}, {Lazaridis},
  {Purver}, {Stappers}, \& {Theureau}}]{2013ApJ...766....5S}
{Shannon}, R.~M., {Cordes}, J.~M., {Metcalfe}, T.~S., {et~al.} 2013, \apj, 766,
  5

\bibitem[{{Shannon} {et~al.}(2014){Shannon}, {Os{\l}owski}, {Dai}, {Bailes},
  {Hobbs}, {Manchester}, {van Straten}, {Raithel}, {Ravi}, {Toomey}, {Bhat},
  {Burke-Spolaor}, {Coles}, {Keith}, {Kerr}, {Levin}, {Sarkissian}, {Wang},
  {Wen}, \& {Zhu}}]{2014MNRAS.443.1463S}
{Shannon}, R.~M., {Os{\l}owski}, S., {Dai}, S., {et~al.} 2014, \mnras, 443,
  1463

\bibitem[{{Shannon} {et~al.}(2016){Shannon}, {Lentati}, {Kerr}, {Bailes},
  {Bhat}, {Coles}, {Dai}, {Dempsey}, {Hobbs}, {Keith}, {Lasky}, {Levin},
  {Manchester}, {Os{\l}owski}, {Ravi}, {Reardon}, {Rosado}, {Spiewak}, {van
  Straten}, {Toomey}, {Wang}, {Wen}, {You}, \& {Zhu}}]{2016ApJ...828L...1S}
{Shannon}, R.~M., {Lentati}, L.~T., {Kerr}, M., {et~al.} 2016, \apjl, 828, L1

\bibitem[{{Stairs} {et~al.}(2000){Stairs}, {Lyne}, \&
  {Shemar}}]{2000Natur.406..484S}
{Stairs}, I.~H., {Lyne}, A.~G., \& {Shemar}, S.~L. 2000, \nat, 406, 484

\bibitem[{Taylor(1992)}]{Taylor117}
Taylor, J.~H. 1992, Philosophical Transactions of the Royal Society of London
  A: Mathematical, Physical and Engineering Sciences, 341, 117

\bibitem[{{Taylor} {et~al.}(1975){Taylor}, {Manchester}, \&
  {Huguenin}}]{1975ApJ...195..513T}
{Taylor}, J.~H., {Manchester}, R.~N., \& {Huguenin}, G.~R. 1975, \apj, 195, 513

\bibitem[{{van Straten}(2004)}]{2004ApJS..152..129V}
{van Straten}, W. 2004, \apjs, 152, 129

\bibitem[{{van Straten}(2006)}]{2006ApJ...642.1004V}
---. 2006, \apj, 642, 1004

\bibitem[{{van Straten}(2013)}]{2013ApJS..204...13V}
---. 2013, \apjs, 204, 13

\bibitem[{{Walker} {et~al.}(2013){Walker}, {Demorest}, \& {van
  Straten}}]{2013ApJ...779...99W}
{Walker}, M.~A., {Demorest}, P.~B., \& {van Straten}, W. 2013, \apj, 779, 99

\end{thebibliography}

\end{document}